\documentclass[aps,prd,amsmath,superscriptaddress,amssymb,showpacs,floatfix,nofootinbib,longbibliography, twocolumn]{revtex4-1} 
\pdfoutput=1
\usepackage{graphicx}
\usepackage{bm}
\usepackage{times}
\usepackage[colorlinks = true,
            linkcolor = blue,
            urlcolor  = blue,
            citecolor = blue,
            anchorcolor = blue]{hyperref}
\usepackage{slashed}
\usepackage{color}
\usepackage{aas_macros}
\usepackage[dvipsnames]{xcolor}
\usepackage{slashed}
\usepackage{lipsum}
\usepackage{subfigure}
\usepackage{multirow}
\usepackage{amsmath}
\usepackage{array} 
\usepackage{varwidth} 
\usepackage{listings}
\usepackage{amssymb}
\usepackage{comment}

\newcommand{\diff}{\mathrm{d}}

\newcommand{\s}{\text{s}}
\newcommand{\cm}{\text{cm}}
\newcommand{\magn}{\text{mag}}
\newcommand{\K}{\text{K}}
\newcommand{\kmsec}{\text{km}\,\text{s}^{-1}}
\newcommand{\km}{\text{km}}
\newcommand{\fWNM}{f_\text{WNM}}

\newcommand{\update}[1]{#1}

\usepackage[normalem]{ulem}

\begin{document}

\title{Weighing the Local Interstellar Medium using Gamma Rays and Dust}
\author{Axel Widmark}
\email{axel.widmark@nbi.ku.dk, ORCID: orcid.org/0000-0001-5686-3743}
\affiliation{Dark Cosmology Centre, Niels Bohr Institute, University of Copenhagen, Jagtvej 128, 2200 Copenhagen N, Denmark}
\author{Michael Korsmeier}
\email{michael.korsmeier@fysik.su.se, ORCID: orcid.org/0000-0003-3478-888X}
\affiliation{Stockholm University and The Oskar Klein Centre for Cosmoparticle Physics,  Alba Nova, 10691 Stockholm, Sweden}
\author{Tim Linden}
\email{linden@fysik.su.se, ORCID: orcid.org/0000-0001-9888-0971}
\affiliation{Stockholm University and The Oskar Klein Centre for Cosmoparticle Physics,  Alba Nova, 10691 Stockholm, Sweden}

\begin{abstract}
Cold gas forms a significant mass fraction of the Milky Way disk, but is its most uncertain baryonic component. The density and distribution of cold gas is of critical importance for Milky Way dynamics, as well as models of stellar and galactic evolution. Previous studies have used correlations between gas and dust to obtain high-resolution measurements of cold gas, but with large normalization uncertainties. We present a novel approach that uses \emph{Fermi}-LAT $\gamma$-ray data to measure the total gas density, achieving a similar precision as previous works, but with independent systematic uncertainties. Notably, our results have sufficient precision to probe the range of results obtained by current world-leading experiments. 
\end{abstract}

\maketitle

Measuring the gas content of the Milky Way and local universe is crucial for many fields in astronomy. Cold gas is the most uncertain baryonic component in Milky Way mass models, hampering the precision of dynamical mass measurements of dark matter \citep{Read2014,2021RPPh...84j4901D}. It is also essential for our theoretical understanding of star and galaxy formation \cite{1976ApJ...205..762S,2022arXiv220200690S}. 

The dominant gas species of the interstellar medium (ISM) are atomic and molecular hydrogen (HI and H$_2$), where especially the latter is difficult to observe directly as it lacks a permanent electric dipole moment. CO observations are often used as a tracer \cite{2015ARA&A..53..583H}. However, the CO-to-H$_2$ ratio is uncertain and depends on the temperature and density of their environment \cite{2013ARA&A..51..207B}. A significant fraction of H$_2$ is known to be CO-dark, but the precise amount is poorly constrained \cite{2005Sci...307.1292G,2010ApJ...716.1191W,2016A&A...593A..42T,2017ApJ...834...63R}.

Gas is additionally traced by dust \cite{2003ARA&A..41..241D}, which is typically easier to observe and serves as a useful proxy for a galaxy's gas content \cite{2007ApJ...657..810D,2014ApJ...783...84S,2014A&A...563A..31R,Gordon_2014,10.1093/mnras/stz2311,2018MNRAS.481.3497S}. Dust is also of direct physical importance for a number of
complex thermal and chemical processes of the ISM, affecting star and planet formation \cite{2016SAAS...43...85K}. In the Milky Way, the dust-to-gas ratio has been estimated using gas observations coming from stellar sight-line UV absorption \citep{1977ApJ...216..291S,1978ApJ...224..132B,shull2021},
soft X-ray scattering \citep{1995A&A...293..889P}, or 21 cm emission \citep{Liszt_2013,2017ApJ...846...38L,2020A&A...639A..26K} (see Appendix~\ref{app:previous_studies} for further details).

\begin{figure}
    \centering
    \includegraphics[width=0.92\columnwidth]{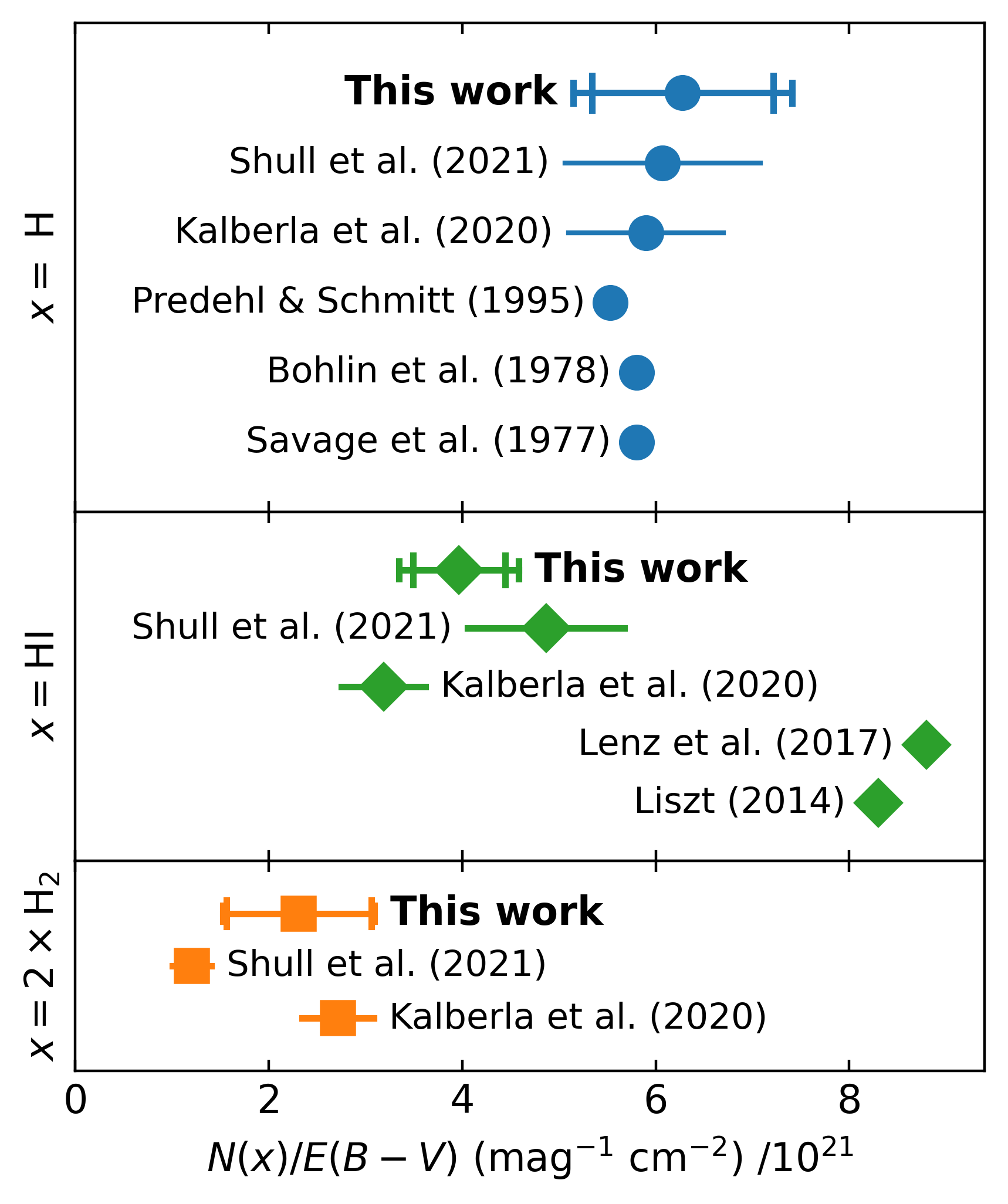}
    \vspace{-0.1cm}
    \caption{Ratio between $E(B-V)$ dust reddening and gas, as determined by our $\gamma$-ray analysis. The three panels depict total hydrogen (H), and its atomic (HI) and molecular (H$_2$) species. The inner error bars of our results correspond to the uncertainty of our fit, while the outer errors bars include a 10\% systematic uncertainty on the $\gamma$-ray cross sections added in quadrature. Our results are consistent with Refs.~\cite{1977ApJ...216..291S,1978ApJ...224..132B,1995A&A...293..889P,2020A&A...639A..26K}, but in tension with Refs.~\cite{Liszt_2013, 2017ApJ...846...38L}.}
    \label{fig:studies_summary}
    \vspace{-0.1cm}
\end{figure}

In this \emph{letter}, we develop a novel method for measuring the \update{local} Milky Way gas content, \update{within roughly 2~kpc}, using $\gamma$-ray data from the \emph{Fermi}-LAT telescope. The diffuse $\gamma$-ray flux is produced primarily by the hadronic interactions of cosmic rays (CRs) with interstellar gas. Because CR measurements ({\it e.g.}, by AMS-02), strongly constrain the local cosmic-ray density, $\gamma$-ray measurements can, in turn, be used to strongly constrain the gas density. Importantly, the hadronic interactions that produce $\gamma$-ray emission are entirely agnostic as to the temperature, spin, or ionization state of the gas targets, making our gas density calculations highly complementary to the techniques described above. Furthermore, we use a new dust map \citep{2019A&A...625A.135L} that was recently produced from observations by the astrometric \emph{Gaia} mission \citep{2018A&A...616A...1G}, and has the novel quality of being three dimensional and (unlike most previous studies) observed in the optical rather than the far infra-red.

\begin{figure*}
    \centering
    \includegraphics[width=.94\textwidth,trim={1cm 16.0cm 9cm 0.5cm},clip]{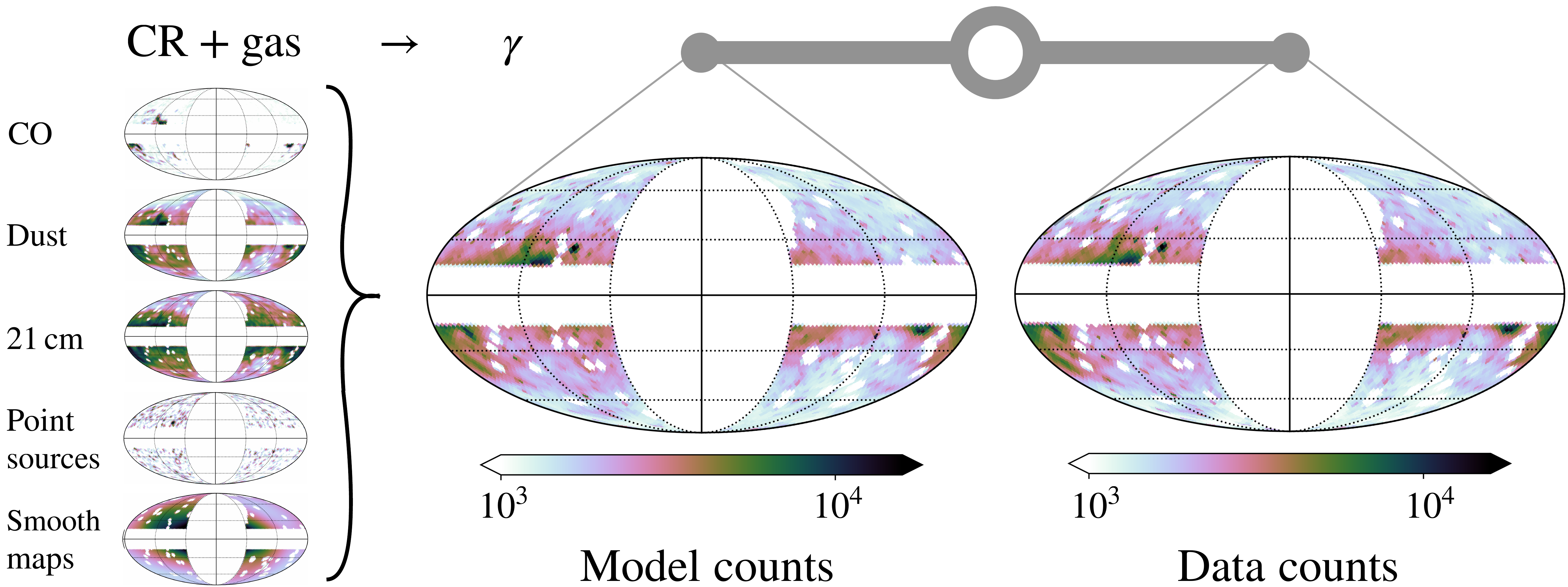}
    \caption{
       $\gamma$-rays are produced by the interaction of CRs with gas, as well as a few other processes like inverse Compton emission or point sources. We perform a template fit of the local $\gamma$-ray sky by comparing model counts and \emph{Fermi}-LAT data. We note that maps are corrected for the \emph{Fermi}-LAT instrumental exposure, which varies across the sky. 
       \update{We use cuts of $|l| < 60^\circ$  to avoid the Galactic center and the \emph{Fermi} bubbles, and $|b| < 16^\circ$ to select nearby emission sources.}
       From the normalization of each template, we infer the conversion factors of the gas tracers to gas densities, most notably, from dust reddening to cold hydrogen gas. 
       }
    \label{fig:setupfig}
\end{figure*}

\update{
Figure~\ref{fig:studies_summary} shows the main results of our study.}
Our measurements of the dust-to-gas ratio have a similar precision as previous work, but are affected by an independent set of systematic uncertainties. Excitingly, our results are precise enough to probe the \update{range of results from} state-of-the-art measurements. 
Furthermore, our study produces a novel and accurate model for high-latitude diffuse $\gamma$-ray emission from the Milky Way,%
\footnote{\update{The templates are made available upon request.}}
with many applications to studies of extragalactic $\gamma$-ray emission and extended $\gamma$-ray sources.

\noindent {\bf $\boldsymbol \gamma$-Ray Emission Models ---} 
$\gamma$-rays provide a unique probe into galactic gas. Here, we focus on 0.1--100~GeV $\gamma$-rays observed at high Galactic latitudes. These are created by several processes, including the interaction of CRs with gas~\cite{Fermi-LAT:2012edv,Fermi-LAT:2014ryh}. The most important process is $\pi^0$-production from hadronic interactions, while the bremsstrahlung of CR electrons provides a $\mathcal{O}$(10\%) contribution below a few GeV \cite{Tibaldo:2021viq,Porter:2017vaa,Johannesson:2018bit,Kissmann:2017ghg,Dundovic:2021ryb}. 

The $\gamma$-ray flux produced from a CR component $i$ and a gas component $j$ is given by the integration of the $\gamma$-ray emission per interaction $\epsilon^{ij}$ and the number density of the gas component $\rho_j$ along the line of sight (l.o.s.) \update{in the direction $l$, $b$ (longitude, latitude)}:
\begin{eqnarray}
    \label{eq:gamma_ray-flux}
    \frac{\diff^2 \phi_\gamma^{ij}}{\diff \Omega \diff E_\gamma} (E_\gamma, l, b) 
    = 
    \int\limits_{\text{l.o.s.}} \;\diff \ell 
    \rho_j(\boldsymbol{x})  \epsilon^{ij}(E_\gamma, \boldsymbol{x}) \, , 
\end{eqnarray}
where the emission per interaction is given by convolving the CR flux $\phi^i_{\rm CR}$ with the $\gamma$-ray production cross section $\sigma_{ij\rightarrow\gamma}$:
\begin{eqnarray}
    \label{eq:gamma_ray-flux_2}
    \epsilon^{ij}(E_\gamma, \boldsymbol{x}) = 
    \int \diff E_i
      \frac{\diff \phi_{\rm CR}^i}{\diff E_i}(E_i, \boldsymbol{x}) 
      \frac{\diff \sigma_{ij\rightarrow\gamma}}{\diff E_\gamma}(E_i, E_\gamma) \,.
\end{eqnarray}
The product $\rho_j(\boldsymbol{x}) \epsilon^{ij}(E_\gamma, \boldsymbol{x})$ is commonly called $\gamma$-ray emissivity.
Thus, the flux depends on the CR flux and the total gas density. 
Importantly, the $\gamma$-ray flux is essentially unaffected by whether the gas is cold or hot, molecular or atomic, ionized or neutral. The dominant CR fluxes are protons and helium, while contributions from heavier nuclei are subdominant. The flux of electrons and positrons is also suppressed compared to protons by 2--3 orders of magnitude, but is important for low-energy bremsstrahlung due to its faster cooling timescale \update{and for the inverse Compton emission}.

We employ the propagation model recently explored in Ref. \cite{Korsmeier:2021brc} and refit it to the CR data of $p$, $\rm He$, $\rm ^3He/^4He$, $\bar{p}/p$, and $e^+$ provided by AMS-02. The $e^-$ data is adjusted in a post-procedure. The CR model includes diffusion, convection, continuous energy losses, and fragmentation losses. Our \update{model is based on the \textsc{Galprop} code~\cite{Strong:1998fr}, which numerically solves the propagation equations. It} includes several improvements compared to previous studies. Details are described in Appendix \ref{app:CRs}.

Besides CR-gas interactions, we model the following processes: (1) Galactic and extragalactic $\gamma$-ray point sources \cite{Fermi-LAT:2019yla} (2) the emission from inverse Compton scattering (ICS), \emph{i.e.} the up-scattering of low-energy photons by CR electrons and positrons, and (3) an isotropic emission that stems from faint and unresolved extragalactic point sources, charged CR contamination, and residual Galactic emission. 

We perform template fits to the  \emph{Fermi}-LAT data constraining the number of $\gamma$-ray photons attributed to each process or gas tracer as sketched in Fig.~\ref{fig:setupfig}. Ultimately, we obtain the conversion factor from the gas tracers (21~cm, dust, CO) to the gas densities (H, HI, H$_2$). The gas tracers and the exact fitting procedure are explained below. \\

\noindent {\bf Gas Tracers --- }The ISM is separated into distinct gas phases, broadly divided into a cold and warm neutral medium (CNM and WNM) \cite{1977ApJ...218..148M,2005ARA&A..43..337C,2011piim.book.....D,2022arXiv220200690S}. The different gas components and species are traced by different maps in our model:

\begin{itemize}

\item {\bf CO --- }Very cold molecular gas is traced by radio observations of \update{rotational} modes in carbon monoxide (CO). We use the two-dimensional full-sky CO data from the Planck survey (\update{the ``\texttt{Type 2}'' map from Ref.}~\cite{2014A&A...571A..13P}). Canonically, the conversion factor between the CO $J = 1\rightarrow 0$ transition signal and the H$_2$ column density is equal to $(2\pm 0.6)\times 10^{20}~\cm^{-2}\,\K^{-1}\,\km^{-1}\,\s$ \cite{1978ApJS...37..407D,2013ARA&A..51..207B}. The component of CO-traced H$_2$ has a small scale height, and thus, due to our cut on Galactic latitude, its relative contribution is small.

\item {\bf Dust --- }We use a three-dimensional dust map to trace molecular and atomic hydrogen in the CNM. This dust map was produced using \emph{Gaia} data \citep{2019A&A...625A.135L}, given in terms of \update{extinction per distance in units $\magn \, \text{pc}^{-1}$,} at a wavelength of 6500~Å (corresponding to the centre of \emph{Gaia}'s photometric $G$-band); this \update{extinction} is proportional to $E(B-V)$ by a factor of $0.88 \times 3.1 = 2.73$ \cite{2019ApJ...887...93G,2019A&A...625A.135L}. The conversion factor between $E(B-V)$ dust reddening and the hydrogen nuclei column density is roughly $6\times 10^{21}~\magn^{-1} \, \cm^{-2}$, although the precise value is highly uncertain (see Fig.~\ref{fig:studies_summary}). The dust distribution of this map has a small scale height. Most of it is contained within 100~pc from the mid-plane, with only weak tails to heights around 300~pc. Hence it is inconsistent with the $\sim 400~\text{pc}$ scale height of the WNM, and we make the interpretation that this dust map only traces the CNM.

\item {\bf 21 cm --- }We use the two-dimensional full sky 21 cm map called HI4PI \cite{2016A&A...594A.116H}, integrated over l.o.s. velocities between $(-90,90)~\kmsec$, to trace atomic hydrogen in the CNM and WNM. This map overlaps with the dust map, since both trace CNM HI. Canonically, the conversion factor between the 21 cm signal and the HI column density is equal to $1.823\times 10^{18}~\cm^{-2}\,\K^{-1}\,\km^{-1}\,\s$ \cite{21cm_conv_fact,2016A&A...594A.116H}. We assume that the 21 cm optical depth is negligible for our Galactic latitude cuts. This assumption breaks down for $|b|\lesssim 10^\circ$ \cite{2018ApJS..238...14M}, which is smaller than our latitude cut.

\item {\bf Analytic WNM distribution --- }As an alternative to the 21 cm map, we consider an analytic model for the WNM, with a density proportional to:
\begin{equation}\label{eq:analytic_WNM}
    f_\text{WNM}(\boldsymbol{x}) \propto \text{sech}^2\left(\frac{z}{400~\text{pc}}\right) \exp\left(-\frac{R}{3.5~\text{kpc}}\right) \, ,
\end{equation}
where $z$ is the height with respect to the disk and $R$ is Galactocentric radius. This functional form reflects the fact that the WNM forms less significant substructure due to its high temperature, such that its distribution in the Solar neighbourhood is well-approximated by a 400~pc scale height and a 3.5~kpc disk scale length \citep{2009ARA&A..47...27K}.
\end{itemize}

\update{The precise relationship between gas components and gas tracers is complex, where some gas components are covered by multiple tracers. In our modeling, we take the CO to trace the surplus of H$_2$ that is otherwise not accounted for by the dust tracer.}
We perform fits for two separate models, labeled A and B, which are summarized in Table~\ref{tab:gas_components}. Model A includes the 21 cm map but not the $f_\text{WNM}$ map, and vice versa for model B. The crucial difference is that the gas tracers have an overlap in terms of the CNM HI component in model A, but not for model B. By comparing their respective results we can estimate the amount of total hydrogen, HI, and CO-dark H$_2$ in the CNM. Model B is expected to perform slightly worse in terms of a $\chi^2$ statistic, since $f_\text{WNM}$ does not capture any WNM substructures. For this reason it is also highly degenerate with other smooth components (the ICS and isotropic background, see below). However, this is not detrimental to our fit, as the purpose of model B is to extract more information about the dust-to-gas conversion factor of the CNM, for which the ignored WNM substructures have a negligible effect.

\update{We employ the canonical assumption that all gas phases are mixed with the same fraction of helium}, constituting 38\% of the hydrogen gas mass, with an additional 4\% mass from heavier elements \cite{1988ApJ...324..248B,2001RvMP...73.1031F,Kramer_2016}. The $\gamma$-ray production per mass is approximately the same for $\pi^0$ while bremsstrahlung from heavier targets is slightly suppressed, as described in Appendix~\ref{app:results}. In our model, we ignore the hot interstellar medium (HIM), which constitutes a sub-percent contribution to the local gas density and, despite its large scale height, a small contribution to the total gas surface density \cite{2015ApJ...814...13M}. The HIM lacks smaller scale spatial structure, so its contribution to the $\gamma$-ray sky is absorbed by other smooth $\gamma$-ray emission components (listed below), without affecting the normalizations of the cold gas in our fit.
\\

\begin{table}
\centering
{\renewcommand{\arraystretch}{1.2}
\begin{tabular}{c|c|c|c|c|c}
    \multicolumn{1}{c}{} & \multicolumn{1}{c}{} & CO H$_2$   & CNM H$_2$     & CNM HI    & WNM HI          \\
    \hline
    \hline
    \multirow{3}{*}{\rotatebox[origin=c]{45}{\parbox[c]{1.2cm}{\centering Model A}}}
    & CO        & \checkmark     & \hspace{1.2cm} & \hspace{1.2cm}  & \hspace{1.2cm} \\
    & Dust      & \hspace{1.2cm} & \checkmark     & \checkmark      &               \\
    & 21 cm     &                &                & \checkmark      & \checkmark    \\
    \hline
    \hline
    \multirow{3}{*}{\rotatebox[origin=c]{45}{\parbox[c]{1.2cm}{\centering Model B}}}
    & CO        & \checkmark     &                &                 &               \\
    & Dust      &                & \checkmark     & \checkmark      &               \\
    & $\fWNM$   &                &                &                 & \checkmark    \\
\end{tabular}
}
\caption{Gas components (top row) and the respective maps (left column) they are traced by in model A and model B.}
\label{tab:gas_components}
\end{table}

\noindent {\bf Other $\gamma$-ray Emission Components --- }While $\gamma$-rays from \mbox{$\pi^0$-decay} and bremsstrahlung both trace Galactic gas, there are several mechanisms that do not depend on the gas density. First, \emph{Fermi}-LAT observations include over 5000~point sources of both Galactic and extragalactic origin, including supernova remnants, pulsars, blazars, and star-forming galaxies. The large number of sources makes it difficult to model the intensity and spectrum of each independently. Thus, we use the default flux and spectra for each source reported in the 4FGL-DR2 catalog~\cite{Ballet:2020hze,Fermi-LAT:2019yla}. This provides an accurate model for non-variable sources. In contrast, we mask the bright and time-variable sources as discussed in Appendix~\ref{app:sources}. 

The second component is the isotropic $\gamma$-ray background, which is primarily produced by the subset of extragalactic sources that are too dim to be individually detected. It also includes a contribution from cosmic rays that are misidentified as $\gamma$-rays by the \emph{Fermi}-LAT. Because both extragalactic sources and cosmic rays are isotropic, the morphological template in each bin matches the \emph{Fermi}-LAT exposure.

The final component is the ICS of starlight by relativistic electrons. Notably, the same electron population produces the ICS and the bremsstrahlung component discussed above. To calculate the convolution of this electron distribution with Galactic radiation, we utilize the 
\update{CR model described in App.~\ref{app:CRs} and} 
the up-to-date interstellar radiation field models given by~Ref.~\cite{Porter:2017vaa}. \\

\noindent {\bf Fitting Procedure --- }We infer our results in a Bayesian framework, with a Poisson count data likelihood and flat box priors on the normalizations of each respective $\gamma$-ray map component, and sample the posterior probability density using Hamiltonian Monte-Carlo. We produce our main results in a joint fit of the 4th--15th energy bin (0.4--100~GeV), which are dominated by hadronic interactions and therefore they are less prone to systematic errors related to bremsstrahlung cross sections, as well as \emph{Fermi}'s angular resolution, which improves at high energies. We employ jackknife sub-sampling by splitting the sky into sub-areas in order to quantify systematic uncertainties, and also perform fits for each separate $\gamma$-ray energy bin. See Appendix~\ref{app:results} for further details.

Because we focus our analysis on the local ISM, we exclude regions of the sky where the $\gamma$-ray data includes significant systematic uncertainties due to distant sources. These include: (1) regions along the Galactic plane with $|b|<16^\circ$, (2) regions near the Galactic center with $|l| < 60^\circ$, which removes contamination from the Fermi bubbles~\cite{Su:2010qj} and Loop I~\cite{2009arXiv0912.3478C}, (3) a $3^\circ$ \update{radius circle} around 17 highly-variable \emph{Fermi}-LAT point sources, for which the quoted 4FGL fluxes may not produce proper flux estimates, and (4) a 5$^\circ$ \update{radius circle} surrounding the position of the Large Magellanic Cloud. \update{The size of these masks are motivated by the \emph{Fermi}-LAT point spread function, which has a $2.93^\circ$ 68\% containment region for our lowest energy bin.} 
\\

\begin{figure}[b]
    \centering
    \includegraphics[width=1.\columnwidth,trim={0cm 0.2cm 0cm 0.2cm},clip]{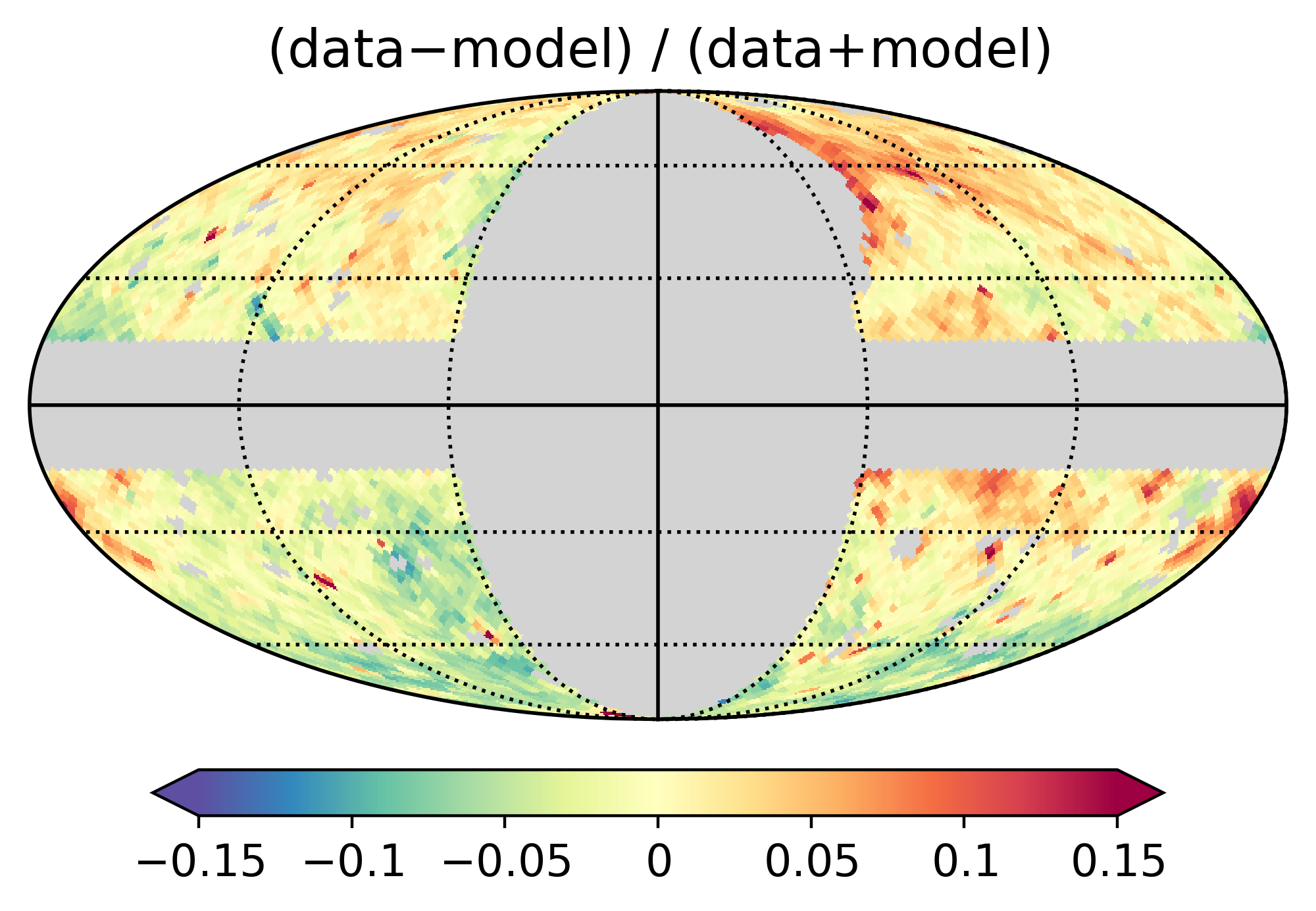}
    \caption{Comparison of our model A with the data counts, from the best fit over the 4th--15th energy bins. Masked regions are gray.}
    \label{fig:data_model_frac}
\end{figure}

\noindent {\bf Results --- }In Fig.~\ref{fig:data_model_frac}, we show a comparison between the data count and the best-fit model A, for a fit of the full non-masked sky over the 4th--15th energy bins \update{(for individual energy bins, see App.~\ref{app:results})}. The higher energy bins have limited statistics, so the fit is mainly driven by roughly the 4th--7th energy bins. Our model provides an overall good description of the data. However, some structures are not fully captured by our model. The mean absolute relative difference between the model and data counts in the non-masked region is 5.9\% and 9.5\% for models A and B, beyond what can be accounted for by statistical noise. Hence, our result is strongly dominated by systematic uncertainties, which we estimate by Jackknife sub-sampling.

The main focus of our results concerns the dust conversion factors, as presented in Fig.~\ref{fig:studies_summary}. A more complete visual representation of our results, including the conversion factors for the 21 cm and CO maps and their energy dependence, can be found in Appendix~\ref{app:results}. The dust-to-H conversion factors differ significantly between the model A and model B fits. This is expected, as the 21 cm map of model A traces CNM HI and thus overlaps with the dust map. However, despite the CNM HI already being accounted for, the dust map is still inferred to contribute significantly to the $\gamma$-ray sky in model A, with a dust-to-H conversion factor of $(2.31 \pm 0.75)\times 10^{21}~\magn^{-1}\,\cm^{-2}$ (posterior mean and standard deviation). Thus the dust map must also trace a component that is not accounted for in the 21 cm or CO maps. We make the interpretation that this additional component is CO-dark H$_2$ mixed in with the CNM. For our model B, which does not include the 21 cm map, the dust map is interpreted to trace the totality of both CNM HI and CNM H$_2$, with a dust-to-H conversion factor of $(6.28 \pm 0.94)\times 10^{21}~\magn^{-1}\,\cm^{-2}$. By taking the dust conversion factor difference between model A and model B, we obtain the amount of CNM HI traced by dust, inferred to be $(3.97 \pm 0.48)\times 10^{21}~\magn^{-1}\,\cm^{-2}$. 
This difference has a smaller uncertainty than the individual conversion factors because we account for correlations in the jackknife sub-sampling. 
The relative uncertainties for the dust-to-H conversion factor is roughly 12\%, on the same order of magnitude as the systematic errors seen in the model and data comparison of Fig.~\ref{fig:data_model_frac}.

For model A, we infer a 21 cm conversion factor of $(1.816 \pm 0.153) \times 10^{18}~\cm^{-2}\,\K^{-1}\,\km^{-1}\,\s$, consistent with the canonical value of $1.823\times 10^{18}~\cm^{-2}\,\K^{-1}\,\km^{-1}\,\s$ \cite{21cm_conv_fact,2016A&A...594A.116H}. This agreement is in general support of the accuracy of our analysis, and does not indicate any significant bias in for example the $\gamma$-ray cross sections. For the CO-to-H$_2$ conversion factor, we infer $(1.67 \pm 0.27)\times 10^{20}$ and $(0.81 \pm 0.45)\times 10^{20}~\cm^{-2}\,\K^{-1}\,\km^{-1}\,\s$ for model A and B, respectively. These results for model A are consistent with the canonical value of $(2 \pm 0.6)\times 10^{20}~\cm^{-2}\,\K^{-1}\,\km^{-1}\,\s$. The CO-to-H$_2$ conversion factor discrepancy between model A and B implies that there is an overlap between the dust map and the CO map; in other words, the CO-bright H$_2$ is also in part traced by dust. The CO-to-H$_2$ conversion factor is potentially not very well understood \cite{2011MNRAS.412..337G}, perhaps further complicated by its interplay with dust as a tracer. To investigate this in more detail would require a more extensive study, preferably including lower Galactic latitudes with a more significant CO signal.

Another important contribution to baryonic models for the Solar neighborhood is that we quantify the amount of CO-dark H$_2$ that is mixed with the CNM HI, to a mass fraction of $36.0 \pm 8.8\%$. This result agrees well with Ref.~\cite{2020A&A...639A..26K}, which uses a completely different method affected by other systematic uncertainties (see Appendix~\ref{app:previous_studies} for details). Outside CO-bright regions, they find a CO-dark H$_2$ mass fraction of 46\% relative to the total amount of hydrogen, although they do not state any uncertainty for this quantity.

We also perform additional tests, described in Appendix~\ref{app:results}, for example by applying an alternative data likelihood that is less sensitive to outliers. In summary, we obtain very similar results for these alternative fits, even with latitude cuts of $|b|>12^\circ$ and $|b|>8^\circ$, as well as at higher or lower angular resolution. Furthermore, in Appendix~\ref{app:surf_dens_comparisons}, \update{we present a model for the cold gas mass density based on our results, and compare this model with estimates from other studies.}
\\

\noindent {\bf Conclusion --- }\update{We weigh the local ISM (within roughly 2~kpc), focusing on its
cold gas components and, as a first work of its kind, do so using $\gamma$-ray data from Fermi-LAT and a dust map
based on \emph{Gaia} data.}
We employ a conservative treatment of uncertainties and potential systematic errors. \update{Our results for the dust-to-gas conversion factor are robust and have sufficient precision to probe the range of results produced by current world-leading experiments.} Furthermore, our method is highly complementary to these other studies, subject to different sources of potential systematic biases, and especially useful for revealing otherwise difficult to observe gas components such as CO-dark H$_2$.


This method for using $\gamma$-rays to probe the ISM of the Solar neighborhood is expected to improve in the near future. The dust maps are getting deeper and more precise, due to better astrometric measurements from \emph{Gaia}, complementary spectro- and photo-astrometric distance information \cite{2021AJ....161..147B,2022A&A...658A..91A}, and more sophisticated modeling of the three-dimensional distribution of dust (\emph{e.g.} employing Gaussian processes \cite{2022arXiv220206797M,2022arXiv220411715L}). Further improvements might be achieved by a better understanding and modeling of the cross sections for hadronic $\gamma$-ray production using the latest data from high-energy experiments.
\\

\section*{Acknowledgements}
We thank Adam Leroy, John Beacom, and Martin Rey for useful discussions. AW is supported by the Carlsberg Foundation via a Semper Ardens grant (CF15-0384). MK and TL are supported by the Swedish Research Council under contract 2019-05135 and the European Research Council under grant 742104.
TL is also supported by the Swedish National Space Agency under contract 117/19.
This project used computing resources from the Swedish National Infrastructure for Computing (SNIC) under project Nos. 2021/3-42, 2021/6-326 and 2021-1-24 partially funded by the Swedish Research Council through grant no. 2018-05973. This work made use of an HPC facility funded by a grant from VILLUM FONDEN (projectnumber 16599).
This work has made use of the following Python packages: 
\textsc{NumPy}       \cite{harris2020array}; 
\textsc{SciPy}       \cite{2020SciPy-NMeth}; 
\textsc{TensorFlow}  \cite{tensorflow2015-whitepaper}; 
\textsc{pymultinest} \cite{2016ascl.soft06005B}; 
\textsc{iminuit}     \cite{iminuit}; 
\textsc{mpi4py}      \cite{DALCIN20051108}.

\clearpage
\newpage
\onecolumngrid

\appendix

\section{Previous studies}
\label{app:previous_studies}

In this Appendix section, we discuss previous works on measuring the ratio between dust and gas in the Milky Way, which are also summarized in Fig.~\ref{fig:studies_summary}. Early works by Refs.~\cite{1977ApJ...216..291S,1978ApJ...224..132B} used Ly$\alpha$ and H$_2$ absorption observations coming from the \emph{Copernicus} orbital observatory in the optical and UV, although they were limited to a small sample size of roughly one hundred early-type stars. These studies already hinted at the presence of CO-dark H$_2$. In Ref.~\cite{1995A&A...293..889P}, they used X-ray observations of 25 point sources and four supernova remnants coming from ROSAT, where cold gas was inferred from absorption spectra and the dust column density was inferred from X-ray haloes produced via soft scattering. A recent study by Ref.~\cite{shull2021} uses data from the FUSE survey, similar to the early works based on \emph{Copernicus} data. Apart from a dust-to-H conversion factor, they also derive a 20\% mass fraction of H$_2$ with respect to the total amount of hydrogen; this number does not have an associated uncertainty, hence the HI and H$_2$ data points in Fig.~\ref{fig:studies_summary} are given by the total relative uncertainty of the dust-to-H conversion factor.

Other studies have used 21 cm radio observations, either from the LAB survey \citep{Liszt_2013} or the HI4PI survey \citep{2017ApJ...846...38L,2020A&A...639A..26K}. All three of these studies use the dust map from Ref.~\cite{1998ApJ...500..525S}, based on far infrared observations from COBE/DIRBE and IRAS \citep{1984ApJ...278L...1N,1993SPIE.2019..180S}. Dust reddening is typically expressed as $E(B-V)$, corresponding to the difference between photometric bands in the optical range. The conversion from the infra-red has been discussed to have systematic errors \citep{2007PASJ...59..205Y,2014ApJ...786...29S}, for example arising from contamination of background galaxies, which could be especially detrimental to studies that focus on areas of the sky with lower dust reddening \citep{Liszt_2013,2017ApJ...846...38L}. In Ref.~\cite{2020A&A...639A..26K}, they improved previous studies by also incorporating the temperature information of the 21 cm sky. They do so under the assumption that dust is a perfect tracer of the total amount of hydrogen gas, and that the relative amount of atomic molecular gas follows the functional form $[N(\text{HI}) + 2N(\text{H}_2)]/N(HI) \propto -\log T_D$, where $T_D$ is the 21 cm Doppler temperature. For the CNM outside CO-bright regions, they infer an H$_2$ mass fraction of 46\% relative to the total amount of hydrogen; this number does not have an associated uncertainty, hence the HI and H$_2$ data points in Fig.~\ref{fig:studies_summary} are given by the total relative uncertainty of the dust-to-H conversion factor. \update{As can be seen in Fig.~\ref{fig:studies_summary}, our results are consistent with Refs.~\cite{2020A&A...639A..26K,shull2021}, but less so with Refs.~\cite{Liszt_2013, 2017ApJ...846...38L} and our own result; apart from the systematic issues discussed above, it is possible that this has a physical cause, since the former studies focus on higher latitudes and sky regions with lower levels of dust extinction, where it could be that the dust-to-H conversion factor is indeed higher, as suggest by e.g. Ref.~\cite{shull2021}.}

\section{Model of Galactic Cosmic Rays}
\label{app:CRs}

Galactic CRs are trapped by and propagate in the turbulent magnetic fields of our Galaxy. 
Effectively, this process can be described by diffusion, with a diffusion coefficient that depends on the rigidity of the CRs, typically described by a power law. Rigidity is defined as the momentum of the CR divided by the absolute value of its charge. The power law index, commonly called $\delta$, depends on the turbulence model. 
\update{The latest AMS-02 data constrain $\delta$ to about $1/2$ from a few GV to 300 GV (pointing towards Kraichnan turbulence) and with a transition to $\delta \sim 1/3$ above 300 GV (pointing towards Kolmogorov) \cite{Korsmeier:2021bkw,Weinrich:2020ftb}.}
At low energies, CRs are affected by several other processes, most importantly, convection, continuous energy losses, and fragmentation losses. In general, low-energetic CRs can also be reaccelerated by the scattering off the Alfv\'en magnetic waves. However, too strong reacceleration would be problematic for the global energy budget of magnetic turbulence \cite{Drury:2016ubm}. The typical propagation time of Galactic CRs at the GeV energy scale is \update{a few million years \cite{Derome:2019jfs}}.
We distinguish between primary and secondary CRs. Primaries are accelerated by astrophysical sources like supernova remnants or pulsars and then injected into the diffusion halo. The diffusion halo extends a few kpc above and below the Galactic plane. During their propagation, the CRs can interact and fragment in interaction with the ISM and produce $\gamma$-rays, as is discussed in the main text. Another option is that heavier nuclei fragment and produce so-called secondary CRs, a typical example being fragmentation of carbon into boron. Furthermore, radioactive nuclei can decay and produce secondary CRs. If the decay time is at the order of the propagation time, those CRs can be used to break the well-known degeneracy between the normalization of the diffusion coefficient and the height of the diffusion halo; a typical example is the decay of $^{10}$Be to $^{10}$B.

The best measurements of Galactic CRs are provided by the AMS-02 spectrometer on the International Space Station. In the last decade, AMS-02 provided CR spectra between 1 GV and a few TV of nuclei from protons to iron with precision at the level of a few percent \cite{Aguilar:2021tos}. Electrons and positrons are measured individually with comparable precision up to an energy of about 1~TeV \cite{Aguilar:2021tos}.
The AMS-02 data has stimulated various analyses focused on CR propagation \cite{Korsmeier:2016kha,Tomassetti:2017hbe,Liu:2018ujp,Genolini:2019ewc,Weinrich:2020ftb,Evoli:2019wwu,Evoli:2019iih,Boschini:2020jty,Luque:2021nxb,DeLaTorreLuque:2021yfq,Schroer:2021ojh,Zhao:2021yzf}.

We employ the propagation model recently explored in \cite{Korsmeier:2021brc, Korsmeier:2021bkw}, which provides a good fit to the data from protons to oxygen. Here we refit this model to the CR data of $p$, $\rm He$, $\rm ^3He/^4He$, $\bar{p}/p$, $e^+$, and $e^-$  provided by the AMS-02 experiment. Furthermore, we use the Voyager data of $p$ and He. Contributions from heavier nuclei are negligible for this analysis. In this sense, our analysis is similar to previous studies that have used $^2$H, $^3$He, and antiprotons (or subsets) to study CR propagation \cite{Coste:2011jc,Korsmeier:2016kha,Wu:2018lqu}. 
Finally, in our model we take diffusion to be homogeneous and isotropic. We assume spherical symmetry and a steady state solution.
We briefly summarize the main ingredients in the following, while further details are available in Refs.~\cite{Korsmeier:2021brc, Korsmeier:2021bkw}.

\begin{itemize}
    \item 
    The diffusion coefficient is taken as a double-broken power law with breaks at about 5 GV and 300 GV. We use the normalization ($D_0$), the three spectral indices ($\delta_l$, $\delta$, $\delta_h$), and the two breaks ($R_{D,0}$, $R_{D,1}$) as free parameters in the fit. We allow for a smoothing of the first break regulated by the parameter $s_D$. The break at low energies can be ascribed to a damping of the magnetic turbulence at low energies, while the break at 300 GeV reflects a phenomenological description of the observed CR spectra which have a break at this rigidity. The reason to attribute this break to the diffusion coefficient rather than the injection spectra of primary CRs is that the break is more pronounced in secondary CR nuclei \cite{Aguilar:2018njt}. 
    \item
    The primary injection spectra of $p$, $\rm He$ and $e^-$ are given by power laws with different spectral indices  ($\gamma_p$, $\gamma_{\rm He}$, $\gamma_{e^-}$). Heavier primary nuclei are taken with the same spectral index as helium.
    \item
    Convection can transport CRs away from the Galactic plane. The convection velocity ($v_{0,\mathrm{c}}$) is assumed to be constant and perpendicular to the Galactic plane.
    \item 
    Reacceleration can be subdominant and neglected if the damping of turbulence is dominant at low energies. We exclude reacceleration from our model which instead uses a break in the diffusion coefficient.
    \item 
    Recent analyses of the Be/B data from AMS-02 \cite{Evoli:2019iih,Weinrich:2020ftb,Korsmeier:2021brc} and the first preliminary data on the $^{10}$Be/$^{9}$Be \cite{DeromeICRC2021} indicate that the half-height of the diffusion halo has to be larger than at least a few kpc. We adopt $z_\mathrm{h}=6$~kpc. \update{The CR flux as function of $z$ depends on the value of $z_\mathrm{h}$. However, we checked that this only has a negligible impact on our results, because the flux at $z=0$ is fixed by the AMS-02 data and the scale height of gas is very small compared to the diffusion halo. For example, when changing $z_\mathrm{h}$ from 4~kpc to 10~kpc the CR proton flux at $z=0.1$~kpc ($z=0.2$~kpc) changes by less than 0.3 percent (1 percent) for a characteristic energy of 10 GeV. This much smaller than the other systematic uncertainties considered in this work.}
    \item
    Uncertainties in the production of secondary nuclei are a crucial ingredient \cite{Genolini:2018ekk,Luque:2021nxb,DeLaTorreLuque:2021yfq}. We include nuisance parameters for the normalization and slope of the $^3$He following the procedure from Ref.~\cite{Korsmeier:2021brc}.
    \item
    Solar modulation is treated in the force-field approximation \cite{Fisk:1976aw}. We use a force-field potential for $\bar{p}$ that differs from the one adopted for the other nuclei and positrons because Solar modulation is charge-sign dependent. We also allow for a free Solar modulation potential for electrons. 
    \item
    We include continuous energy losses like ionization, bremsstrahlung, synchrotron, \update{and inverse Compton,} as well as catastrophic energy losses like inelastic scattering leading to the production of secondary protons or tertiary antiprotons. 
    \item
    Electrons have a primary and a secondary component as well as a contribution from pulsars. We assume that the source distribution of primary electrons follows the source distribution of SNRs from Ref.~\cite{Green:2015isa}. 
\end{itemize}

The novelty with respect to Refs.~\cite{Korsmeier:2021brc, Korsmeier:2021bkw} is that here we also consider electrons and positrons. For positrons, we include two sources. Below 30 GeV, positrons are dominated by secondary CRs, while at higher energies an additional component is required to fit the data. We follow the most common assumption and attribute this component to primary positrons from pulsars. In detail, we implement a smooth distribution of pulsars with the spatial distribution from  Ref.~\cite{astro-ph/0308501} and the energy spectrum from Ref.~\cite{Hooper:2008kg}. The normalization ($N_\mathrm{pls}$), spectral index ($\gamma_\mathrm{pls}$), and cutoff energy ($E_{c,\mathrm{pls}}$) of the pulsars' energy spectrum are included as free parameters. 

In contrast, electrons are not only produced as secondary CRs and by pulsars but also by supernova remnants. As a consequence, there is a lot of freedom in the source term of electrons and they cannot be used to further constrain propagation. 
Hence, we follow a two-step procedure: In the first step, we perform a joint fit on the CR data of $p$ \cite{Aguilar:2015ooa, 2013Sci...341..150S}, $\rm He$ \cite{Aguilar:2017hno, 2013Sci...341..150S}, $\rm ^3He/^4He$ \cite{Aguilar:2019eiz}, $\bar{p}/p$ \cite{Aguilar:2016kjl}, and $e^+$ \cite{AMS:2014xys,AMS:2019rhg}, in order to obtain joint source spectrum and propagation constraints. We note that these nuclei and positrons form a self-consistent subset, in the sense that the secondary CRs $\bar p$, $^3$He, and $e^+$ are predominantly produced by $p$ and He. 
Then, in the second step, we fix the best-fit parameters and only adjust the injection spectrum of $e^-$ from supernova remnants, finding that a single power law is sufficient to describe the AMS-02 data. We adopt the supernova-remnant distribution from \cite{Green:2015isa}.
The electron data is taken from \cite{AMS:2014xys}. For secondary electrons and positrons, we use the default production cross sections from \textsc{Galprop} \cite{Strong:1998fr}. The normalization of the secondary positron cross section ($A_{{\rm XS},e+}$) is treated as a nuisance parameter in the fit.

\begin{figure}[t]
    \centering
    \includegraphics[width=0.5\columnwidth,trim={3.3cm 1.7cm 3.5cm 2.7cm},clip]{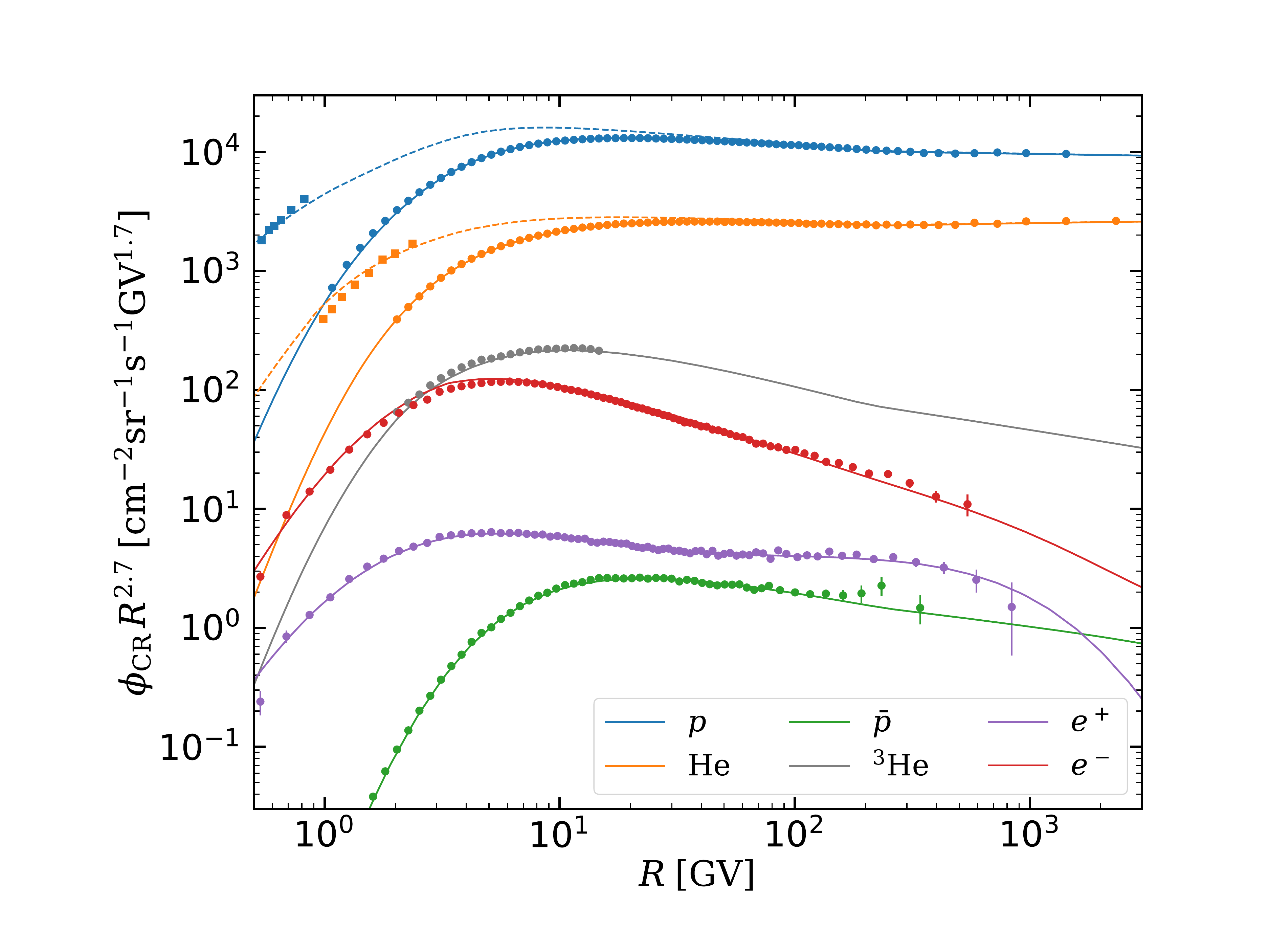}
    \caption{
        Summary of the CR spectra relevant for this analysis. Solid lines show the spectra after Solar modulation which are compared to the AMS-02 data while the dotted lines for proton and helium show the interstellar spectra which are compared to Voyager data at low energies.
    }
    \label{fig:CR_summary}
\end{figure}

\begin{table}[b]
    \centering
    \renewcommand{\arraystretch}{1.6}
    \begin{tabular}{l c c l c c l c }
\hline \hline
parameter & value & $\qquad$$\qquad$ &
parameter & value & $\qquad$$\qquad$ &
parameter & value  \\ \hline
   $ \gamma_{1,p}                                                  $    &  $   2.36    ^{  + 0.008  }_{  - 0.003 }  $ & &
    $ R_{D,1} \, \mathrm{[10^2 GeV]}                                $    &  $   2.4     ^{  + 0.34   }_{  - 0.25  }  $ & &
     $ (\phi_{\bar{p}}-\phi)_{\rm AMS-02}\, \mathrm{[GeV]}           $    &  $   0.12    ^{  + 0.01   }_{  - 0.08  }  $ \\
   $ \gamma_1                                                      $    &  $   2.30    ^{  + 0.006  }_{  - 0.003 }  $ & &
    $ s_D                                                           $    &  $   0.35    ^{  + 0.01   }_{  - 0.03  }  $ & &
     $ A_p                                                           $    &  $   1.00    ^{  + 0.004  }_{  - 0.003 }  $ \\
   $ D_0 \, \mathrm{[10^{28} cm^2/s]}                              $    &  $   4.8     ^{  + 0.3    }_{  - 0.3   }  $ & &
    $ v_{0,{\rm c}} \, \mathrm{km/s}                                $    &  $   0.4     ^{  + 1.7    }_{  - 0.37  }  $ & &
     $ A_{\rm He}                                                    $    &  $   1.05    ^{  + 0.001  }_{  - 0.008 }  $ \\
   $ \delta_l                                                      $    &  $  -0.65    ^{  + 0.036  }_{  - 0.059 }  $ & &
    $ \log_{10}(\dot{N}_\mathrm{pls}/{\rm (100 y)^{-1}})            $    &  $  -0.41    ^{  + 0.13   }_{  - 0.16  }  $ & &
     $ A_{\rm XS,^3He}                                               $    &  $   1.14    ^{  + 0.05   }_{  - 0.01  }  $ \\
   $ \delta                                                        $    &  $   0.51    ^{  + 0.003  }_{  - 0.011 }  $ & &
    $ \gamma_\mathrm{pls}                                           $    &  $   1.63    ^{  + 0.06   }_{  - 0.07  }  $ & &
     $ \delta_{\rm XS,^3He}                                          $    &  $  -0.02    ^{  + 0.02   }_{  - 0.01  }  $ \\
   $ \delta_h - \delta                                             $    &  $  -0.13    ^{  + 0.010  }_{  - 0.016 }  $ & &
    $ \log_{10}(E_{c,\mathrm{pls}}/{\rm GeV})                       $    &  $   2.93    ^{  + 0.07   }_{  - 0.08  }  $ & &
     $ A_{{\rm XS},e+}                                               $    &  $   1.09    ^{  + 0.11   }_{  - 0.025 }  $ \\
   $ R_{D,0} \, \mathrm{[GeV]}                                     $    &  $   4.3     ^{  + 0.24   }_{  - 0.14  }  $ & &
    $ \phi_{\rm AMS-02}   \, \mathrm{[GeV]}                         $    &  $   0.54    ^{  + 0.02   }_{  - 0.01  }  $ \vspace{0.5cm} \\
   $ \log_{10}(A_{e^-}/(\mathrm{MeV\,s\,sr\,cm^2}))                $    &  $  -9.368   ^{  + 0.002  }_{  - 0.002 }  $ & &
    $ \gamma_{e^-}                                                  $    &  $   2.803   ^{  + 0.007  }_{  - 0.008 }  $ & &
     $ \phi_{e^- \rm AMS-02}  \, \mathrm{[GeV]}                      $    &  $   0.79    ^{  + 0.01   }_{  - 0.02  }  $ \\
\hline
    \end{tabular}
    \caption{
      Best-fit parameters of the CR fit. Uncertainties are stated at the 68\% C.L. in frequentist interpretation. The parameters in the last row are adjusted in the second fit step while all other parameters are adjusted in the first step.
    }
    \label{tab:CR_results}
    \renewcommand{\arraystretch}{1.0}
\end{table}

All the relevant processes for CR propagation can be described by a chain of coupled diffusion equations which we solve numerically using the \textsc{Galprop} code \cite{Strong:1998fr}%
\footnote{
   Specifically, we use \textsc{Galprop} version~56.0.2870 combined with \textsc{Galtoollibs}~855 \url{https://galprop.stanford.edu/download.php} including some custom modifications detailed in \cite{Korsmeier:2021brc}.
}. 
In order to sample the large parameter space of 20 parameters efficiently we use the \textsc{MultiNest} code \cite{Feroz:2008xx}.%
\footnote{
    Some parameters like an overall normalization or Solar modulation potentials do not depend on the \textsc{Galprop} evaluation. To speed up the fit those parameters are not sampled with \textsc{MultiNest} and instead profiled for each evaluation of \textsc{Galprop}. For details see \cite{Korsmeier:2021brc, Korsmeier:2021bkw}.  
}

The results of our fit are summarized in Tab.~\ref{tab:CR_results}. We state the best-fit parameters and the $1\sigma$ uncertainties (frequentist interpretation). The best-fit spectra are compared to the data of AMS-02 and Voyager in Fig.~\ref{fig:CR_summary}. Our model provides a good description of the CR data with an overall $\chi^2$/ndf converging to 210/362. In particular, our model fits the $p$ and He as well as the electron and positron data very well. In the context of this work, those are the important CR spectra for the hadronic and bremsstrahlung emission of $\gamma$-rays, respectively. \update{The energy of the $\gamma$-rays is typically one or two orders of magnitude smaller than the one of the parent nuclei. Thus, our CR model is focused on the range from about 1 GeV to a few TeV.}
\update{Furthermore, electrons and positrons produce inverse Compton emission, which depends on the interstellar radiation field. We adopt the model from Ref. \cite{Porter:2017vaa}. In principle, the interstellar radiation field and consequently also the inverse Compton emission is subject to systematic uncertainties. However, we marginalize over the normalization of the inverse Compton emission. Since it is morphologically very different from the gas components, it does not contribute significantly to systematic uncertainty of our final results.}

Finally, we note that the analysis of the diffuse $\gamma$-ray emission revealed a hardening of the $\gamma$-ray spectrum towards the Galactic center \cite{Fermi-LAT:2016zaq,Yang:2016jda,Pothast:2018bvh}. Such a hardening could e.g. be explained by the hardening of CR hadronic spectra towards the Galactic center. We do not model this effect, because our study focuses on high latitudes and, therefore, relatively local gas densities. 

\section{$\gamma$-ray analysis}

Improving our knowledge of the diffuse $\gamma$-ray background is the aim of several studies. On the one hand, the diffuse $\gamma$-ray forms the background when studying other targets, such as point sources \cite{Fermi-LAT:2014ryh} or the Galactic center \cite{Fermi-LAT:2015sau, Carlson:2016iis}. On the other hand, the diffuse $\gamma$-ray background is an interesting target in itself, for example, providing information about CR propagation and interactions \cite{Tibaldo:2021viq}.
Diffuse emission models have been provided and tested against data using the CR propagation codes \textsc{GALPROP} \cite{Porter:2017vaa,Johannesson:2018bit},
\textsc{DRAGON} combined with \textsc{HERMES} \cite{Dundovic:2021ryb}, and \textsc{PICARD} \cite{Kissmann:2017ghg}.

\update{In general, the diffuse $\gamma$-ray emission is calculated from Eq.~\eqref{eq:gamma_ray-flux}. 
However, two of our gas tracers, the 21~cm and CO maps, do not include three-dimensional information and instead provide only the column density, which is the integral of the gas density along the l.o.s. }
To obtain $\gamma$-ray maps, we convolve those gas column densities with the CR flux at the Solar position ($\boldsymbol{x}_\odot$), thus neglecting CR spatial variations, which is reasonable since our Galactic latitude cut ensures that most of the gas is in the immediate Solar neighborhood.
In this case, Eq.~\eqref{eq:gamma_ray-flux} simplifies to:
\begin{eqnarray}
    \label{eq:gamma_ray-flux_3}
    \frac{\diff^2 \phi_\gamma^{ij}}{\diff \Omega \diff E_\gamma} (E_\gamma, l, b) 
    = 
    n_j(l, b) \, \epsilon^{ij}(E_\gamma, \boldsymbol{x}_\odot) \, ,
\end{eqnarray}
where $n_j(l, b)$ is the column density of the gas component $j$.

Our analysis provides an accurate $\gamma$-ray emission model, especially at high latitudes. This model is affected by different systematic uncertainties because we use dust as a tracer for the CNM, while other models typically use only 21~cm and CO data to calibrate their gas models. Dust maps as a tracer have the clear advantage that they provide reliable three-dimensional information, which is not directly available for 21~cm or CO. However, there are attempts to reconstruct three-dimensional HI maps using the velocity information 21~cm maps \cite{Bissantz:2003ak,Pohl:2007dz,Mertsch:2022oee}. In contrast, one disadvantage of dust is the lower angular resolution compared to other tracers. With dust, we are limited to an angular resolution of a couple of degrees, because the dust map is provided on a grid of 5~pc and with a resolution of 25~pc \cite{2019A&A...625A.135L}. The effective resolution is significantly larger than the resolution of \emph{Fermi}-LAT.

\begin{figure}[b]
    \centering
    \includegraphics[width=.5\columnwidth,trim={1.5cm 0.8cm 1.8cm 1.8cm},clip]{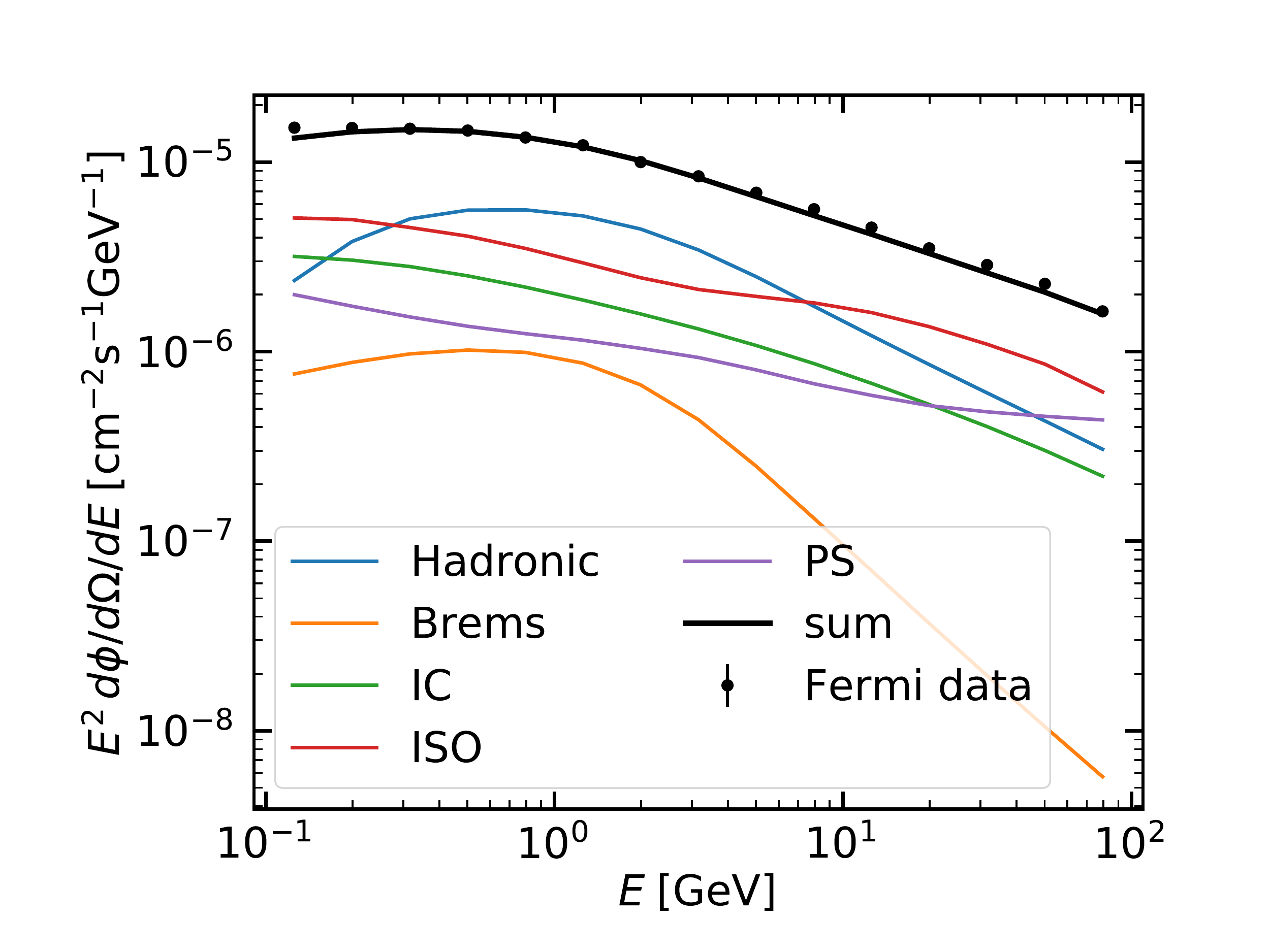}
    \caption{
        $\gamma$-ray flux as function of energy split into different physical origins. The flux is integrated above our region of interest. The lines correspond the the best fit of model A meaning that the hadronic and bremsstrahlung lines include the dust, 21~cm, and CO maps.
    }
    \label{fig:gamm_ray_flux}
\end{figure}

We model different physical processes contributing to the $\gamma$-ray sky. They can be distinguished by their spatial morphology as well as by their spectral behavior. In Fig.~\ref{fig:gamm_ray_flux}, we show the contribution from various physical processes as a function of energy obtained for the non-masked region. Hadronic emission dominates between a few 100~MeV and a few GeV. At higher and lower energies, isotropic emission, ICS, and point sources are more important. Bremsstrahlung is subdominant at all energies. 
Morphologically, we can group the different processes into three classes: 
\begin{itemize}
    \item 
    Hadronic emission and bremsstrahlung are both induced by CR-gas interaction and, therefore, trace the gas distribution.
    \item
    Point sources have their own, very distinct morphology.
    \item
    ICS and isotropic emission follow a very smooth distribution in the sky. 
\end{itemize}

The relative contribution of bremsstrahlung with respect to the hadronic emission drops as a function of energy, which is consistent with the spectral behavior of the parent CR spectra from leptons and nuclei, respectively. In order to reduce the impact of the bremsstrahlung component, we focus on higher energies and exclude the lowest $\gamma$-ray energies from our analysis. One important reason to reduce the impact of CR leptons is that they are strongly affected by energy losses making them a more local CR species than nuclei. 
Furthermore, we note that all gas components produce both hadronic emission and bremsstrahlung. For a given gas component, the ratio between them is fixed to vary in unison in our fits, according to the same re-normalization factor.

\subsection{Data Selection}

We utilize 13 years of \emph{Fermi}-LAT \texttt{ULTRACLEAN} class data (MET: 239557417 -- 655368598) spanning the energy range from 100~MeV to 100~GeV. The ultraclean class of events is designed to provide the highest level of cosmic-ray background rejection, a selection choice that is important because our result is sensitive to the overall normalization of the diffuse cosmic-ray flux. We include both front- and back-converting events observed at a zenith angle of less than 90$^\circ$. We place standard cuts on the \emph{Fermi}-LAT instrumental status, ensuring that the data is taken while the instrument lies outside the South Atlantic Anomaly.  

We compute the instrumental exposure using the most recent \texttt{P8R3\_ULTRACLEAN\_V3} instrumental response functions, which include best-fit calibrations for the instrumental exposure, point-spread function, and energy dispersion. 
We model the point sources in our analysis using the standard \emph{Fermi}-LAT tools {\tt gtsrcmaps} and {\tt gtmodel} in order to produce an exposure-corrected counts map of the 4FGL point source population binned at Healpix order 8 and in 15 energy bins, which corresponds to the parameters that are chosen in the rest of our analysis. More details are given in Sec.~\ref{app:sources} below.

We smooth the model maps by the \emph{Fermi} point spread function (PSF), according to the mid-point energy of each separate energy bin. The PSF decreases with energy; the 1st, 4th and 15th energy bins have a 68\% containment within $2.93^\circ$, $0.99^\circ$, and $0.07^\circ$. We are also limited by the resolution of the dust map \cite{2019A&A...625A.135L}, which is given on a three-dimensional Cartesian grid with a bin side length of 5~pc and a resolution of 25~pc (i.e. to $1.4^\circ$ at a distance of 1~kpc). We then reduce the model sky maps to HEALPix \update{order 5}, corresponding to a pixel size of $1.83^\circ$. 

We then perform a likelihood fit on the data, first binning the data into 5 logarithmic energy bins per decade. We perform the fit independently in each energy bin, and compare the spectrum of each component to a theoretically modeled spectrum in post-processing. This provides an important sanity check for the fidelity of our $\gamma$-ray model. Our model components include the gas and dust maps described in the main text, a Galprop-based model for the ICS component, an isotropic component which produces a counts map identical to the \emph{Fermi}-LAT exposure map, a point source template that is described in detail below.

We mask several regions of the sky where bright emission or systematic errors make it difficult to accurately extract the diffuse components. The first is the region along the Galactic plane with a latitude of $|b|$~$<$~16$^\circ$. The second region of $|l|$~$<$~60$^\circ$, which is designed to remove emission corresponding to the Galactic bulge, the Fermi bubbles~\cite{Su:2010qj}, and Loop I. We note that Loop I slightly extends beyond this latitude cut, so we additionally use a standard template from Ref.~\cite{2009arXiv0912.3478C} to remove the remaining emission.

\subsection{Point Source Models}
\label{app:sources}

Over 12 years of operation, the \emph{Fermi}-LAT has detected more than 5000 $\gamma$-ray point sources, which are listed in the 4FGL-DR3 catalog and combine to produce a substantial fraction of the total $\gamma$-ray flux from the Universe~\cite{Fermi-LAT:2019yla, Fermi-LAT:2022byn}. 
There are two approaches that can account for the bright $\gamma$-ray emission from known $\gamma$-ray point sources: masking regions of the sky that surround the brightest sources, or modeling the $\gamma$-ray emission from each source. We follow a hybrid strategy between modeling and masking. This allows us to retain a large region of interest while at the same time reducing the systematic uncertainty. The large quantity of point sources makes it unfeasible to fit the flux and spectrum of each source individually. Instead, we rely on a collective treatment of all point sources at the same time. In detail, we use the fluxes of the 4FGL-DR2 catalog~\cite{Ballet:2020hze,Fermi-LAT:2019yla} to produce a single map of all point sources. This procedure is expected to work well for stable point sources.
However, this treatment can be problematic for time-varying sources, because the catalog is based on 12 years of data while we exploit 13 years of \emph{Fermi}-data. Hence, we mask all point sources that have both a  \texttt{Variability\_Index} larger than $10^{2.5}$ and a flux (\texttt{Flux100}) brighter than $10^{-8.5}~\cm^{-2}\,\s^{-1}$~\cite{Ballet:2020hze,Fermi-LAT:2019yla}. While this would require the masking of 105 variable point sources, the majority of them lie near the Galactic plane, in regions that are already masked in our analysis. Within the unmasked region of our standard analysis ($|b|<16^\circ$ and $|l|<60^\circ$) we masked only 17 variable point sources, using a 3$^\circ$ \update{radius circle} around each source.

We finally note that point sources have a very distinct morphology making it rather easy to separate them from the other components in the $\gamma$-ray sky. Importantly, the morphology of individually modeled point sources is not degenerate with any of our diffuse maps. This means that if an individual point source (or set of point sources) is mismodeled in our analysis, it will not greatly affect the normalization of any of the diffuse templates and instead will show up as a point-like $\gamma$-ray residual. This limits the effect of point source mismodeling on the results shown in this paper.

\section{Further Results and Additional Figures}\label{app:results}

\begin{figure}
    \centering
    \setlength{\unitlength}{1\textwidth}
    \begin{picture}(1,1.2)
      \put(-0.00 , 0.000){\includegraphics[width=.48\columnwidth,trim={0cm 0cm 0cm 0cm},clip]{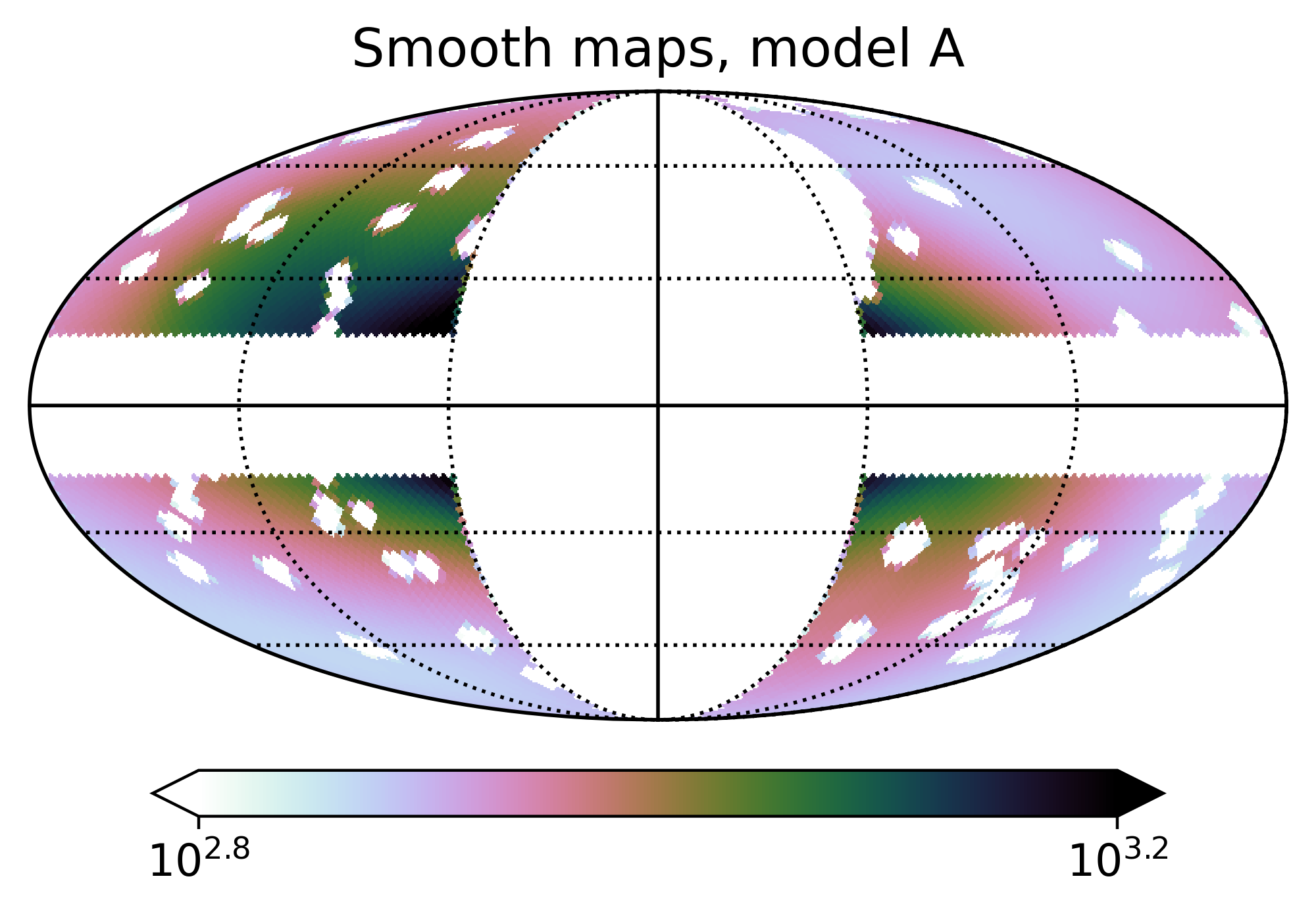}}
      \put( 0.54 , 0.000){\includegraphics[width=.48\columnwidth,trim={0cm 0cm 0cm 0cm},clip]{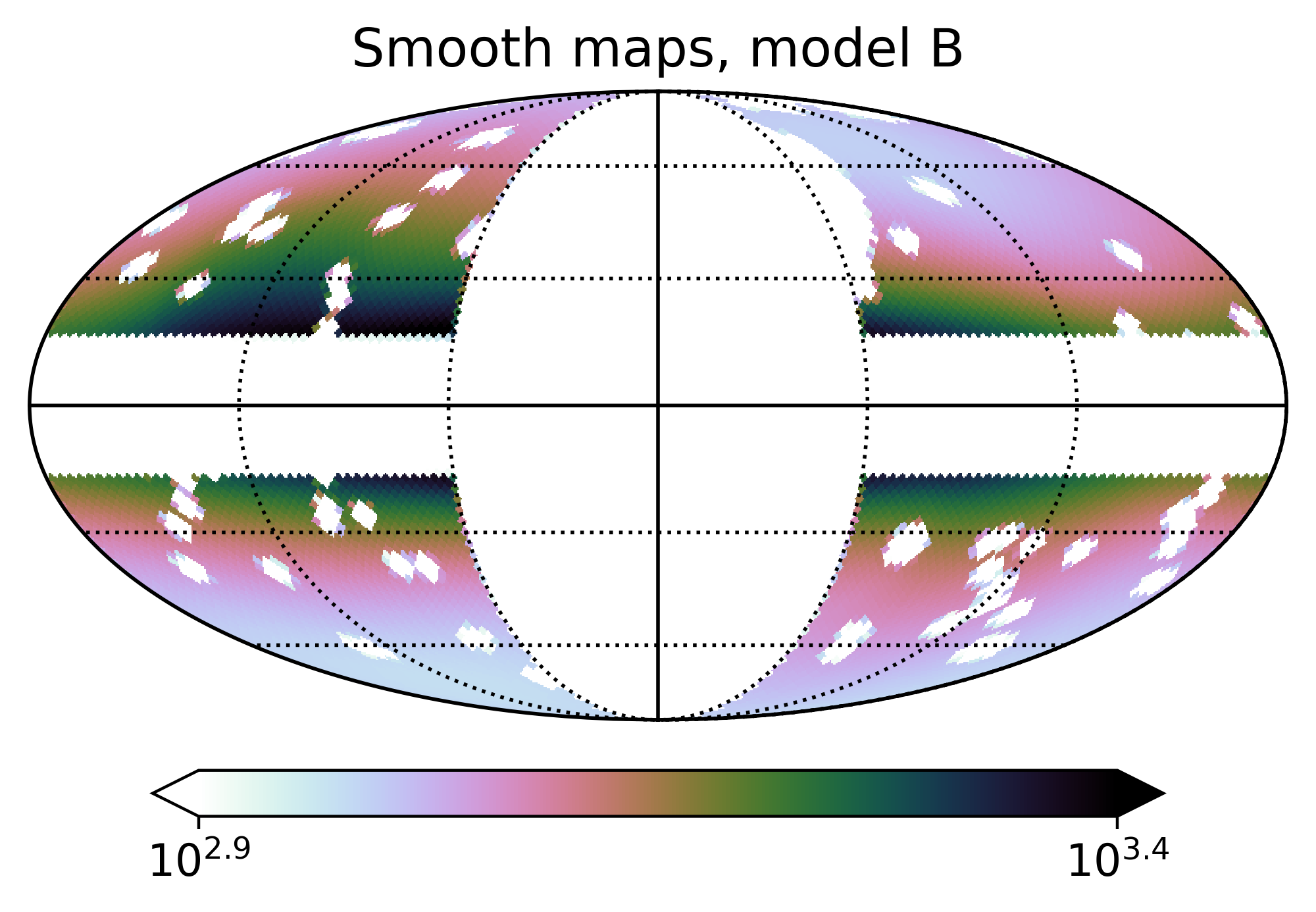}}
      \put(-0.00 , 0.350){\includegraphics[width=.48\columnwidth,trim={0cm 0cm 0cm 0cm},clip]{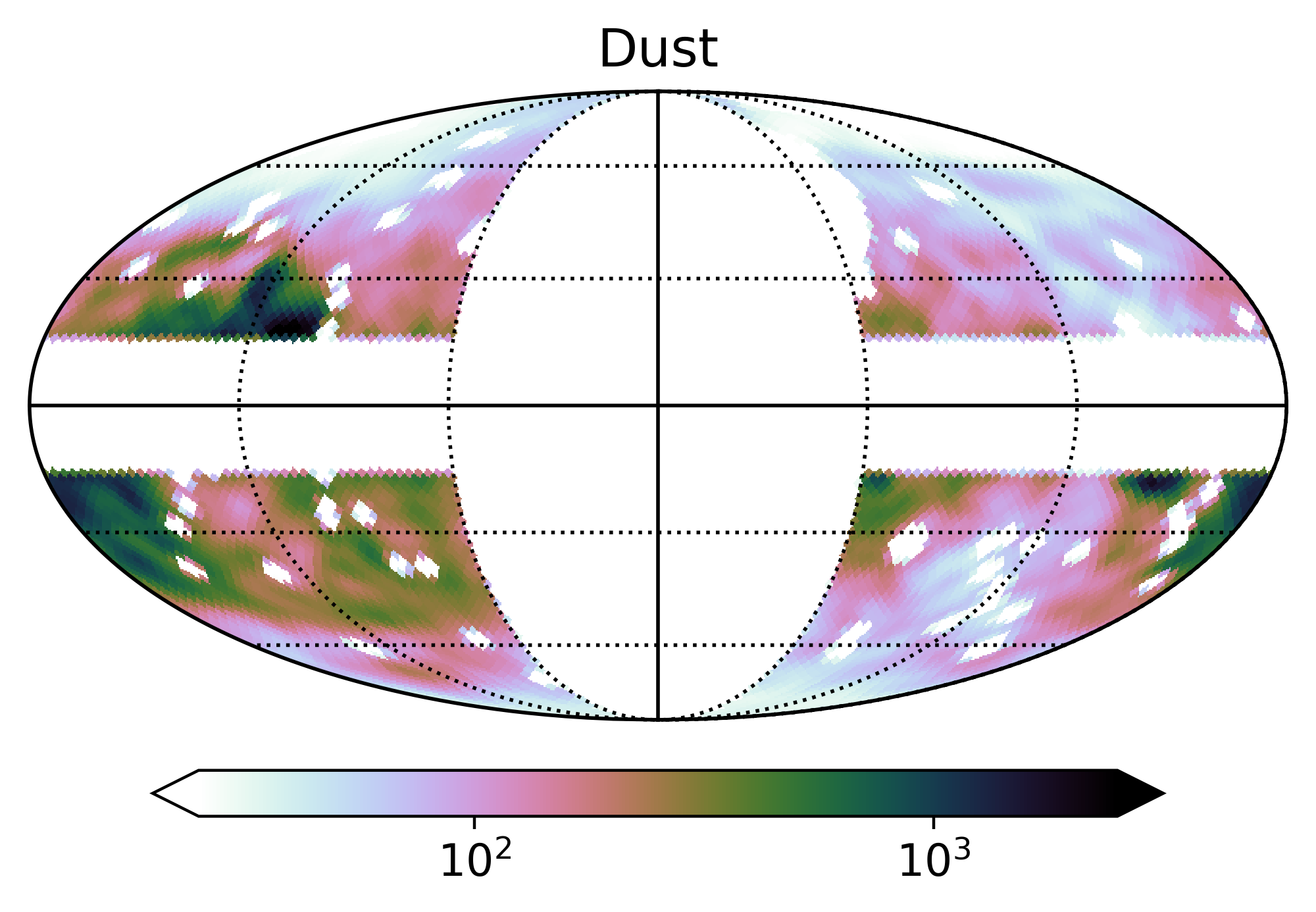}}
      \put( 0.54 , 0.350){\includegraphics[width=.48\columnwidth,trim={0cm 0cm 0cm 0cm},clip]{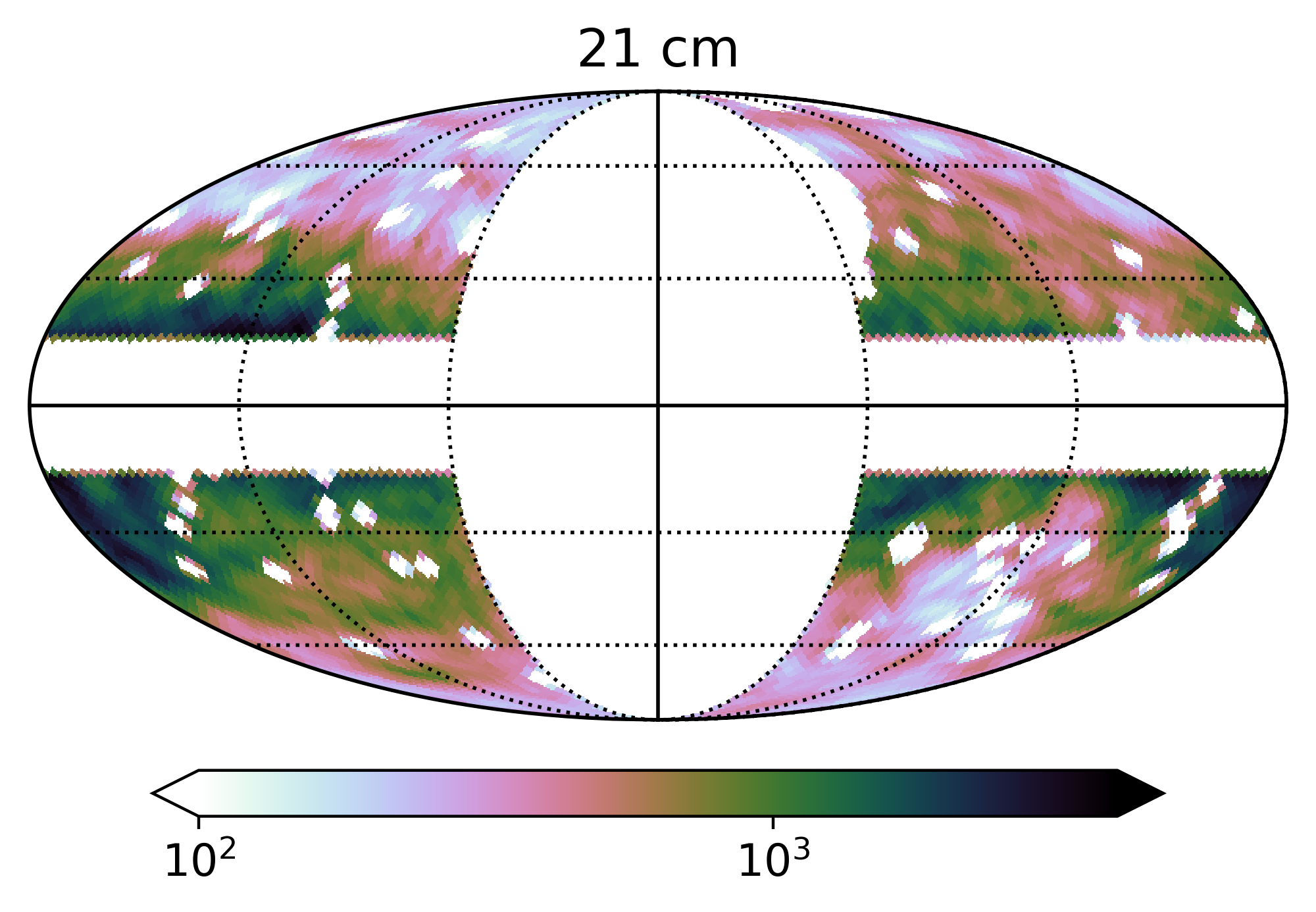}}
      \put(-0.00 , 0.700){\includegraphics[width=.48\columnwidth,trim={0cm 0cm 0cm 0cm},clip]{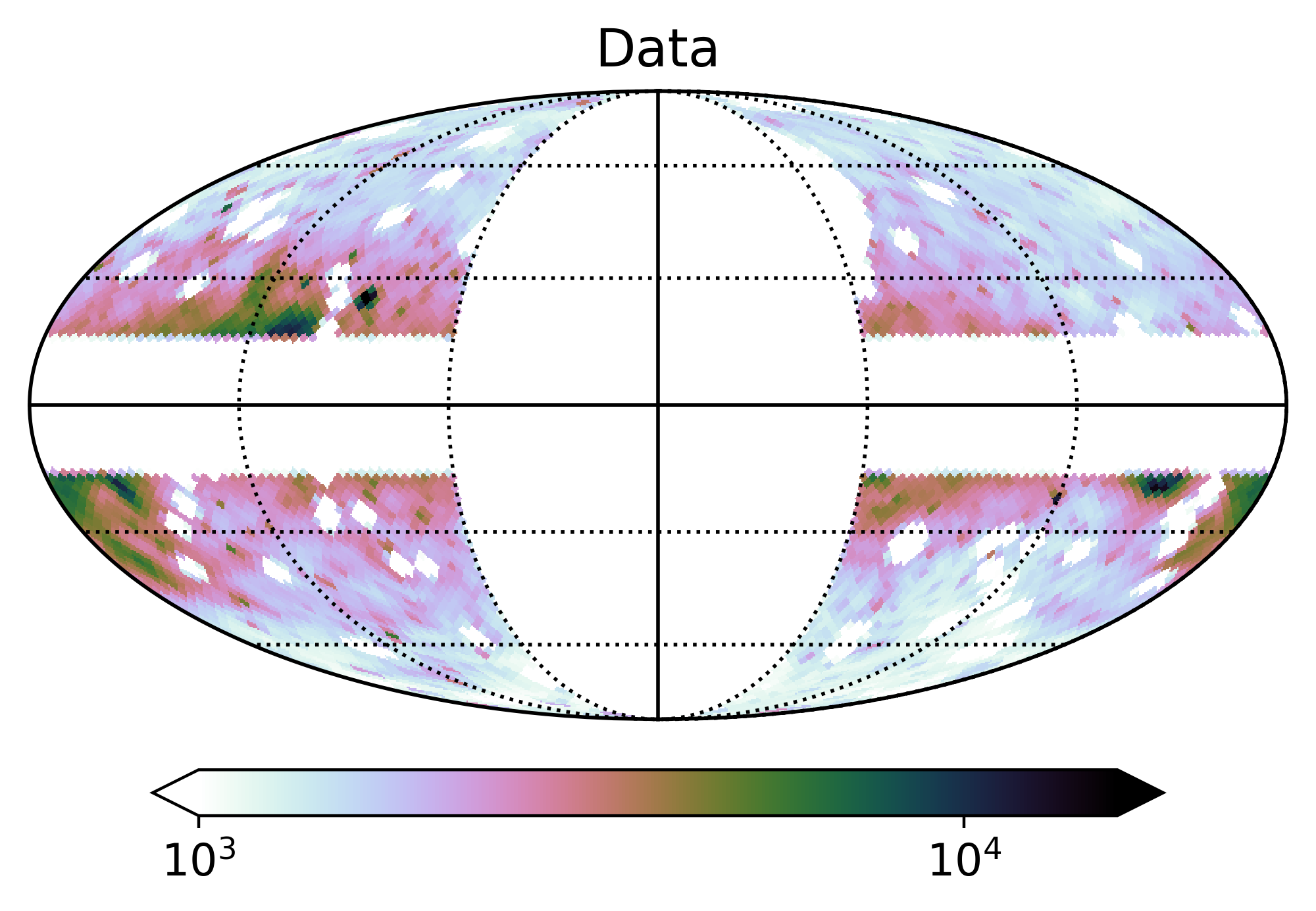}}
      \put( 0.54 , 0.700){\includegraphics[width=.48\columnwidth,trim={0cm 0cm 0cm 0cm},clip]{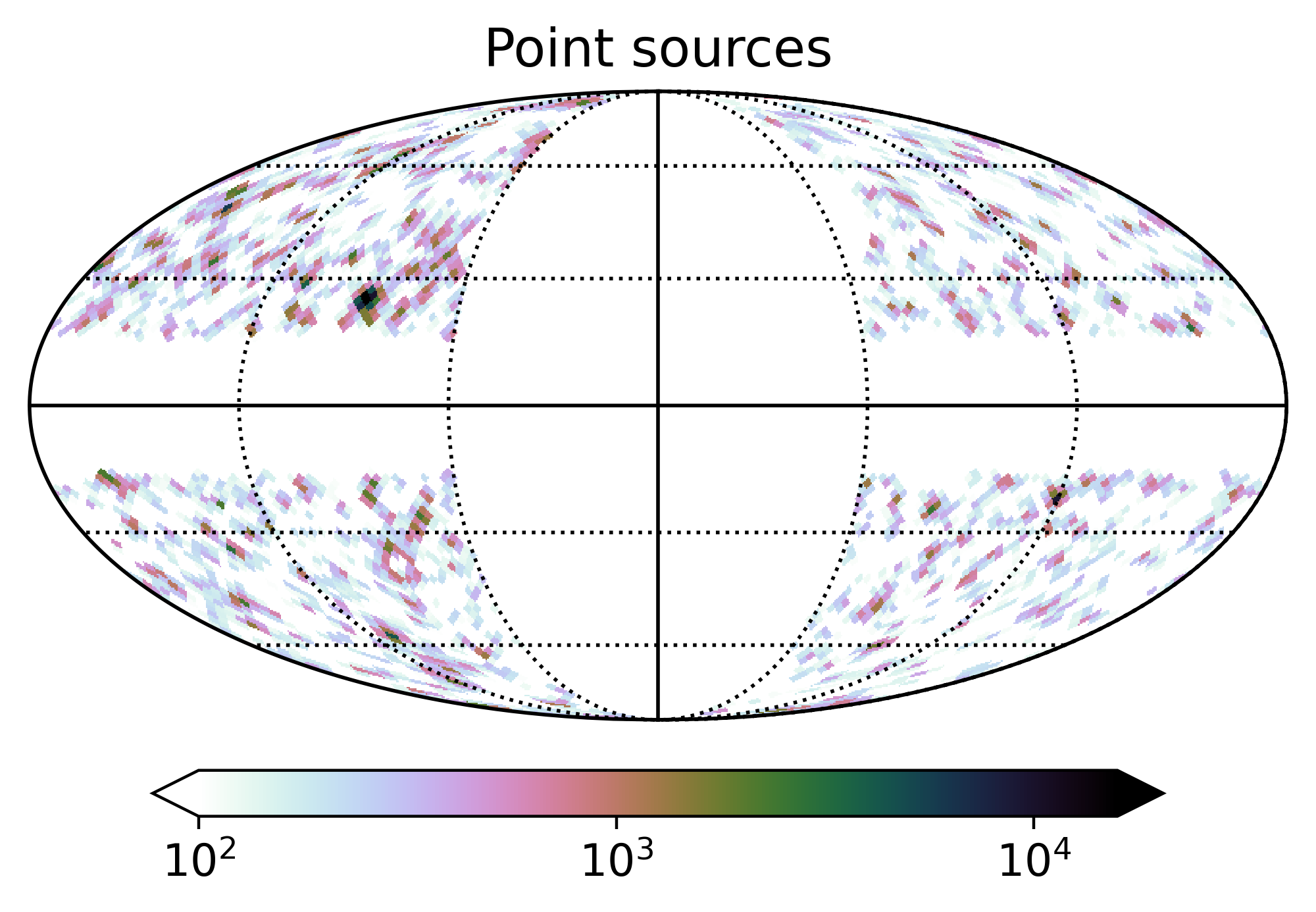}}
    \end{picture}
    \caption{
       Template and data maps. Counts are summed over the 4th--15th energy bins. The count maps are produced by convolving the flux map for each component with the instrumental exposure of the \emph{Fermi}-LAT, which creates small, $\mathcal{O}$(10\%), asymmetries most clearly visible in the smooth maps.
    \label{fig:data_map_and_templates}}
\end{figure}

In Fig.~\ref{fig:data_map_and_templates} we show the data map as well as all the templates we are using in this analysis. Templates are normalized to the median posterior values of model A, except for the analytic WNM which is normalized to the median posterior values of model B. There are degeneracies between the smooth $\gamma$-ray components (isotropic, ICS, and $f_\text{WNM}$), which make them prone to systematic errors and difficult to disentangle. However, the components we are most interested in, which trace gas, have a clear and unique sky morphology, different from the smooth maps and the point-source map, and produce robust results.

\subsection{Hadronic $\gamma$-ray production cross section }

\begin{figure}[b]
    \centering
    \includegraphics[width=.5\columnwidth,trim={0.5cm 0.8cm 1.8cm 1.8cm},clip]{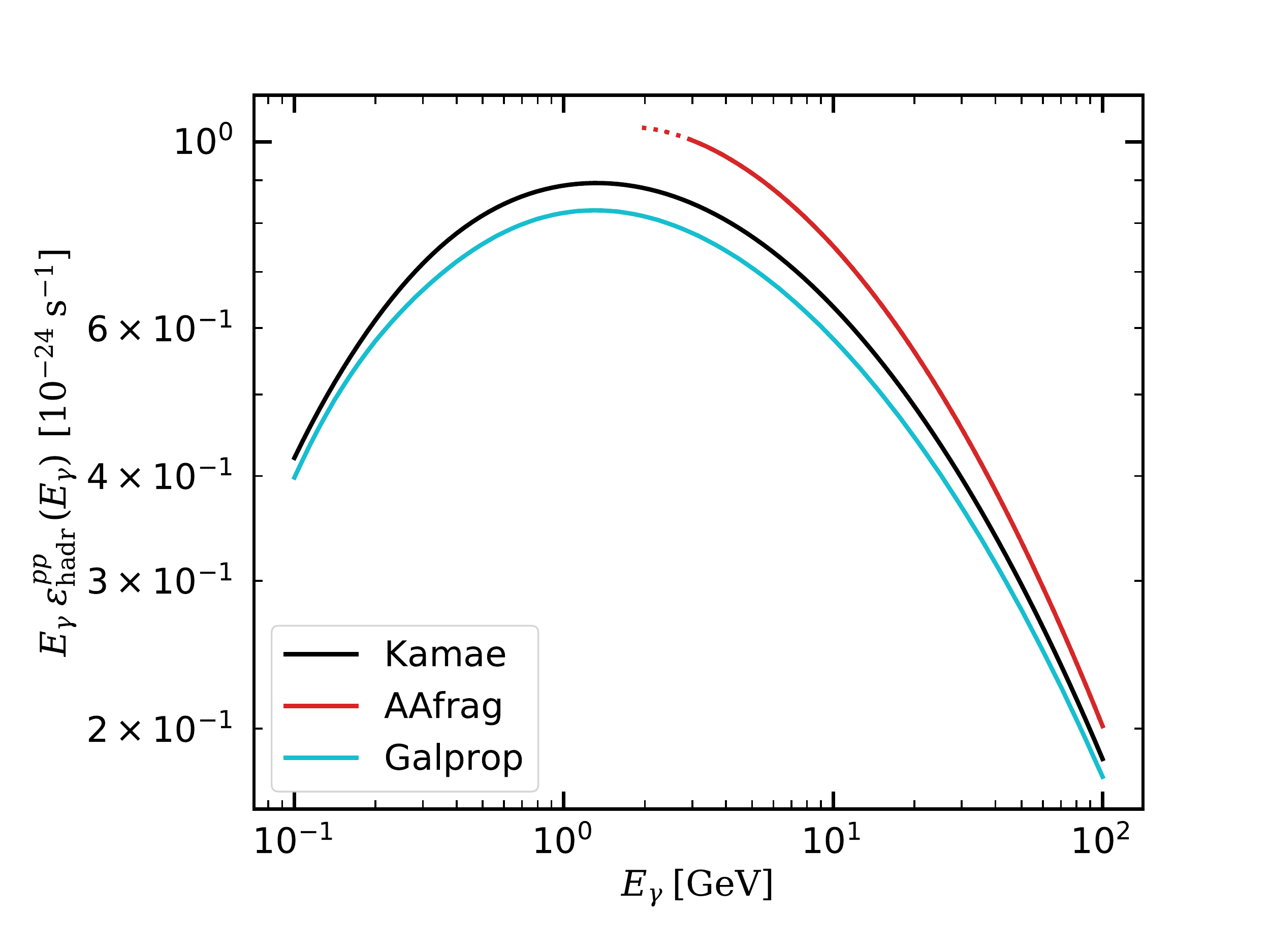}
    \caption{
        Impact of different hadronic $\gamma$-ray cross section. We show the $\gamma$-ray emission per proton-proton interaction, $\epsilon^{pp}_\mathrm{(hadr)}$. Hadronic emission means that in Eq.~\ref{eq:gamma_ray-flux_2} we use only the hadronic and not the bremsstrahlung cross section.
    }
    \label{fig:had_XS}
\end{figure}

One important systematic uncertainty revolves around the cross sections in nuclear interactions. Both the hadronic and bremsstrahlung cross sections are uncertain, but the $\gamma$-ray emission from the latter is subdominant, contributing less than 10\% for energies above a few GeV. Hence, we focus on the hadronic cross sections and examine the impact of three different parametrizations used in the literature.
In our default setup, we employ the cross section from Kamae et al. \emph{et al.}~\cite{Kamae:2004xx,Kamae:2006bf}, which is derived using the \textsc{PYTHIA} event generator~\cite{Sjostrand:2001yu}. The total $p+p$ cross sections, including resonances, are carefully modeled and the tuned event generator can reproduce the available data of the $\pi^0$ multiplicity from a momentum of 1 GeV to 1 TeV. 
As an alternative, we check the cross sections from \textsc{AAfrag}~\cite{Kachelriess:2019ifk}, which are based on the \textsc{QGSJET-II-04} Monte Carlo event generator \cite{Ostapchenko:2010vb}. The hadronic parameters of the generator are tuned to recent data from the LHCf, LHCb, and NA61 experiments. \textsc{AAfrag} provides the cross sections for projectile energies above 3~GeV per nucleon. In general, we note that 
\textsc{QGSJET-II-04} is developed for interactions at high energies. The data of NA61 extends down to projectile energies of 20 GeV such that \textsc{AAfrag} is expected to give reliable predictions above about 10 GeV. 
Finally, we use the scaling model from \textsc{Galprop}~\cite{Moskalenko:1997gh} that relies on the cross section prediction of neutral pions from Ref.~\cite{1981Ap&SS..76..213S}.
The impact of the different cross sections on the $\gamma$-ray emission per proton-proton interaction is shown in Fig.~\ref{fig:had_XS}. The emission derived with the \textsc{Galprop} cross section lies $\sim8$\% below the one with Kamae cross section. In contrast, the cross section from \textsc{AAfrag} lies 18\% above Kamae at 10 GeV dropping to 9\% at 100 GeV. The uncertainty of the $\gamma$-ray emission directly propagates into the uncertainty of the $\gamma$-ray maps and our final result. Consequently, we adopt a $\pm10$\% systematic uncertainty due to cross section on the dust-to-gas ratio, as shown in Fig.~\ref{fig:studies_summary}.

Next to proton-proton interactions we also consider nuclei interactions up to $^4\mathrm{He}$. We assume that the hadronic cross sections scale with the atomic mass number. For bremsstrahlung from electrons and positrons on He in the ISM we adopt the cross section scaling from Galprop, $\sigma_{e^-{\rm He}\rightarrow\gamma}^{\rm(brem)} = 2.9 \sigma_{e^-p\rightarrow\gamma}^{\rm(brem)}$.

\subsection{Spatial distribution of CRs }

Another potential source of uncertainty is the spatial distribution of CRs. The derivation of the $\gamma$-ray maps relies on the knowledge of the CR density in our Galactic neighborhood, while experimental measurements by AMS-02 and Voyager only probe the CR density at the Solar position. However, CR nuclei diffuse over long times and large distances, in general propagating for millions of years while traversing tens of kiloparsecs. This gives rise to a relatively smooth distribution of galactic CRs, making the observed CR density at the solar position a good first-order approximation for the CR density throughout the solar neighborhood. In practice, we have to know the CR density where we expect the majority of the Galactic gas, \emph{i.e.} a couple of hundred pc above and below the Galactic plane. 
Due to this, if our analysis included smaller latitude values $|b|$, it would be sensitive to more distant regions of the disk. For our default cut of $|b|>16^\circ$ radial distances up to about 1~kpc matter. 

As a default, we employ the CR densities from our \textsc{Galprop}-based CR model including 3D spatial information.
We have checked that our results are not affected by using only the CR density at the Solar position, discarding the spatial information from \textsc{Galprop}. Our main conclusions are not affected by changing the latitude cut to $8^\circ$ or $12^\circ$, discussed further in Sec.~\ref{app:add_tests} below.

\subsection{MCMC and jackknife sub-sampling }

\begin{figure}[b]
    \centering
    \includegraphics[width=1.\columnwidth,trim={0cm 0cm 0cm 0cm},clip]{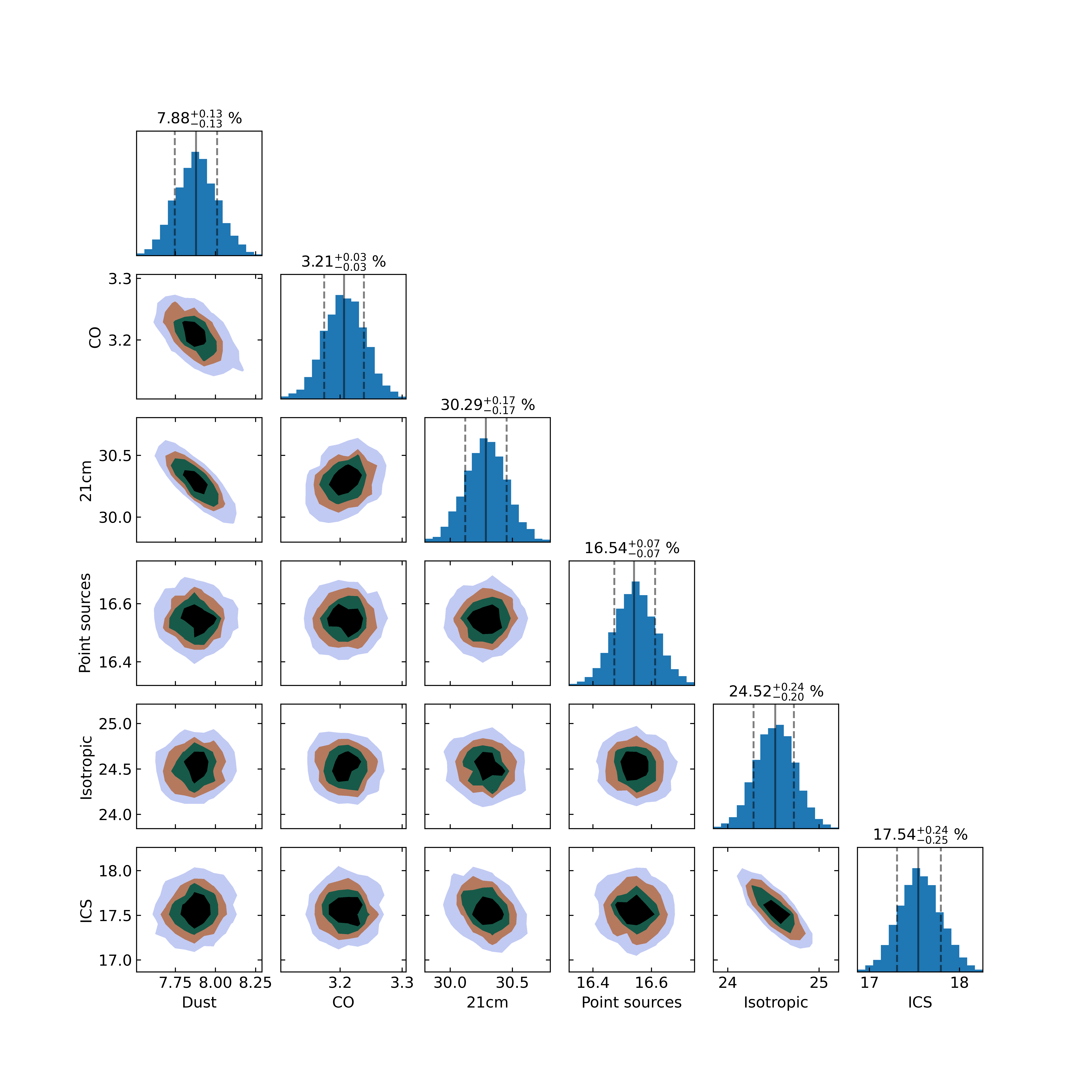}
    \caption{Marginalized posterior densities for the sixth energy bin and a Galactic latitude cut of $|b|>16^\circ$, for model A, without any jackknife sub-sampling. The respective model components are shown in terms of their percentage number count with respect to the total data number count in the unmasked sky region. The vertical lines in the one-dimensional histograms show the 16th, 50th, and 84th percentiles of the posterior, which are also written above the panels. The contour levels of the two-dimensional histograms cover 20\%, 40\%, 60\%, and 80\% of the marginalized posterior density.
    }
    \label{fig:corner}
\end{figure}

For all our fits, we use Markov chain Monte Carlo (MCMC) sampling in order to trace the posterior probability densities, by which we obtain the purely statistical uncertainty of our fits. An example is shown in Fig.~\ref{fig:corner}. This is for only the 6th energy bin, and despite the relatively small subset of the total data, the statistical uncertainty is small. For our main fit over the 4th--15th energy bins, the purely statistical uncertainty is significantly smaller and practically negligible.

However, there are some structures in the data count map that our models do not account for, as seen in Fig.~\ref{fig:data_model_frac}. The average $\chi^2$ of our main fit is 12.5 and 31.3 for models A and B, so the statistical strain is high. In order to estimate the significance of these spatially dependent systematic errors and inflate the posterior density distributions, we employ the technique of delete-$d$ jackknife sub-sampling \citep{efron1982jackknife}. With this technique, the uncertainty of an inferred parameter $X$, represented by a standard deviation $\text{Std}(X)$, is proportional to the standard deviation of the results of the respective sub-samples, written $\text{Std}^*(X)$, according to
\begin{equation}
	\text{Std}(X) = \sqrt{\frac{n-d}{d}} \text{Std}^*(X)\,,
\end{equation}
where $n$ is the number of data points in the complete data set and $d$ is the number of deleted data points.

Ideally, one would iterate over all possible sub-samples, but this is not computationally feasible for us. Instead, we divide the sky into eight different sub-regions, which we label N1--N4 and S1--S4. The letter denotes the Galactic north or south ($b>16^\circ$ or $b<-16^\circ$), and the number denotes a $60^\circ$ longitude bin in the range 60$^\circ$--320$^\circ$. We run a total of eight jackknife iterations. For each iteration, we construct a half sky region using four out of eight sub-regions, using the following sub-region combinations: (N1--N4), (S1--S4), (N1, N2, S1, S2), (N3, N4, S3, S4), (N1, N4, S1, S4), (N2, N3, S2, S3), (N1, N3, S2, S4), and (N2, N4, S1, S3). Because we are splitting the total number of data points in half, such that $(n-d)/d=1$, we have a direct relationship between the standard deviation of our jackknife iterations and the total systematic uncertainty. Conservatively, we use both the statistical uncertainty of each separate sub-sample, obtained using MCMC sampling, in addition to the systematic uncertainty evaluated via jackknife sub-sampling. However, the latter completely dominates over the former.

\update{We perform a test of convergence for our jackknife sub-sampling, using our fiducial fit with $|b| < 16^\circ$. For this test, we double the number of half-sky sub-regions to 16 using similar angular cuts, but with longitude edges on $l = 30^\circ+i \times 60^\circ$ (instead of $l = i \times 60^\circ$). We obtain results with a negligible difference with respect to the standard jackknife with eight sub-regions described above: the dust-to-H conversion factor mean and standard deviation differ by 1.9\% and 3.5\%, respectively.}

\begin{figure}
    \centering
    \includegraphics[width=1\textwidth,trim={0cm 0cm 0cm 0cm},clip]{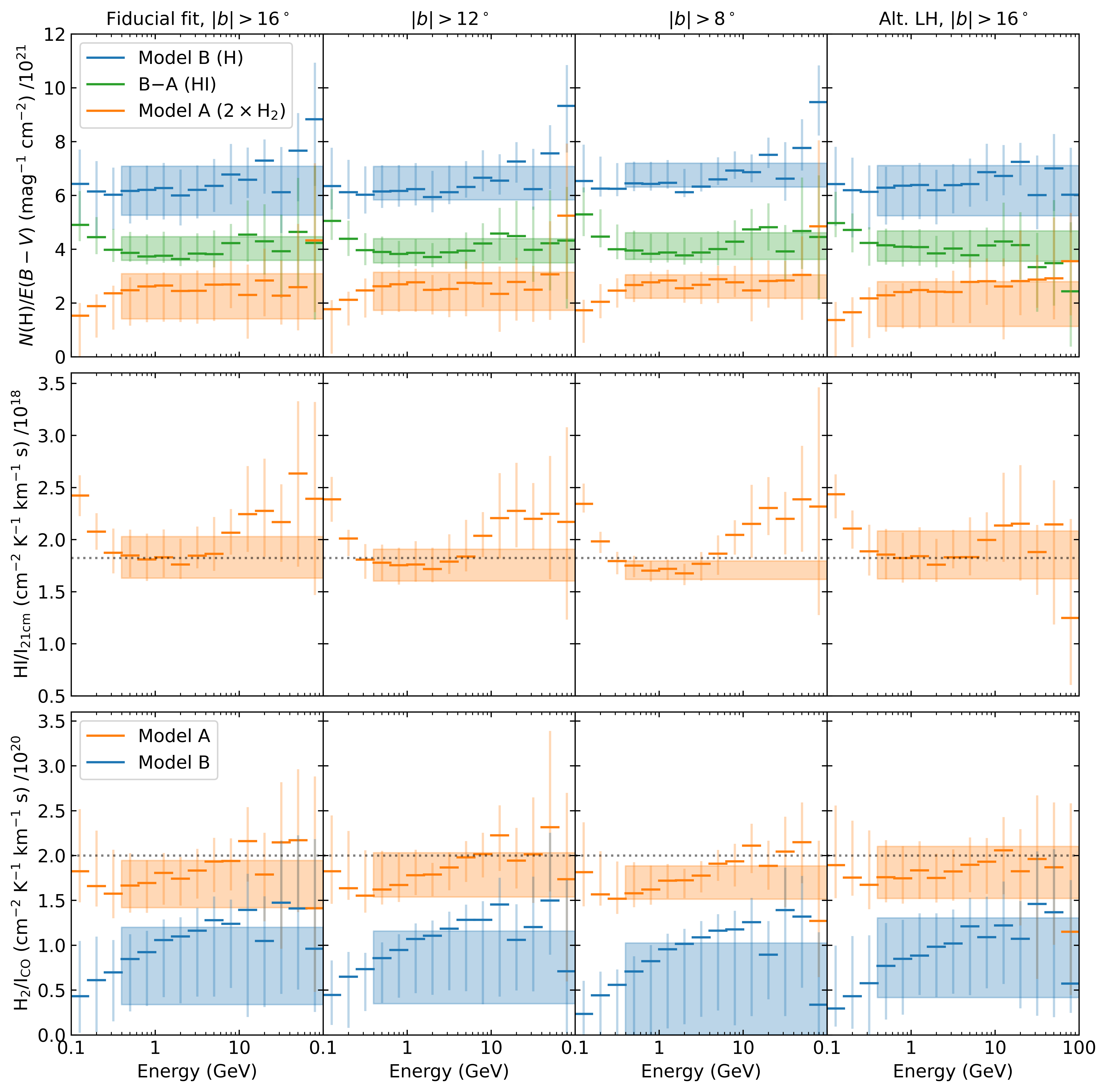}
    \caption{\update{Conversion factors for the gas components traced by dust, 21 cm, and CO. The colored bands show the joint fit of the 4th--15th energy bins, whose vertical range corresponds to the 16th and 84th posterior percentile. The solid horizontal lines correspond to the median value of each energy bin, with their 16th and 84th percentile range shown as vertical lines. The horizontal dotted lines in the central and right rows mark the canonical conversion factor values for 21 cm and CO ($1.823\times 10^{18}~\cm^{-2}\,\text{K}^{-1}\text{km}^{-1}\,\text{s}$ and $2\times 10^{20}~\cm^{-2}\,\text{K}^{-1}\text{km}^{-1}\,\text{s}$, respectively).
    The four columns correspond to, from left to right, our fiducial fit with a Galactic latitude cut of $|b|>16^\circ$, a cut of $|b|>12^\circ$, a cut of $|b|>8^\circ$, and finally a fit using an alternative data likelihood (described in Appendix~\ref{app:add_tests}) less sensitive to outliers with $|b|>16^\circ$.}}
    \label{fig:conv_factors}
\end{figure}

The gas conversion factor results of our main fit can be seen in Fig.~\ref{fig:conv_factors}. We see generally robust results, but also some weaker systematic trends with respect to energy. At the lowest energies, there is a slight slope for the gas-to-dust conversion factors, which could be related to, for example, systematic errors in the $\gamma$-ray cross sections or hydrogen gas ionization levels. The mid and high energy range, where we perform our main fit, is stable, although the highest energies suffer from less statistics and have a high uncertainty.

\subsection{Additional tests}\label{app:add_tests}

We perform a few additional tests in order to diagnose potential systematic errors. We modify the angular resolution of our fitted $\gamma$-ray sky, to half and double the resolution of our main fit. We test the significance of CR spatial variations by convolving all gas components only with the CR flux at the Solar position, rather than the full 3D distribution. For both of these tests, we obtain only minimal differences in the inferred gas conversion factors.

We also perform a test where we replace the Poisson number count with an alternative data likelihood that is less sensitive to outliers. This alternative likelihood is given by
\begin{equation}
    \log \mathcal{L}(\text{data} \, | \, \text{model}) =
    -\bigg(\frac{5}{2}\bigg)^2 \, \text{log} \bigg\{ \text{cosh} \bigg[ \frac{2}{5} \sqrt{ \frac{(\text{data}-\text{model})^2}{\text{model}} } \bigg] \bigg\} + \text{constant}\,.
\end{equation}
This functional form behaves like a Gaussian 2-norm close to zero and a 1-norm at large outlier values, with a smooth transition between them at $2.5\sigma$. The inferred gas conversion factors when using the alternative data likelihood are shown in Fig.~\ref{fig:conv_factors} \update{(fourth column)}. They are very well consistent with our main result, indicating that data outliers do not significantly bias our results.


Furthermore, we perform fits where we lower the limit in Galactic latitude, thus including more of the $\gamma$-ray sky close to the Galactic mid-plane. In Fig~\ref{fig:conv_factors} \update{(second and third column)}, we show the gas conversion results for $|b|>12^\circ$ and $|b|>8^\circ$. We see consistent results, indicating that our main choice of $|b|>16^\circ$ gives robust results. The uncertainties are lower when including a larger sky area and more statistical power; however, we still consider $|b|>16^\circ$ to be a cautious choice, as the lower latitude cuts are more likely to be affected by uncontrolled systematic biases.

Finally, we note that the conversion factors obtained here are specific to the dust and {21~cm} map employed in this work. In order to prevent a potential bias when extrapolating to a new dust or {21~cm} tracer map, one should first compare the new maps to the maps employed in this analysis.



\subsection{Model for the cold gas surface densities}\label{app:surf_dens_comparisons}

Based on our results, we recommend the following two-component model for the cold gas in the Solar neighborhood, traced by dust and CO. The dust-to-hydrogen conversion factor is $(6.28 \pm 0.94)\times 10^{21}~\magn^{-1}\,\cm^{-2}$ (consisting of both HI and H$_2$), and the CO-to-H$_2$ conversion factor is $(0.81 \pm 0.45)\times 10^{20}~\cm^{-2}\,\K^{-1}\,\km^{-1}\,\s$. The latter is lower than the canonical value because of the overlap between the two components (\emph{i.e.} the CO-bright H$_2$ is also in part traced by dust). Both components are also mixed with helium and heavier elements, constituting an additional \update{amount of gas equal to 42\% of the total hydrogen mass}. Furthermore, we recommend conservatively adding a relative 10\% uncertainty in quadrature, due to a potential systematic bias in the $\gamma$-ray cross sections.

From the dust-to-gas conversion factors, we can calculate the surface density of the CNM. We improve on previous models for the local gas densities, in the sense that we account for its three-dimensional distribution. Because the dust distribution is non-uniform, we need to average over some disk plane area in order to calculate the surface density. Given a 500~pc or 800~pc cylindrical radius around the Solar position, the total hydrogen traced by dust has a surface density of $\Sigma_\text{H} = 9.57\pm 1.42~\text{M}_\odot \text{pc}^{-2}$ or $7.53 \pm 1.12~\text{M}_\odot \text{pc}^{-2}$, where we have also included helium and heavier elements constituting 42\% of the total hydrogen density. We compare this with the baryonic model for the Solar neighborhood by Ref.~\cite{Schutz:2017tfp}, with gas components from Ref.~\cite{Kramer_2016}, which is an update of the model from Ref.~\cite{2015ApJ...814...13M}. They quote a CO-traced H$_2$ surface density of $\Sigma_{\text{H}_2} = 1.33 \pm 0.40~\text{M}_\odot \text{pc}^{-2}$ and a CNM HI surface density of $\Sigma_{\text{HI}} = 7.14 \pm 1.37~\text{M}_\odot \text{pc}^{-2}$ (both also including 42\% He). We are roughly consistent with these values, although a straight forward comparison is difficult given that our model accounts for three-dimensional spatial variations.

\bibliographystyle{apsrev4-1}
\bibliography{main}

\begin{thebibliography}{121}%
\makeatletter
\providecommand \@ifxundefined [1]{%
 \@ifx{#1\undefined}
}%
\providecommand \@ifnum [1]{%
 \ifnum #1\expandafter \@firstoftwo
 \else \expandafter \@secondoftwo
 \fi
}%
\providecommand \@ifx [1]{%
 \ifx #1\expandafter \@firstoftwo
 \else \expandafter \@secondoftwo
 \fi
}%
\providecommand \natexlab [1]{#1}%
\providecommand \enquote  [1]{``#1''}%
\providecommand \bibnamefont  [1]{#1}%
\providecommand \bibfnamefont [1]{#1}%
\providecommand \citenamefont [1]{#1}%
\providecommand \href@noop [0]{\@secondoftwo}%
\providecommand \href [0]{\begingroup \@sanitize@url \@href}%
\providecommand \@href[1]{\@@startlink{#1}\@@href}%
\providecommand \@@href[1]{\endgroup#1\@@endlink}%
\providecommand \@sanitize@url [0]{\catcode `\\12\catcode `\$12\catcode
  `\&12\catcode `\#12\catcode `\^12\catcode `\_12\catcode `\%12\relax}%
\providecommand \@@startlink[1]{}%
\providecommand \@@endlink[0]{}%
\providecommand \url  [0]{\begingroup\@sanitize@url \@url }%
\providecommand \@url [1]{\endgroup\@href {#1}{\urlprefix }}%
\providecommand \urlprefix  [0]{URL }%
\providecommand \Eprint [0]{\href }%
\providecommand \doibase [0]{http://dx.doi.org/}%
\providecommand \selectlanguage [0]{\@gobble}%
\providecommand \bibinfo  [0]{\@secondoftwo}%
\providecommand \bibfield  [0]{\@secondoftwo}%
\providecommand \translation [1]{[#1]}%
\providecommand \BibitemOpen [0]{}%
\providecommand \bibitemStop [0]{}%
\providecommand \bibitemNoStop [0]{.\EOS\space}%
\providecommand \EOS [0]{\spacefactor3000\relax}%
\providecommand \BibitemShut  [1]{\csname bibitem#1\endcsname}%
\let\auto@bib@innerbib\@empty
\bibitem [{\citenamefont {{Read}}(2014)}]{Read2014}%
  \BibitemOpen
  \bibfield  {author} {\bibinfo {author} {\bibfnamefont {J.~I.}\ \bibnamefont
  {{Read}}},\ }\href {\doibase 10.1088/0954-3899/41/6/063101} {\bibfield
  {journal} {\bibinfo  {journal} {Journal of Physics G Nuclear Physics}\
  }\textbf {\bibinfo {volume} {41}},\ \bibinfo {eid} {063101} (\bibinfo {year}
  {2014})},\ \Eprint {http://arxiv.org/abs/1404.1938} {arXiv:1404.1938}
  \BibitemShut {NoStop}%
\bibitem [{\citenamefont {{de Salas}}\ and\ \citenamefont
  {{Widmark}}(2021)}]{2021RPPh...84j4901D}%
  \BibitemOpen
  \bibfield  {author} {\bibinfo {author} {\bibfnamefont {P.~F.}\ \bibnamefont
  {{de Salas}}}\ and\ \bibinfo {author} {\bibfnamefont {A.}~\bibnamefont
  {{Widmark}}},\ }\href {\doibase 10.1088/1361-6633/ac24e7} {\bibfield
  {journal} {\bibinfo  {journal} {Reports on Progress in Physics}\ }\textbf
  {\bibinfo {volume} {84}},\ \bibinfo {eid} {104901} (\bibinfo {year}
  {2021})},\ \Eprint {http://arxiv.org/abs/2012.11477} {arXiv:2012.11477
  [astro-ph.GA]} \BibitemShut {NoStop}%
\bibitem [{\citenamefont {{Shapiro}}\ and\ \citenamefont
  {{Field}}(1976)}]{1976ApJ...205..762S}%
  \BibitemOpen
  \bibfield  {author} {\bibinfo {author} {\bibfnamefont {P.~R.}\ \bibnamefont
  {{Shapiro}}}\ and\ \bibinfo {author} {\bibfnamefont {G.~B.}\ \bibnamefont
  {{Field}}},\ }\href {\doibase 10.1086/154332} {\bibfield  {journal} {\bibinfo
   {journal} {\apj}\ }\textbf {\bibinfo {volume} {205}},\ \bibinfo {pages}
  {762} (\bibinfo {year} {1976})}\BibitemShut {NoStop}%
\bibitem [{\citenamefont {{Saintonge}}\ and\ \citenamefont
  {{Catinella}}(2022)}]{2022arXiv220200690S}%
  \BibitemOpen
  \bibfield  {author} {\bibinfo {author} {\bibfnamefont {A.}~\bibnamefont
  {{Saintonge}}}\ and\ \bibinfo {author} {\bibfnamefont {B.}~\bibnamefont
  {{Catinella}}},\ }\href@noop {} {\bibfield  {journal} {\bibinfo  {journal}
  {arXiv e-prints}\ ,\ \bibinfo {eid} {arXiv:2202.00690}} (\bibinfo {year}
  {2022})},\ \Eprint {http://arxiv.org/abs/2202.00690} {arXiv:2202.00690
  [astro-ph.GA]} \BibitemShut {NoStop}%
\bibitem [{\citenamefont {{Heyer}}\ and\ \citenamefont
  {{Dame}}(2015)}]{2015ARA&A..53..583H}%
  \BibitemOpen
  \bibfield  {author} {\bibinfo {author} {\bibfnamefont {M.}~\bibnamefont
  {{Heyer}}}\ and\ \bibinfo {author} {\bibfnamefont {T.~M.}\ \bibnamefont
  {{Dame}}},\ }\href {\doibase 10.1146/annurev-astro-082214-122324} {\bibfield
  {journal} {\bibinfo  {journal} {\araa}\ }\textbf {\bibinfo {volume} {53}},\
  \bibinfo {pages} {583} (\bibinfo {year} {2015})}\BibitemShut {NoStop}%
\bibitem [{\citenamefont {{Bolatto}}\ \emph {et~al.}(2013)\citenamefont
  {{Bolatto}}, \citenamefont {{Wolfire}},\ and\ \citenamefont
  {{Leroy}}}]{2013ARA&A..51..207B}%
  \BibitemOpen
  \bibfield  {author} {\bibinfo {author} {\bibfnamefont {A.~D.}\ \bibnamefont
  {{Bolatto}}}, \bibinfo {author} {\bibfnamefont {M.}~\bibnamefont
  {{Wolfire}}}, \ and\ \bibinfo {author} {\bibfnamefont {A.~K.}\ \bibnamefont
  {{Leroy}}},\ }\href {\doibase 10.1146/annurev-astro-082812-140944} {\bibfield
   {journal} {\bibinfo  {journal} {\araa}\ }\textbf {\bibinfo {volume} {51}},\
  \bibinfo {pages} {207} (\bibinfo {year} {2013})},\ \Eprint
  {http://arxiv.org/abs/1301.3498} {arXiv:1301.3498 [astro-ph.GA]} \BibitemShut
  {NoStop}%
\bibitem [{\citenamefont {{Grenier}}\ \emph {et~al.}(2005)\citenamefont
  {{Grenier}}, \citenamefont {{Casandjian}},\ and\ \citenamefont
  {{Terrier}}}]{2005Sci...307.1292G}%
  \BibitemOpen
  \bibfield  {author} {\bibinfo {author} {\bibfnamefont {I.~A.}\ \bibnamefont
  {{Grenier}}}, \bibinfo {author} {\bibfnamefont {J.-M.}\ \bibnamefont
  {{Casandjian}}}, \ and\ \bibinfo {author} {\bibfnamefont {R.}~\bibnamefont
  {{Terrier}}},\ }\href {\doibase 10.1126/science.1106924} {\bibfield
  {journal} {\bibinfo  {journal} {Science}\ }\textbf {\bibinfo {volume}
  {307}},\ \bibinfo {pages} {1292} (\bibinfo {year} {2005})}\BibitemShut
  {NoStop}%
\bibitem [{\citenamefont {{Wolfire}}\ \emph {et~al.}(2010)\citenamefont
  {{Wolfire}}, \citenamefont {{Hollenbach}},\ and\ \citenamefont
  {{McKee}}}]{2010ApJ...716.1191W}%
  \BibitemOpen
  \bibfield  {author} {\bibinfo {author} {\bibfnamefont {M.~G.}\ \bibnamefont
  {{Wolfire}}}, \bibinfo {author} {\bibfnamefont {D.}~\bibnamefont
  {{Hollenbach}}}, \ and\ \bibinfo {author} {\bibfnamefont {C.~F.}\
  \bibnamefont {{McKee}}},\ }\href {\doibase 10.1088/0004-637X/716/2/1191}
  {\bibfield  {journal} {\bibinfo  {journal} {\apj}\ }\textbf {\bibinfo
  {volume} {716}},\ \bibinfo {pages} {1191} (\bibinfo {year} {2010})},\ \Eprint
  {http://arxiv.org/abs/1004.5401} {arXiv:1004.5401 [astro-ph.GA]} \BibitemShut
  {NoStop}%
\bibitem [{\citenamefont {{Tang}}\ \emph {et~al.}(2016)\citenamefont {{Tang}},
  \citenamefont {{Li}}, \citenamefont {{Heiles}}, \citenamefont {{Wang}},
  \citenamefont {{Pan}},\ and\ \citenamefont {{Wang}}}]{2016A&A...593A..42T}%
  \BibitemOpen
  \bibfield  {author} {\bibinfo {author} {\bibfnamefont {N.}~\bibnamefont
  {{Tang}}}, \bibinfo {author} {\bibfnamefont {D.}~\bibnamefont {{Li}}},
  \bibinfo {author} {\bibfnamefont {C.}~\bibnamefont {{Heiles}}}, \bibinfo
  {author} {\bibfnamefont {S.}~\bibnamefont {{Wang}}}, \bibinfo {author}
  {\bibfnamefont {Z.}~\bibnamefont {{Pan}}}, \ and\ \bibinfo {author}
  {\bibfnamefont {J.-J.}\ \bibnamefont {{Wang}}},\ }\href {\doibase
  10.1051/0004-6361/201528055} {\bibfield  {journal} {\bibinfo  {journal}
  {\aap}\ }\textbf {\bibinfo {volume} {593}},\ \bibinfo {eid} {A42} (\bibinfo
  {year} {2016})},\ \Eprint {http://arxiv.org/abs/1606.00949} {arXiv:1606.00949
  [astro-ph.GA]} \BibitemShut {NoStop}%
\bibitem [{\citenamefont {{Reach}}\ \emph {et~al.}(2017)\citenamefont
  {{Reach}}, \citenamefont {{Heiles}},\ and\ \citenamefont
  {{Bernard}}}]{2017ApJ...834...63R}%
  \BibitemOpen
  \bibfield  {author} {\bibinfo {author} {\bibfnamefont {W.~T.}\ \bibnamefont
  {{Reach}}}, \bibinfo {author} {\bibfnamefont {C.}~\bibnamefont {{Heiles}}}, \
  and\ \bibinfo {author} {\bibfnamefont {J.-P.}\ \bibnamefont {{Bernard}}},\
  }\href {\doibase 10.3847/1538-4357/834/1/63} {\bibfield  {journal} {\bibinfo
  {journal} {\apj}\ }\textbf {\bibinfo {volume} {834}},\ \bibinfo {eid} {63}
  (\bibinfo {year} {2017})},\ \Eprint {http://arxiv.org/abs/1611.04475}
  {arXiv:1611.04475 [astro-ph.GA]} \BibitemShut {NoStop}%
\bibitem [{\citenamefont {{Draine}}(2003)}]{2003ARA&A..41..241D}%
  \BibitemOpen
  \bibfield  {author} {\bibinfo {author} {\bibfnamefont {B.~T.}\ \bibnamefont
  {{Draine}}},\ }\href {\doibase 10.1146/annurev.astro.41.011802.094840}
  {\bibfield  {journal} {\bibinfo  {journal} {\araa}\ }\textbf {\bibinfo
  {volume} {41}},\ \bibinfo {pages} {241} (\bibinfo {year} {2003})},\ \Eprint
  {http://arxiv.org/abs/astro-ph/0304489} {arXiv:astro-ph/0304489 [astro-ph]}
  \BibitemShut {NoStop}%
\bibitem [{\citenamefont {{Draine}}\ and\ \citenamefont
  {{Li}}(2007)}]{2007ApJ...657..810D}%
  \BibitemOpen
  \bibfield  {author} {\bibinfo {author} {\bibfnamefont {B.~T.}\ \bibnamefont
  {{Draine}}}\ and\ \bibinfo {author} {\bibfnamefont {A.}~\bibnamefont
  {{Li}}},\ }\href {\doibase 10.1086/511055} {\bibfield  {journal} {\bibinfo
  {journal} {\apj}\ }\textbf {\bibinfo {volume} {657}},\ \bibinfo {pages} {810}
  (\bibinfo {year} {2007})},\ \Eprint {http://arxiv.org/abs/astro-ph/0608003}
  {arXiv:astro-ph/0608003 [astro-ph]} \BibitemShut {NoStop}%
\bibitem [{\citenamefont {{Scoville}}\ \emph {et~al.}(2014)\citenamefont
  {{Scoville}}, \citenamefont {{Aussel}}, \citenamefont {{Sheth}},
  \citenamefont {{Scott}}, \citenamefont {{Sanders}}, \citenamefont {{Ivison}},
  \citenamefont {{Pope}}, \citenamefont {{Capak}}, \citenamefont {{Vanden
  Bout}}, \citenamefont {{Manohar}}, \citenamefont {{Kartaltepe}},
  \citenamefont {{Robertson}},\ and\ \citenamefont
  {{Lilly}}}]{2014ApJ...783...84S}%
  \BibitemOpen
  \bibfield  {author} {\bibinfo {author} {\bibfnamefont {N.}~\bibnamefont
  {{Scoville}}}, \bibinfo {author} {\bibfnamefont {H.}~\bibnamefont
  {{Aussel}}}, \bibinfo {author} {\bibfnamefont {K.}~\bibnamefont {{Sheth}}},
  \bibinfo {author} {\bibfnamefont {K.~S.}\ \bibnamefont {{Scott}}}, \bibinfo
  {author} {\bibfnamefont {D.}~\bibnamefont {{Sanders}}}, \bibinfo {author}
  {\bibfnamefont {R.}~\bibnamefont {{Ivison}}}, \bibinfo {author}
  {\bibfnamefont {A.}~\bibnamefont {{Pope}}}, \bibinfo {author} {\bibfnamefont
  {P.}~\bibnamefont {{Capak}}}, \bibinfo {author} {\bibfnamefont
  {P.}~\bibnamefont {{Vanden Bout}}}, \bibinfo {author} {\bibfnamefont
  {S.}~\bibnamefont {{Manohar}}}, \bibinfo {author} {\bibfnamefont
  {J.}~\bibnamefont {{Kartaltepe}}}, \bibinfo {author} {\bibfnamefont
  {B.}~\bibnamefont {{Robertson}}}, \ and\ \bibinfo {author} {\bibfnamefont
  {S.}~\bibnamefont {{Lilly}}},\ }\href {\doibase 10.1088/0004-637X/783/2/84}
  {\bibfield  {journal} {\bibinfo  {journal} {\apj}\ }\textbf {\bibinfo
  {volume} {783}},\ \bibinfo {eid} {84} (\bibinfo {year} {2014})},\ \Eprint
  {http://arxiv.org/abs/1401.2987} {arXiv:1401.2987 [astro-ph.GA]} \BibitemShut
  {NoStop}%
\bibitem [{\citenamefont {{R{\'e}my-Ruyer}}\ \emph {et~al.}(2014)\citenamefont
  {{R{\'e}my-Ruyer}}, \citenamefont {{Madden}}, \citenamefont {{Galliano}},
  \citenamefont {{Galametz}}, \citenamefont {{Takeuchi}}, \citenamefont
  {{Asano}}, \citenamefont {{Zhukovska}}, \citenamefont {{Lebouteiller}},\ and\
  \citenamefont {{et al.}}}]{2014A&A...563A..31R}%
  \BibitemOpen
  \bibfield  {author} {\bibinfo {author} {\bibfnamefont {A.}~\bibnamefont
  {{R{\'e}my-Ruyer}}}, \bibinfo {author} {\bibfnamefont {S.~C.}\ \bibnamefont
  {{Madden}}}, \bibinfo {author} {\bibfnamefont {F.}~\bibnamefont
  {{Galliano}}}, \bibinfo {author} {\bibfnamefont {M.}~\bibnamefont
  {{Galametz}}}, \bibinfo {author} {\bibfnamefont {T.~T.}\ \bibnamefont
  {{Takeuchi}}}, \bibinfo {author} {\bibfnamefont {R.~S.}\ \bibnamefont
  {{Asano}}}, \bibinfo {author} {\bibfnamefont {S.}~\bibnamefont
  {{Zhukovska}}}, \bibinfo {author} {\bibfnamefont {V.}~\bibnamefont
  {{Lebouteiller}}}, \ and\ \bibinfo {author} {\bibnamefont {{et al.}}},\
  }\href {\doibase 10.1051/0004-6361/201322803} {\bibfield  {journal} {\bibinfo
   {journal} {\aap}\ }\textbf {\bibinfo {volume} {563}},\ \bibinfo {eid} {A31}
  (\bibinfo {year} {2014})},\ \Eprint {http://arxiv.org/abs/1312.3442}
  {arXiv:1312.3442 [astro-ph.GA]} \BibitemShut {NoStop}%
\bibitem [{\citenamefont {Gordon}\ \emph {et~al.}(2014)\citenamefont {Gordon},
  \citenamefont {Roman-Duval}, \citenamefont {Bot}, \citenamefont {Meixner},
  \citenamefont {Babler}, \citenamefont {Bernard}, \citenamefont {Bolatto},
  \citenamefont {Boyer},\ and\ \citenamefont {{et al.}}}]{Gordon_2014}%
  \BibitemOpen
  \bibfield  {author} {\bibinfo {author} {\bibfnamefont {K.~D.}\ \bibnamefont
  {Gordon}}, \bibinfo {author} {\bibfnamefont {J.}~\bibnamefont {Roman-Duval}},
  \bibinfo {author} {\bibfnamefont {C.}~\bibnamefont {Bot}}, \bibinfo {author}
  {\bibfnamefont {M.}~\bibnamefont {Meixner}}, \bibinfo {author} {\bibfnamefont
  {B.}~\bibnamefont {Babler}}, \bibinfo {author} {\bibfnamefont {J.-P.}\
  \bibnamefont {Bernard}}, \bibinfo {author} {\bibfnamefont {A.}~\bibnamefont
  {Bolatto}}, \bibinfo {author} {\bibfnamefont {M.~L.}\ \bibnamefont {Boyer}},
  \ and\ \bibinfo {author} {\bibnamefont {{et al.}}},\ }\href {\doibase
  10.1088/0004-637x/797/2/85} {\bibfield  {journal} {\bibinfo  {journal} {ApJ}\
  }\textbf {\bibinfo {volume} {797}},\ \bibinfo {pages} {85} (\bibinfo {year}
  {2014})}\BibitemShut {NoStop}%
\bibitem [{\citenamefont {Lamperti}\ \emph {et~al.}(2019)\citenamefont
  {Lamperti}, \citenamefont {Saintonge}, \citenamefont {De Looze},
  \citenamefont {Accurso}, \citenamefont {Clark}, \citenamefont {Smith},
  \citenamefont {Wilson}, \citenamefont {Brinks},\ and\ \citenamefont {{et
  al.}}}]{10.1093/mnras/stz2311}%
  \BibitemOpen
  \bibfield  {author} {\bibinfo {author} {\bibfnamefont {I.}~\bibnamefont
  {Lamperti}}, \bibinfo {author} {\bibfnamefont {A.}~\bibnamefont {Saintonge}},
  \bibinfo {author} {\bibfnamefont {I.}~\bibnamefont {De Looze}}, \bibinfo
  {author} {\bibfnamefont {G.}~\bibnamefont {Accurso}}, \bibinfo {author}
  {\bibfnamefont {C.~J.~R.}\ \bibnamefont {Clark}}, \bibinfo {author}
  {\bibfnamefont {M.~W.~L.}\ \bibnamefont {Smith}}, \bibinfo {author}
  {\bibfnamefont {C.~D.}\ \bibnamefont {Wilson}}, \bibinfo {author}
  {\bibfnamefont {E.}~\bibnamefont {Brinks}}, \ and\ \bibinfo {author}
  {\bibnamefont {{et al.}}},\ }\href@noop {} {\bibfield  {journal} {\bibinfo
  {journal} {Monthly Notices of the Royal Astronomical Society}\ }\textbf
  {\bibinfo {volume} {489}},\ \bibinfo {pages} {4389} (\bibinfo {year}
  {2019})}\BibitemShut {NoStop}%
\bibitem [{\citenamefont {{Saintonge}}\ \emph {et~al.}(2018)\citenamefont
  {{Saintonge}}, \citenamefont {{Wilson}}, \citenamefont {{Xiao}},
  \citenamefont {{Lin}}, \citenamefont {{Hwang}}, \citenamefont {{Tosaki}},
  \citenamefont {{Bureau}}, \citenamefont {{Cigan}}, \citenamefont {{Clark}},
  \citenamefont {{Clements}} \emph {et~al.}}]{2018MNRAS.481.3497S}%
  \BibitemOpen
  \bibfield  {author} {\bibinfo {author} {\bibfnamefont {A.}~\bibnamefont
  {{Saintonge}}}, \bibinfo {author} {\bibfnamefont {C.~D.}\ \bibnamefont
  {{Wilson}}}, \bibinfo {author} {\bibfnamefont {T.}~\bibnamefont {{Xiao}}},
  \bibinfo {author} {\bibfnamefont {L.}~\bibnamefont {{Lin}}}, \bibinfo
  {author} {\bibfnamefont {H.~S.}\ \bibnamefont {{Hwang}}}, \bibinfo {author}
  {\bibfnamefont {T.}~\bibnamefont {{Tosaki}}}, \bibinfo {author}
  {\bibfnamefont {M.}~\bibnamefont {{Bureau}}}, \bibinfo {author}
  {\bibfnamefont {P.~J.}\ \bibnamefont {{Cigan}}}, \bibinfo {author}
  {\bibfnamefont {C.~J.~R.}\ \bibnamefont {{Clark}}}, \bibinfo {author}
  {\bibfnamefont {D.~L.}\ \bibnamefont {{Clements}}},  \emph {et~al.},\ }\href
  {\doibase 10.1093/mnras/sty2499} {\bibfield  {journal} {\bibinfo  {journal}
  {\mnras}\ }\textbf {\bibinfo {volume} {481}},\ \bibinfo {pages} {3497}
  (\bibinfo {year} {2018})},\ \Eprint {http://arxiv.org/abs/1809.07336}
  {arXiv:1809.07336 [astro-ph.GA]} \BibitemShut {NoStop}%
\bibitem [{\citenamefont {{Klessen}}\ and\ \citenamefont
  {{Glover}}(2016)}]{2016SAAS...43...85K}%
  \BibitemOpen
  \bibfield  {author} {\bibinfo {author} {\bibfnamefont {R.~S.}\ \bibnamefont
  {{Klessen}}}\ and\ \bibinfo {author} {\bibfnamefont {S.~C.~O.}\ \bibnamefont
  {{Glover}}},\ }\href {\doibase 10.1007/978-3-662-47890-5\_2} {\bibfield
  {journal} {\bibinfo  {journal} {Saas-Fee Advanced Course}\ }\textbf {\bibinfo
  {volume} {43}},\ \bibinfo {pages} {85} (\bibinfo {year} {2016})},\ \Eprint
  {http://arxiv.org/abs/1412.5182} {arXiv:1412.5182 [astro-ph.GA]} \BibitemShut
  {NoStop}%
\bibitem [{\citenamefont {{Savage}}\ \emph {et~al.}(1977)\citenamefont
  {{Savage}}, \citenamefont {{Bohlin}}, \citenamefont {{Drake}},\ and\
  \citenamefont {{Budich}}}]{1977ApJ...216..291S}%
  \BibitemOpen
  \bibfield  {author} {\bibinfo {author} {\bibfnamefont {B.~D.}\ \bibnamefont
  {{Savage}}}, \bibinfo {author} {\bibfnamefont {R.~C.}\ \bibnamefont
  {{Bohlin}}}, \bibinfo {author} {\bibfnamefont {J.~F.}\ \bibnamefont
  {{Drake}}}, \ and\ \bibinfo {author} {\bibfnamefont {W.}~\bibnamefont
  {{Budich}}},\ }\href {\doibase 10.1086/155471} {\bibfield  {journal}
  {\bibinfo  {journal} {\apj}\ }\textbf {\bibinfo {volume} {216}},\ \bibinfo
  {pages} {291} (\bibinfo {year} {1977})}\BibitemShut {NoStop}%
\bibitem [{\citenamefont {{Bohlin}}\ \emph {et~al.}(1978)\citenamefont
  {{Bohlin}}, \citenamefont {{Savage}},\ and\ \citenamefont
  {{Drake}}}]{1978ApJ...224..132B}%
  \BibitemOpen
  \bibfield  {author} {\bibinfo {author} {\bibfnamefont {R.~C.}\ \bibnamefont
  {{Bohlin}}}, \bibinfo {author} {\bibfnamefont {B.~D.}\ \bibnamefont
  {{Savage}}}, \ and\ \bibinfo {author} {\bibfnamefont {J.~F.}\ \bibnamefont
  {{Drake}}},\ }\href {\doibase 10.1086/156357} {\bibfield  {journal} {\bibinfo
   {journal} {\apj}\ }\textbf {\bibinfo {volume} {224}},\ \bibinfo {pages}
  {132} (\bibinfo {year} {1978})}\BibitemShut {NoStop}%
\bibitem [{\citenamefont {{Shull}}\ \emph {et~al.}(2021)\citenamefont
  {{Shull}}, \citenamefont {{Danforth}},\ and\ \citenamefont
  {{Anderson}}}]{shull2021}%
  \BibitemOpen
  \bibfield  {author} {\bibinfo {author} {\bibfnamefont {J.~M.}\ \bibnamefont
  {{Shull}}}, \bibinfo {author} {\bibfnamefont {C.~W.}\ \bibnamefont
  {{Danforth}}}, \ and\ \bibinfo {author} {\bibfnamefont {K.~L.}\ \bibnamefont
  {{Anderson}}},\ }\href {\doibase 10.3847/1538-4357/abe707} {\bibfield
  {journal} {\bibinfo  {journal} {\apj}\ }\textbf {\bibinfo {volume} {911}},\
  \bibinfo {eid} {55} (\bibinfo {year} {2021})},\ \Eprint
  {http://arxiv.org/abs/2102.11301} {arXiv:2102.11301 [astro-ph.GA]}
  \BibitemShut {NoStop}%
\bibitem [{\citenamefont {{Predehl}}\ and\ \citenamefont
  {{Schmitt}}(1995)}]{1995A&A...293..889P}%
  \BibitemOpen
  \bibfield  {author} {\bibinfo {author} {\bibfnamefont {P.}~\bibnamefont
  {{Predehl}}}\ and\ \bibinfo {author} {\bibfnamefont {J.~H.~M.~M.}\
  \bibnamefont {{Schmitt}}},\ }\href@noop {} {\bibfield  {journal} {\bibinfo
  {journal} {\aap}\ }\textbf {\bibinfo {volume} {293}},\ \bibinfo {pages} {889}
  (\bibinfo {year} {1995})}\BibitemShut {NoStop}%
\bibitem [{\citenamefont {Liszt}(2013)}]{Liszt_2013}%
  \BibitemOpen
  \bibfield  {author} {\bibinfo {author} {\bibfnamefont {H.}~\bibnamefont
  {Liszt}},\ }\href {\doibase 10.1088/0004-637x/780/1/10} {\bibfield  {journal}
  {\bibinfo  {journal} {The Astrophysical Journal}\ }\textbf {\bibinfo {volume}
  {780}},\ \bibinfo {pages} {10} (\bibinfo {year} {2013})}\BibitemShut
  {NoStop}%
\bibitem [{\citenamefont {{Lenz}}\ \emph {et~al.}(2017)\citenamefont {{Lenz}},
  \citenamefont {{Hensley}},\ and\ \citenamefont
  {{Dor{\'e}}}}]{2017ApJ...846...38L}%
  \BibitemOpen
  \bibfield  {author} {\bibinfo {author} {\bibfnamefont {D.}~\bibnamefont
  {{Lenz}}}, \bibinfo {author} {\bibfnamefont {B.~S.}\ \bibnamefont
  {{Hensley}}}, \ and\ \bibinfo {author} {\bibfnamefont {O.}~\bibnamefont
  {{Dor{\'e}}}},\ }\href {\doibase 10.3847/1538-4357/aa84af} {\bibfield
  {journal} {\bibinfo  {journal} {\apj}\ }\textbf {\bibinfo {volume} {846}},\
  \bibinfo {eid} {38} (\bibinfo {year} {2017})},\ \Eprint
  {http://arxiv.org/abs/1706.00011} {arXiv:1706.00011 [astro-ph.GA]}
  \BibitemShut {NoStop}%
\bibitem [{\citenamefont {{Kalberla}}\ \emph {et~al.}(2020)\citenamefont
  {{Kalberla}}, \citenamefont {{Kerp}},\ and\ \citenamefont
  {{Haud}}}]{2020A&A...639A..26K}%
  \BibitemOpen
  \bibfield  {author} {\bibinfo {author} {\bibfnamefont {P.~M.~W.}\
  \bibnamefont {{Kalberla}}}, \bibinfo {author} {\bibfnamefont
  {J.}~\bibnamefont {{Kerp}}}, \ and\ \bibinfo {author} {\bibfnamefont
  {U.}~\bibnamefont {{Haud}}},\ }\href {\doibase 10.1051/0004-6361/202037602}
  {\bibfield  {journal} {\bibinfo  {journal} {\aap}\ }\textbf {\bibinfo
  {volume} {639}},\ \bibinfo {eid} {A26} (\bibinfo {year} {2020})},\ \Eprint
  {http://arxiv.org/abs/2004.14630} {arXiv:2004.14630 [astro-ph.GA]}
  \BibitemShut {NoStop}%
\bibitem [{\citenamefont {{Lallement}}\ \emph {et~al.}(2019)\citenamefont
  {{Lallement}}, \citenamefont {{Babusiaux}}, \citenamefont {{Vergely}},
  \citenamefont {{Katz}}, \citenamefont {{Arenou}}, \citenamefont {{Valette}},
  \citenamefont {{Hottier}},\ and\ \citenamefont
  {{Capitanio}}}]{2019A&A...625A.135L}%
  \BibitemOpen
  \bibfield  {author} {\bibinfo {author} {\bibfnamefont {R.}~\bibnamefont
  {{Lallement}}}, \bibinfo {author} {\bibfnamefont {C.}~\bibnamefont
  {{Babusiaux}}}, \bibinfo {author} {\bibfnamefont {J.~L.}\ \bibnamefont
  {{Vergely}}}, \bibinfo {author} {\bibfnamefont {D.}~\bibnamefont {{Katz}}},
  \bibinfo {author} {\bibfnamefont {F.}~\bibnamefont {{Arenou}}}, \bibinfo
  {author} {\bibfnamefont {B.}~\bibnamefont {{Valette}}}, \bibinfo {author}
  {\bibfnamefont {C.}~\bibnamefont {{Hottier}}}, \ and\ \bibinfo {author}
  {\bibfnamefont {L.}~\bibnamefont {{Capitanio}}},\ }\href {\doibase
  10.1051/0004-6361/201834695} {\bibfield  {journal} {\bibinfo  {journal}
  {\aap}\ }\textbf {\bibinfo {volume} {625}},\ \bibinfo {eid} {A135} (\bibinfo
  {year} {2019})},\ \Eprint {http://arxiv.org/abs/1902.04116} {arXiv:1902.04116
  [astro-ph.GA]} \BibitemShut {NoStop}%
\bibitem [{\citenamefont {{Gaia Collaboration}}(2018)}]{2018A&A...616A...1G}%
  \BibitemOpen
  \bibfield  {author} {\bibinfo {author} {\bibnamefont {{Gaia
  Collaboration}}},\ }\href {\doibase 10.1051/0004-6361/201833051} {\bibfield
  {journal} {\bibinfo  {journal} {\aap}\ }\textbf {\bibinfo {volume} {616}},\
  \bibinfo {eid} {A1} (\bibinfo {year} {2018})},\ \Eprint
  {http://arxiv.org/abs/1804.09365} {arXiv:1804.09365} \BibitemShut {NoStop}%
\bibitem [{\citenamefont {Ackermann}\ \emph {et~al.}(2012)\citenamefont
  {Ackermann} \emph {et~al.}}]{Fermi-LAT:2012edv}%
  \BibitemOpen
  \bibfield  {author} {\bibinfo {author} {\bibfnamefont {M.}~\bibnamefont
  {Ackermann}} \emph {et~al.} (\bibinfo {collaboration} {Fermi-LAT}),\ }\href
  {\doibase 10.1088/0004-637X/750/1/3} {\bibfield  {journal} {\bibinfo
  {journal} {Astrophys. J.}\ }\textbf {\bibinfo {volume} {750}},\ \bibinfo
  {pages} {3} (\bibinfo {year} {2012})},\ \Eprint
  {http://arxiv.org/abs/1202.4039} {arXiv:1202.4039 [astro-ph.HE]} \BibitemShut
  {NoStop}%
\bibitem [{\citenamefont {Ackermann}\ \emph {et~al.}(2015)\citenamefont
  {Ackermann} \emph {et~al.}}]{Fermi-LAT:2014ryh}%
  \BibitemOpen
  \bibfield  {author} {\bibinfo {author} {\bibfnamefont {M.}~\bibnamefont
  {Ackermann}} \emph {et~al.} (\bibinfo {collaboration} {Fermi-LAT}),\ }\href
  {\doibase 10.1088/0004-637X/799/1/86} {\bibfield  {journal} {\bibinfo
  {journal} {Astrophys. J.}\ }\textbf {\bibinfo {volume} {799}},\ \bibinfo
  {pages} {86} (\bibinfo {year} {2015})},\ \Eprint
  {http://arxiv.org/abs/1410.3696} {arXiv:1410.3696 [astro-ph.HE]} \BibitemShut
  {NoStop}%
\bibitem [{\citenamefont {Tibaldo}\ \emph {et~al.}(2021)\citenamefont
  {Tibaldo}, \citenamefont {Gaggero},\ and\ \citenamefont
  {Martin}}]{Tibaldo:2021viq}%
  \BibitemOpen
  \bibfield  {author} {\bibinfo {author} {\bibfnamefont {L.}~\bibnamefont
  {Tibaldo}}, \bibinfo {author} {\bibfnamefont {D.}~\bibnamefont {Gaggero}}, \
  and\ \bibinfo {author} {\bibfnamefont {P.}~\bibnamefont {Martin}},\ }\href
  {\doibase 10.3390/universe7050141} {\bibfield  {journal} {\bibinfo  {journal}
  {Universe}\ }\textbf {\bibinfo {volume} {7}},\ \bibinfo {pages} {141}
  (\bibinfo {year} {2021})},\ \Eprint {http://arxiv.org/abs/2103.16423}
  {arXiv:2103.16423 [astro-ph.HE]} \BibitemShut {NoStop}%
\bibitem [{\citenamefont {Porter}\ \emph {et~al.}(2017)\citenamefont {Porter},
  \citenamefont {Johannesson},\ and\ \citenamefont
  {Moskalenko}}]{Porter:2017vaa}%
  \BibitemOpen
  \bibfield  {author} {\bibinfo {author} {\bibfnamefont {T.~A.}\ \bibnamefont
  {Porter}}, \bibinfo {author} {\bibfnamefont {G.}~\bibnamefont {Johannesson}},
  \ and\ \bibinfo {author} {\bibfnamefont {I.~V.}\ \bibnamefont {Moskalenko}},\
  }\href {\doibase 10.3847/1538-4357/aa844d} {\bibfield  {journal} {\bibinfo
  {journal} {Astrophys. J.}\ }\textbf {\bibinfo {volume} {846}},\ \bibinfo
  {pages} {67} (\bibinfo {year} {2017})},\ \Eprint
  {http://arxiv.org/abs/1708.00816} {arXiv:1708.00816 [astro-ph.HE]}
  \BibitemShut {NoStop}%
\bibitem [{\citenamefont {J\'ohannesson}\ \emph {et~al.}(2018)\citenamefont
  {J\'ohannesson}, \citenamefont {Porter},\ and\ \citenamefont
  {Moskalenko}}]{Johannesson:2018bit}%
  \BibitemOpen
  \bibfield  {author} {\bibinfo {author} {\bibfnamefont {G.}~\bibnamefont
  {J\'ohannesson}}, \bibinfo {author} {\bibfnamefont {T.~A.}\ \bibnamefont
  {Porter}}, \ and\ \bibinfo {author} {\bibfnamefont {I.~V.}\ \bibnamefont
  {Moskalenko}},\ }\href {\doibase 10.3847/1538-4357/aab26e} {\bibfield
  {journal} {\bibinfo  {journal} {Astrophys. J.}\ }\textbf {\bibinfo {volume}
  {856}},\ \bibinfo {pages} {45} (\bibinfo {year} {2018})},\ \Eprint
  {http://arxiv.org/abs/1802.08646} {arXiv:1802.08646 [astro-ph.HE]}
  \BibitemShut {NoStop}%
\bibitem [{\citenamefont {Kissmann}\ \emph {et~al.}(2017)\citenamefont
  {Kissmann}, \citenamefont {Niederwanger}, \citenamefont {Reimer},\ and\
  \citenamefont {Strong}}]{Kissmann:2017ghg}%
  \BibitemOpen
  \bibfield  {author} {\bibinfo {author} {\bibfnamefont {R.}~\bibnamefont
  {Kissmann}}, \bibinfo {author} {\bibfnamefont {F.}~\bibnamefont
  {Niederwanger}}, \bibinfo {author} {\bibfnamefont {O.}~\bibnamefont
  {Reimer}}, \ and\ \bibinfo {author} {\bibfnamefont {A.~W.}\ \bibnamefont
  {Strong}},\ }\href {\doibase 10.1063/1.4969008} {\bibfield  {journal}
  {\bibinfo  {journal} {AIP Conf. Proc.}\ }\textbf {\bibinfo {volume} {1792}},\
  \bibinfo {pages} {070011} (\bibinfo {year} {2017})},\ \Eprint
  {http://arxiv.org/abs/1701.07285} {arXiv:1701.07285 [astro-ph.HE]}
  \BibitemShut {NoStop}%
\bibitem [{\citenamefont {Dundovic}\ \emph {et~al.}(2021)\citenamefont
  {Dundovic}, \citenamefont {Evoli}, \citenamefont {Gaggero},\ and\
  \citenamefont {Grasso}}]{Dundovic:2021ryb}%
  \BibitemOpen
  \bibfield  {author} {\bibinfo {author} {\bibfnamefont {A.}~\bibnamefont
  {Dundovic}}, \bibinfo {author} {\bibfnamefont {C.}~\bibnamefont {Evoli}},
  \bibinfo {author} {\bibfnamefont {D.}~\bibnamefont {Gaggero}}, \ and\
  \bibinfo {author} {\bibfnamefont {D.}~\bibnamefont {Grasso}},\ }\href
  {\doibase 10.1051/0004-6361/202140801} {\bibfield  {journal} {\bibinfo
  {journal} {Astron. Astrophys.}\ }\textbf {\bibinfo {volume} {653}},\ \bibinfo
  {pages} {A18} (\bibinfo {year} {2021})},\ \Eprint
  {http://arxiv.org/abs/2105.13165} {arXiv:2105.13165 [astro-ph.HE]}
  \BibitemShut {NoStop}%
\bibitem [{\citenamefont {Korsmeier}\ and\ \citenamefont
  {Cuoco}(2021)}]{Korsmeier:2021brc}%
  \BibitemOpen
  \bibfield  {author} {\bibinfo {author} {\bibfnamefont {M.}~\bibnamefont
  {Korsmeier}}\ and\ \bibinfo {author} {\bibfnamefont {A.}~\bibnamefont
  {Cuoco}},\ }\href {\doibase 10.1103/PhysRevD.103.103016} {\bibfield
  {journal} {\bibinfo  {journal} {Phys. Rev. D}\ }\textbf {\bibinfo {volume}
  {103}},\ \bibinfo {pages} {103016} (\bibinfo {year} {2021})},\ \Eprint
  {http://arxiv.org/abs/2103.09824} {arXiv:2103.09824 [astro-ph.HE]}
  \BibitemShut {NoStop}%
\bibitem [{\citenamefont {Strong}\ \emph {et~al.}(2000)\citenamefont {Strong},
  \citenamefont {Moskalenko},\ and\ \citenamefont {Reimer}}]{Strong:1998fr}%
  \BibitemOpen
  \bibfield  {author} {\bibinfo {author} {\bibfnamefont {A.}~\bibnamefont
  {Strong}}, \bibinfo {author} {\bibfnamefont {I.}~\bibnamefont {Moskalenko}},
  \ and\ \bibinfo {author} {\bibfnamefont {O.}~\bibnamefont {Reimer}},\ }\href
  {\doibase 10.1086/309038} {\bibfield  {journal} {\bibinfo  {journal}
  {Astrophys. J.}\ }\textbf {\bibinfo {volume} {537}},\ \bibinfo {pages} {763}
  (\bibinfo {year} {2000})},\ \bibinfo {note} {[Erratum: ApJ. 541, 1109
  (2000)]},\ \Eprint {http://arxiv.org/abs/astro-ph/9811296}
  {arXiv:astro-ph/9811296} \BibitemShut {NoStop}%
\bibitem [{\citenamefont {Abdollahi}\ \emph {et~al.}(2020)\citenamefont
  {Abdollahi} \emph {et~al.}}]{Fermi-LAT:2019yla}%
  \BibitemOpen
  \bibfield  {author} {\bibinfo {author} {\bibfnamefont {S.}~\bibnamefont
  {Abdollahi}} \emph {et~al.} (\bibinfo {collaboration} {Fermi-LAT}),\ }\href
  {\doibase 10.3847/1538-4365/ab6bcb} {\bibfield  {journal} {\bibinfo
  {journal} {Astrophys. J. Suppl.}\ }\textbf {\bibinfo {volume} {247}},\
  \bibinfo {pages} {33} (\bibinfo {year} {2020})},\ \Eprint
  {http://arxiv.org/abs/1902.10045} {arXiv:1902.10045 [astro-ph.HE]}
  \BibitemShut {NoStop}%
\bibitem [{\citenamefont {{McKee}}\ and\ \citenamefont
  {{Ostriker}}(1977)}]{1977ApJ...218..148M}%
  \BibitemOpen
  \bibfield  {author} {\bibinfo {author} {\bibfnamefont {C.~F.}\ \bibnamefont
  {{McKee}}}\ and\ \bibinfo {author} {\bibfnamefont {J.~P.}\ \bibnamefont
  {{Ostriker}}},\ }\href {\doibase 10.1086/155667} {\bibfield  {journal}
  {\bibinfo  {journal} {\apj}\ }\textbf {\bibinfo {volume} {218}},\ \bibinfo
  {pages} {148} (\bibinfo {year} {1977})}\BibitemShut {NoStop}%
\bibitem [{\citenamefont {{Cox}}(2005)}]{2005ARA&A..43..337C}%
  \BibitemOpen
  \bibfield  {author} {\bibinfo {author} {\bibfnamefont {D.~P.}\ \bibnamefont
  {{Cox}}},\ }\href {\doibase 10.1146/annurev.astro.43.072103.150615}
  {\bibfield  {journal} {\bibinfo  {journal} {\araa}\ }\textbf {\bibinfo
  {volume} {43}},\ \bibinfo {pages} {337} (\bibinfo {year} {2005})}\BibitemShut
  {NoStop}%
\bibitem [{\citenamefont {{Draine}}(2011)}]{2011piim.book.....D}%
  \BibitemOpen
  \bibfield  {author} {\bibinfo {author} {\bibfnamefont {B.~T.}\ \bibnamefont
  {{Draine}}},\ }\href@noop {} {\emph {\bibinfo {title} {{Physics of the
  Interstellar and Intergalactic Medium}}}}\ (\bibinfo {year}
  {2011})\BibitemShut {NoStop}%
\bibitem [{\citenamefont {{Planck Collaboration}}(2014)}]{2014A&A...571A..13P}%
  \BibitemOpen
  \bibfield  {author} {\bibinfo {author} {\bibnamefont {{Planck
  Collaboration}}},\ }\href {\doibase 10.1051/0004-6361/201321553} {\bibfield
  {journal} {\bibinfo  {journal} {\aap}\ }\textbf {\bibinfo {volume} {571}},\
  \bibinfo {eid} {A13} (\bibinfo {year} {2014})},\ \Eprint
  {http://arxiv.org/abs/1303.5073} {arXiv:1303.5073 [astro-ph.GA]} \BibitemShut
  {NoStop}%
\bibitem [{\citenamefont {{Dickman}}(1978)}]{1978ApJS...37..407D}%
  \BibitemOpen
  \bibfield  {author} {\bibinfo {author} {\bibfnamefont {R.~L.}\ \bibnamefont
  {{Dickman}}},\ }\href {\doibase 10.1086/190535} {\bibfield  {journal}
  {\bibinfo  {journal} {\apjs}\ }\textbf {\bibinfo {volume} {37}},\ \bibinfo
  {pages} {407} (\bibinfo {year} {1978})}\BibitemShut {NoStop}%
\bibitem [{\citenamefont {{Green}}\ \emph {et~al.}(2019)\citenamefont
  {{Green}}, \citenamefont {{Schlafly}}, \citenamefont {{Zucker}},
  \citenamefont {{Speagle}},\ and\ \citenamefont
  {{Finkbeiner}}}]{2019ApJ...887...93G}%
  \BibitemOpen
  \bibfield  {author} {\bibinfo {author} {\bibfnamefont {G.~M.}\ \bibnamefont
  {{Green}}}, \bibinfo {author} {\bibfnamefont {E.}~\bibnamefont {{Schlafly}}},
  \bibinfo {author} {\bibfnamefont {C.}~\bibnamefont {{Zucker}}}, \bibinfo
  {author} {\bibfnamefont {J.~S.}\ \bibnamefont {{Speagle}}}, \ and\ \bibinfo
  {author} {\bibfnamefont {D.}~\bibnamefont {{Finkbeiner}}},\ }\href {\doibase
  10.3847/1538-4357/ab5362} {\bibfield  {journal} {\bibinfo  {journal} {\apj}\
  }\textbf {\bibinfo {volume} {887}},\ \bibinfo {eid} {93} (\bibinfo {year}
  {2019})},\ \Eprint {http://arxiv.org/abs/1905.02734} {arXiv:1905.02734
  [astro-ph.GA]} \BibitemShut {NoStop}%
\bibitem [{\citenamefont {{HI4PI Collaboration}}(2016)}]{2016A&A...594A.116H}%
  \BibitemOpen
  \bibfield  {author} {\bibinfo {author} {\bibnamefont {{HI4PI
  Collaboration}}},\ }\href {\doibase 10.1051/0004-6361/201629178} {\bibfield
  {journal} {\bibinfo  {journal} {\aap}\ }\textbf {\bibinfo {volume} {594}},\
  \bibinfo {eid} {A116} (\bibinfo {year} {2016})},\ \Eprint
  {http://arxiv.org/abs/1610.06175} {arXiv:1610.06175 [astro-ph.GA]}
  \BibitemShut {NoStop}%
\bibitem [{\citenamefont {{Wilson}}\ \emph {et~al.}(2013)\citenamefont
  {{Wilson}}, \citenamefont {{Rohlfs}},\ and\ \citenamefont
  {{H{\"u}ttemeister}}}]{21cm_conv_fact}%
  \BibitemOpen
  \bibfield  {author} {\bibinfo {author} {\bibfnamefont {T.~L.}\ \bibnamefont
  {{Wilson}}}, \bibinfo {author} {\bibfnamefont {K.}~\bibnamefont {{Rohlfs}}},
  \ and\ \bibinfo {author} {\bibfnamefont {S.}~\bibnamefont
  {{H{\"u}ttemeister}}},\ }\href {\doibase 10.1007/978-3-642-39950-3} {\emph
  {\bibinfo {title} {{Tools of Radio Astronomy}}}}\ (\bibinfo {year}
  {2013})\BibitemShut {NoStop}%
\bibitem [{\citenamefont {{Murray}}\ \emph {et~al.}(2018)\citenamefont
  {{Murray}}, \citenamefont {{Stanimirovi{\'c}}}, \citenamefont {{Goss}},
  \citenamefont {{Heiles}}, \citenamefont {{Dickey}}, \citenamefont
  {{Babler}},\ and\ \citenamefont {{Kim}}}]{2018ApJS..238...14M}%
  \BibitemOpen
  \bibfield  {author} {\bibinfo {author} {\bibfnamefont {C.}~\bibnamefont
  {{Murray}}}, \bibinfo {author} {\bibfnamefont {S.}~\bibnamefont
  {{Stanimirovi{\'c}}}}, \bibinfo {author} {\bibfnamefont {W.}~\bibnamefont
  {{Goss}}}, \bibinfo {author} {\bibfnamefont {C.}~\bibnamefont {{Heiles}}},
  \bibinfo {author} {\bibfnamefont {J.}~\bibnamefont {{Dickey}}}, \bibinfo
  {author} {\bibfnamefont {B.}~\bibnamefont {{Babler}}}, \ and\ \bibinfo
  {author} {\bibfnamefont {C.-G.}\ \bibnamefont {{Kim}}},\ }\href {\doibase
  10.3847/1538-4365/aad81a} {\bibfield  {journal} {\bibinfo  {journal} {\apjs}\
  }\textbf {\bibinfo {volume} {238}},\ \bibinfo {eid} {14} (\bibinfo {year}
  {2018})},\ \Eprint {http://arxiv.org/abs/1806.06065} {arXiv:1806.06065}
  \BibitemShut {NoStop}%
\bibitem [{\citenamefont {{Kalberla}}\ and\ \citenamefont
  {{Kerp}}(2009)}]{2009ARA&A..47...27K}%
  \BibitemOpen
  \bibfield  {author} {\bibinfo {author} {\bibfnamefont {P.~M.~W.}\
  \bibnamefont {{Kalberla}}}\ and\ \bibinfo {author} {\bibfnamefont
  {J.}~\bibnamefont {{Kerp}}},\ }\href {\doibase
  10.1146/annurev-astro-082708-101823} {\bibfield  {journal} {\bibinfo
  {journal} {\araa}\ }\textbf {\bibinfo {volume} {47}},\ \bibinfo {pages} {27}
  (\bibinfo {year} {2009})}\BibitemShut {NoStop}%
\bibitem [{\citenamefont {{Bronfman}}\ \emph {et~al.}(1988)\citenamefont
  {{Bronfman}}, \citenamefont {{Cohen}}, \citenamefont {{Alvarez}},
  \citenamefont {{May}},\ and\ \citenamefont
  {{Thaddeus}}}]{1988ApJ...324..248B}%
  \BibitemOpen
  \bibfield  {author} {\bibinfo {author} {\bibfnamefont {L.}~\bibnamefont
  {{Bronfman}}}, \bibinfo {author} {\bibfnamefont {R.~S.}\ \bibnamefont
  {{Cohen}}}, \bibinfo {author} {\bibfnamefont {H.}~\bibnamefont {{Alvarez}}},
  \bibinfo {author} {\bibfnamefont {J.}~\bibnamefont {{May}}}, \ and\ \bibinfo
  {author} {\bibfnamefont {P.}~\bibnamefont {{Thaddeus}}},\ }\href {\doibase
  10.1086/165892} {\bibfield  {journal} {\bibinfo  {journal} {\apj}\ }\textbf
  {\bibinfo {volume} {324}},\ \bibinfo {pages} {248} (\bibinfo {year}
  {1988})}\BibitemShut {NoStop}%
\bibitem [{\citenamefont {{Ferri{\`e}re}}(2001)}]{2001RvMP...73.1031F}%
  \BibitemOpen
  \bibfield  {author} {\bibinfo {author} {\bibfnamefont {K.~M.}\ \bibnamefont
  {{Ferri{\`e}re}}},\ }\href {\doibase 10.1103/RevModPhys.73.1031} {\bibfield
  {journal} {\bibinfo  {journal} {Reviews of Modern Physics}\ }\textbf
  {\bibinfo {volume} {73}},\ \bibinfo {pages} {1031} (\bibinfo {year}
  {2001})},\ \Eprint {http://arxiv.org/abs/astro-ph/0106359}
  {arXiv:astro-ph/0106359 [astro-ph]} \BibitemShut {NoStop}%
\bibitem [{\citenamefont {Kramer}\ and\ \citenamefont
  {Randall}(2016)}]{Kramer_2016}%
  \BibitemOpen
  \bibfield  {author} {\bibinfo {author} {\bibfnamefont {E.~D.}\ \bibnamefont
  {Kramer}}\ and\ \bibinfo {author} {\bibfnamefont {L.}~\bibnamefont
  {Randall}},\ }\href {\doibase 10.3847/0004-637x/829/2/126} {\bibfield
  {journal} {\bibinfo  {journal} {The Astrophysical Journal}\ }\textbf
  {\bibinfo {volume} {829}},\ \bibinfo {pages} {126} (\bibinfo {year}
  {2016})}\BibitemShut {NoStop}%
\bibitem [{\citenamefont {{McKee}}\ \emph {et~al.}(2015)\citenamefont
  {{McKee}}, \citenamefont {{Parravano}},\ and\ \citenamefont
  {{Hollenbach}}}]{2015ApJ...814...13M}%
  \BibitemOpen
  \bibfield  {author} {\bibinfo {author} {\bibfnamefont {C.~F.}\ \bibnamefont
  {{McKee}}}, \bibinfo {author} {\bibfnamefont {A.}~\bibnamefont
  {{Parravano}}}, \ and\ \bibinfo {author} {\bibfnamefont {D.~J.}\ \bibnamefont
  {{Hollenbach}}},\ }\href {\doibase 10.1088/0004-637X/814/1/13} {\bibfield
  {journal} {\bibinfo  {journal} {\apj}\ }\textbf {\bibinfo {volume} {814}},\
  \bibinfo {eid} {13} (\bibinfo {year} {2015})},\ \Eprint
  {http://arxiv.org/abs/1509.05334} {arXiv:1509.05334 [astro-ph.GA]}
  \BibitemShut {NoStop}%
\bibitem [{\citenamefont {Ballet}\ \emph {et~al.}(2020)\citenamefont {Ballet},
  \citenamefont {Burnett}, \citenamefont {Digel},\ and\ \citenamefont
  {Lott}}]{Ballet:2020hze}%
  \BibitemOpen
  \bibfield  {author} {\bibinfo {author} {\bibfnamefont {J.}~\bibnamefont
  {Ballet}}, \bibinfo {author} {\bibfnamefont {T.~H.}\ \bibnamefont {Burnett}},
  \bibinfo {author} {\bibfnamefont {S.~W.}\ \bibnamefont {Digel}}, \ and\
  \bibinfo {author} {\bibfnamefont {B.}~\bibnamefont {Lott}} (\bibinfo
  {collaboration} {Fermi-LAT}),\ }\href@noop {} {\  (\bibinfo {year} {2020})},\
  \Eprint {http://arxiv.org/abs/2005.11208} {arXiv:2005.11208 [astro-ph.HE]}
  \BibitemShut {NoStop}%
\bibitem [{\citenamefont {Su}\ \emph {et~al.}(2010)\citenamefont {Su},
  \citenamefont {Slatyer},\ and\ \citenamefont {Finkbeiner}}]{Su:2010qj}%
  \BibitemOpen
  \bibfield  {author} {\bibinfo {author} {\bibfnamefont {M.}~\bibnamefont
  {Su}}, \bibinfo {author} {\bibfnamefont {T.~R.}\ \bibnamefont {Slatyer}}, \
  and\ \bibinfo {author} {\bibfnamefont {D.~P.}\ \bibnamefont {Finkbeiner}},\
  }\href {\doibase 10.1088/0004-637X/724/2/1044} {\bibfield  {journal}
  {\bibinfo  {journal} {Astrophys. J.}\ }\textbf {\bibinfo {volume} {724}},\
  \bibinfo {pages} {1044} (\bibinfo {year} {2010})},\ \Eprint
  {http://arxiv.org/abs/1005.5480} {arXiv:1005.5480 [astro-ph.HE]} \BibitemShut
  {NoStop}%
\bibitem [{\citenamefont {{Casandjian}}\ and\ \citenamefont
  {{Grenier}}(2009)}]{2009arXiv0912.3478C}%
  \BibitemOpen
  \bibfield  {author} {\bibinfo {author} {\bibfnamefont {J.-M.}\ \bibnamefont
  {{Casandjian}}}\ and\ \bibinfo {author} {\bibfnamefont {I.}~\bibnamefont
  {{Grenier}}},\ }\href@noop {} {\bibfield  {journal} {\bibinfo  {journal}
  {arXiv e-prints}\ ,\ \bibinfo {eid} {arXiv:0912.3478}} (\bibinfo {year}
  {2009})},\ \Eprint {http://arxiv.org/abs/0912.3478} {arXiv:0912.3478
  [astro-ph.HE]} \BibitemShut {NoStop}%
\bibitem [{\citenamefont {{Glover}}\ and\ \citenamefont {{Mac
  Low}}(2011)}]{2011MNRAS.412..337G}%
  \BibitemOpen
  \bibfield  {author} {\bibinfo {author} {\bibfnamefont {S.~C.~O.}\
  \bibnamefont {{Glover}}}\ and\ \bibinfo {author} {\bibfnamefont {M.~M.}\
  \bibnamefont {{Mac Low}}},\ }\href {\doibase
  10.1111/j.1365-2966.2010.17907.x} {\bibfield  {journal} {\bibinfo  {journal}
  {\mnras}\ }\textbf {\bibinfo {volume} {412}},\ \bibinfo {pages} {337}
  (\bibinfo {year} {2011})},\ \Eprint {http://arxiv.org/abs/1003.1340}
  {arXiv:1003.1340 [astro-ph.GA]} \BibitemShut {NoStop}%
\bibitem [{\citenamefont {{Bailer-Jones}}\ \emph {et~al.}(2021)\citenamefont
  {{Bailer-Jones}}, \citenamefont {{Rybizki}}, \citenamefont {{Fouesneau}},
  \citenamefont {{Demleitner}},\ and\ \citenamefont
  {{Andrae}}}]{2021AJ....161..147B}%
  \BibitemOpen
  \bibfield  {author} {\bibinfo {author} {\bibfnamefont {C.~A.~L.}\
  \bibnamefont {{Bailer-Jones}}}, \bibinfo {author} {\bibfnamefont
  {J.}~\bibnamefont {{Rybizki}}}, \bibinfo {author} {\bibfnamefont
  {M.}~\bibnamefont {{Fouesneau}}}, \bibinfo {author} {\bibfnamefont
  {M.}~\bibnamefont {{Demleitner}}}, \ and\ \bibinfo {author} {\bibfnamefont
  {R.}~\bibnamefont {{Andrae}}},\ }\href {\doibase 10.3847/1538-3881/abd806}
  {\bibfield  {journal} {\bibinfo  {journal} {\aj}\ }\textbf {\bibinfo {volume}
  {161}},\ \bibinfo {eid} {147} (\bibinfo {year} {2021})},\ \Eprint
  {http://arxiv.org/abs/2012.05220} {arXiv:2012.05220 [astro-ph.SR]}
  \BibitemShut {NoStop}%
\bibitem [{\citenamefont {{Anders}}\ \emph {et~al.}(2022)\citenamefont
  {{Anders}}, \citenamefont {{Khalatyan}}, \citenamefont {{Queiroz}},
  \citenamefont {{Chiappini}}, \citenamefont {{Ard{\`e}vol}}, \citenamefont
  {{Casamiquela}}, \citenamefont {{Figueras}}, \citenamefont
  {{Jim{\'e}nez-Arranz}},\ and\ \citenamefont {{et
  al.}}}]{2022A&A...658A..91A}%
  \BibitemOpen
  \bibfield  {author} {\bibinfo {author} {\bibfnamefont {F.}~\bibnamefont
  {{Anders}}}, \bibinfo {author} {\bibfnamefont {A.}~\bibnamefont
  {{Khalatyan}}}, \bibinfo {author} {\bibfnamefont {A.~B.~A.}\ \bibnamefont
  {{Queiroz}}}, \bibinfo {author} {\bibfnamefont {C.}~\bibnamefont
  {{Chiappini}}}, \bibinfo {author} {\bibfnamefont {J.}~\bibnamefont
  {{Ard{\`e}vol}}}, \bibinfo {author} {\bibfnamefont {L.}~\bibnamefont
  {{Casamiquela}}}, \bibinfo {author} {\bibfnamefont {F.}~\bibnamefont
  {{Figueras}}}, \bibinfo {author} {\bibfnamefont {{\'O}.}~\bibnamefont
  {{Jim{\'e}nez-Arranz}}}, \ and\ \bibinfo {author} {\bibnamefont {{et al.}}},\
  }\href {\doibase 10.1051/0004-6361/202142369} {\bibfield  {journal} {\bibinfo
   {journal} {\aap}\ }\textbf {\bibinfo {volume} {658}},\ \bibinfo {eid} {A91}
  (\bibinfo {year} {2022})},\ \Eprint {http://arxiv.org/abs/2111.01860}
  {arXiv:2111.01860 [astro-ph.GA]} \BibitemShut {NoStop}%
\bibitem [{\citenamefont {{Miller}}\ \emph {et~al.}(2022)\citenamefont
  {{Miller}}, \citenamefont {{Anderson}}, \citenamefont {{Leistedt}},
  \citenamefont {{Cunningham}}, \citenamefont {{Hogg}},\ and\ \citenamefont
  {{Blei}}}]{2022arXiv220206797M}%
  \BibitemOpen
  \bibfield  {author} {\bibinfo {author} {\bibfnamefont {A.~C.}\ \bibnamefont
  {{Miller}}}, \bibinfo {author} {\bibfnamefont {L.}~\bibnamefont
  {{Anderson}}}, \bibinfo {author} {\bibfnamefont {B.}~\bibnamefont
  {{Leistedt}}}, \bibinfo {author} {\bibfnamefont {J.~P.}\ \bibnamefont
  {{Cunningham}}}, \bibinfo {author} {\bibfnamefont {D.~W.}\ \bibnamefont
  {{Hogg}}}, \ and\ \bibinfo {author} {\bibfnamefont {D.~M.}\ \bibnamefont
  {{Blei}}},\ }\href@noop {} {\  (\bibinfo {year} {2022})},\ \Eprint
  {http://arxiv.org/abs/2202.06797} {arXiv:2202.06797 [astro-ph.GA]}
  \BibitemShut {NoStop}%
\bibitem [{\citenamefont {{Leike}}\ \emph {et~al.}(2022)\citenamefont
  {{Leike}}, \citenamefont {{Edenhofer}}, \citenamefont {{Knollm{\"u}ller}},
  \citenamefont {{Alig}}, \citenamefont {{Frank}},\ and\ \citenamefont
  {{En{\ss}lin}}}]{2022arXiv220411715L}%
  \BibitemOpen
  \bibfield  {author} {\bibinfo {author} {\bibfnamefont {R.~H.}\ \bibnamefont
  {{Leike}}}, \bibinfo {author} {\bibfnamefont {G.}~\bibnamefont
  {{Edenhofer}}}, \bibinfo {author} {\bibfnamefont {J.}~\bibnamefont
  {{Knollm{\"u}ller}}}, \bibinfo {author} {\bibfnamefont {C.}~\bibnamefont
  {{Alig}}}, \bibinfo {author} {\bibfnamefont {P.}~\bibnamefont {{Frank}}}, \
  and\ \bibinfo {author} {\bibfnamefont {T.~A.}\ \bibnamefont {{En{\ss}lin}}},\
  }\href@noop {} {\  (\bibinfo {year} {2022})},\ \Eprint
  {http://arxiv.org/abs/2204.11715} {arXiv:2204.11715 [astro-ph.GA]}
  \BibitemShut {NoStop}%
\bibitem [{\citenamefont {Harris}\ \emph {et~al.}(2020)\citenamefont {Harris},
  \citenamefont {Millman}, \citenamefont {van~der Walt}, \citenamefont
  {Gommers}, \citenamefont {Virtanen}, \citenamefont {Cournapeau},
  \citenamefont {Wieser}, \citenamefont {Taylor}, \citenamefont {Berg},
  \citenamefont {Smith}, \citenamefont {Kern}, \citenamefont {Picus},
  \citenamefont {Hoyer}, \citenamefont {van Kerkwijk}, \citenamefont {Brett},
  \citenamefont {Haldane}, \citenamefont {del R{\'{i}}o}, \citenamefont
  {Wiebe}, \citenamefont {Peterson}, \citenamefont {G{\'{e}}rard-Marchant},
  \citenamefont {Sheppard}, \citenamefont {Reddy}, \citenamefont {Weckesser},
  \citenamefont {Abbasi}, \citenamefont {Gohlke},\ and\ \citenamefont
  {Oliphant}}]{harris2020array}%
  \BibitemOpen
  \bibfield  {author} {\bibinfo {author} {\bibfnamefont {C.~R.}\ \bibnamefont
  {Harris}}, \bibinfo {author} {\bibfnamefont {K.~J.}\ \bibnamefont {Millman}},
  \bibinfo {author} {\bibfnamefont {S.~J.}\ \bibnamefont {van~der Walt}},
  \bibinfo {author} {\bibfnamefont {R.}~\bibnamefont {Gommers}}, \bibinfo
  {author} {\bibfnamefont {P.}~\bibnamefont {Virtanen}}, \bibinfo {author}
  {\bibfnamefont {D.}~\bibnamefont {Cournapeau}}, \bibinfo {author}
  {\bibfnamefont {E.}~\bibnamefont {Wieser}}, \bibinfo {author} {\bibfnamefont
  {J.}~\bibnamefont {Taylor}}, \bibinfo {author} {\bibfnamefont
  {S.}~\bibnamefont {Berg}}, \bibinfo {author} {\bibfnamefont {N.~J.}\
  \bibnamefont {Smith}}, \bibinfo {author} {\bibfnamefont {R.}~\bibnamefont
  {Kern}}, \bibinfo {author} {\bibfnamefont {M.}~\bibnamefont {Picus}},
  \bibinfo {author} {\bibfnamefont {S.}~\bibnamefont {Hoyer}}, \bibinfo
  {author} {\bibfnamefont {M.~H.}\ \bibnamefont {van Kerkwijk}}, \bibinfo
  {author} {\bibfnamefont {M.}~\bibnamefont {Brett}}, \bibinfo {author}
  {\bibfnamefont {A.}~\bibnamefont {Haldane}}, \bibinfo {author} {\bibfnamefont
  {J.~F.}\ \bibnamefont {del R{\'{i}}o}}, \bibinfo {author} {\bibfnamefont
  {M.}~\bibnamefont {Wiebe}}, \bibinfo {author} {\bibfnamefont
  {P.}~\bibnamefont {Peterson}}, \bibinfo {author} {\bibfnamefont
  {P.}~\bibnamefont {G{\'{e}}rard-Marchant}}, \bibinfo {author} {\bibfnamefont
  {K.}~\bibnamefont {Sheppard}}, \bibinfo {author} {\bibfnamefont
  {T.}~\bibnamefont {Reddy}}, \bibinfo {author} {\bibfnamefont
  {W.}~\bibnamefont {Weckesser}}, \bibinfo {author} {\bibfnamefont
  {H.}~\bibnamefont {Abbasi}}, \bibinfo {author} {\bibfnamefont
  {C.}~\bibnamefont {Gohlke}}, \ and\ \bibinfo {author} {\bibfnamefont {T.~E.}\
  \bibnamefont {Oliphant}},\ }\href {\doibase 10.1038/s41586-020-2649-2}
  {\bibfield  {journal} {\bibinfo  {journal} {Nature}\ }\textbf {\bibinfo
  {volume} {585}},\ \bibinfo {pages} {357} (\bibinfo {year}
  {2020})}\BibitemShut {NoStop}%
\bibitem [{\citenamefont {Virtanen}\ \emph {et~al.}(2020)\citenamefont
  {Virtanen}, \citenamefont {Gommers}, \citenamefont {Oliphant}, \citenamefont
  {Haberland}, \citenamefont {Reddy}, \citenamefont {Cournapeau}, \citenamefont
  {Burovski}, \citenamefont {Peterson}, \citenamefont {Weckesser},
  \citenamefont {Bright}, \citenamefont {{van der Walt}}, \citenamefont
  {Brett}, \citenamefont {Wilson}, \citenamefont {Millman}, \citenamefont
  {Mayorov}, \citenamefont {Nelson}, \citenamefont {Jones}, \citenamefont
  {Kern}, \citenamefont {Larson}, \citenamefont {Carey}, \citenamefont {Polat},
  \citenamefont {Feng}, \citenamefont {Moore}, \citenamefont {{VanderPlas}},
  \citenamefont {Laxalde}, \citenamefont {Perktold}, \citenamefont {Cimrman},
  \citenamefont {Henriksen}, \citenamefont {Quintero}, \citenamefont {Harris},
  \citenamefont {Archibald}, \citenamefont {Ribeiro}, \citenamefont
  {Pedregosa}, \citenamefont {{van Mulbregt}},\ and\ \citenamefont {{SciPy 1.0
  Contributors}}}]{2020SciPy-NMeth}%
  \BibitemOpen
  \bibfield  {author} {\bibinfo {author} {\bibfnamefont {P.}~\bibnamefont
  {Virtanen}}, \bibinfo {author} {\bibfnamefont {R.}~\bibnamefont {Gommers}},
  \bibinfo {author} {\bibfnamefont {T.~E.}\ \bibnamefont {Oliphant}}, \bibinfo
  {author} {\bibfnamefont {M.}~\bibnamefont {Haberland}}, \bibinfo {author}
  {\bibfnamefont {T.}~\bibnamefont {Reddy}}, \bibinfo {author} {\bibfnamefont
  {D.}~\bibnamefont {Cournapeau}}, \bibinfo {author} {\bibfnamefont
  {E.}~\bibnamefont {Burovski}}, \bibinfo {author} {\bibfnamefont
  {P.}~\bibnamefont {Peterson}}, \bibinfo {author} {\bibfnamefont
  {W.}~\bibnamefont {Weckesser}}, \bibinfo {author} {\bibfnamefont
  {J.}~\bibnamefont {Bright}}, \bibinfo {author} {\bibfnamefont {S.~J.}\
  \bibnamefont {{van der Walt}}}, \bibinfo {author} {\bibfnamefont
  {M.}~\bibnamefont {Brett}}, \bibinfo {author} {\bibfnamefont
  {J.}~\bibnamefont {Wilson}}, \bibinfo {author} {\bibfnamefont {K.~J.}\
  \bibnamefont {Millman}}, \bibinfo {author} {\bibfnamefont {N.}~\bibnamefont
  {Mayorov}}, \bibinfo {author} {\bibfnamefont {A.~R.~J.}\ \bibnamefont
  {Nelson}}, \bibinfo {author} {\bibfnamefont {E.}~\bibnamefont {Jones}},
  \bibinfo {author} {\bibfnamefont {R.}~\bibnamefont {Kern}}, \bibinfo {author}
  {\bibfnamefont {E.}~\bibnamefont {Larson}}, \bibinfo {author} {\bibfnamefont
  {C.~J.}\ \bibnamefont {Carey}}, \bibinfo {author} {\bibfnamefont
  {{\.I}.}~\bibnamefont {Polat}}, \bibinfo {author} {\bibfnamefont
  {Y.}~\bibnamefont {Feng}}, \bibinfo {author} {\bibfnamefont {E.~W.}\
  \bibnamefont {Moore}}, \bibinfo {author} {\bibfnamefont {J.}~\bibnamefont
  {{VanderPlas}}}, \bibinfo {author} {\bibfnamefont {D.}~\bibnamefont
  {Laxalde}}, \bibinfo {author} {\bibfnamefont {J.}~\bibnamefont {Perktold}},
  \bibinfo {author} {\bibfnamefont {R.}~\bibnamefont {Cimrman}}, \bibinfo
  {author} {\bibfnamefont {I.}~\bibnamefont {Henriksen}}, \bibinfo {author}
  {\bibfnamefont {E.~A.}\ \bibnamefont {Quintero}}, \bibinfo {author}
  {\bibfnamefont {C.~R.}\ \bibnamefont {Harris}}, \bibinfo {author}
  {\bibfnamefont {A.~M.}\ \bibnamefont {Archibald}}, \bibinfo {author}
  {\bibfnamefont {A.~H.}\ \bibnamefont {Ribeiro}}, \bibinfo {author}
  {\bibfnamefont {F.}~\bibnamefont {Pedregosa}}, \bibinfo {author}
  {\bibfnamefont {P.}~\bibnamefont {{van Mulbregt}}}, \ and\ \bibinfo {author}
  {\bibnamefont {{SciPy 1.0 Contributors}}},\ }\href {\doibase
  10.1038/s41592-019-0686-2} {\bibfield  {journal} {\bibinfo  {journal} {Nature
  Methods}\ }\textbf {\bibinfo {volume} {17}},\ \bibinfo {pages} {261}
  (\bibinfo {year} {2020})}\BibitemShut {NoStop}%
\bibitem [{\citenamefont {Abadi}\ \emph {et~al.}(2015)\citenamefont {Abadi},
  \citenamefont {Agarwal}, \citenamefont {Barham}, \citenamefont {Brevdo},
  \citenamefont {Chen}, \citenamefont {Citro}, \citenamefont {Corrado},
  \citenamefont {Davis}, \citenamefont {Dean}, \citenamefont {Devin},
  \citenamefont {Ghemawat}, \citenamefont {Goodfellow}, \citenamefont {Harp},
  \citenamefont {Irving}, \citenamefont {Isard}, \citenamefont {Jia},
  \citenamefont {Jozefowicz}, \citenamefont {Kaiser}, \citenamefont {Kudlur},
  \citenamefont {Levenberg}, \citenamefont {Man\'{e}}, \citenamefont {Monga},
  \citenamefont {Moore}, \citenamefont {Murray}, \citenamefont {Olah},
  \citenamefont {Schuster}, \citenamefont {Shlens}, \citenamefont {Steiner},
  \citenamefont {Sutskever}, \citenamefont {Talwar}, \citenamefont {Tucker},
  \citenamefont {Vanhoucke}, \citenamefont {Vasudevan}, \citenamefont
  {Vi\'{e}gas}, \citenamefont {Vinyals}, \citenamefont {Warden}, \citenamefont
  {Wattenberg}, \citenamefont {Wicke}, \citenamefont {Yu},\ and\ \citenamefont
  {Zheng}}]{tensorflow2015-whitepaper}%
  \BibitemOpen
  \bibfield  {author} {\bibinfo {author} {\bibfnamefont {M.}~\bibnamefont
  {Abadi}}, \bibinfo {author} {\bibfnamefont {A.}~\bibnamefont {Agarwal}},
  \bibinfo {author} {\bibfnamefont {P.}~\bibnamefont {Barham}}, \bibinfo
  {author} {\bibfnamefont {E.}~\bibnamefont {Brevdo}}, \bibinfo {author}
  {\bibfnamefont {Z.}~\bibnamefont {Chen}}, \bibinfo {author} {\bibfnamefont
  {C.}~\bibnamefont {Citro}}, \bibinfo {author} {\bibfnamefont {G.~S.}\
  \bibnamefont {Corrado}}, \bibinfo {author} {\bibfnamefont {A.}~\bibnamefont
  {Davis}}, \bibinfo {author} {\bibfnamefont {J.}~\bibnamefont {Dean}},
  \bibinfo {author} {\bibfnamefont {M.}~\bibnamefont {Devin}}, \bibinfo
  {author} {\bibfnamefont {S.}~\bibnamefont {Ghemawat}}, \bibinfo {author}
  {\bibfnamefont {I.}~\bibnamefont {Goodfellow}}, \bibinfo {author}
  {\bibfnamefont {A.}~\bibnamefont {Harp}}, \bibinfo {author} {\bibfnamefont
  {G.}~\bibnamefont {Irving}}, \bibinfo {author} {\bibfnamefont
  {M.}~\bibnamefont {Isard}}, \bibinfo {author} {\bibfnamefont
  {Y.}~\bibnamefont {Jia}}, \bibinfo {author} {\bibfnamefont {R.}~\bibnamefont
  {Jozefowicz}}, \bibinfo {author} {\bibfnamefont {L.}~\bibnamefont {Kaiser}},
  \bibinfo {author} {\bibfnamefont {M.}~\bibnamefont {Kudlur}}, \bibinfo
  {author} {\bibfnamefont {J.}~\bibnamefont {Levenberg}}, \bibinfo {author}
  {\bibfnamefont {D.}~\bibnamefont {Man\'{e}}}, \bibinfo {author}
  {\bibfnamefont {R.}~\bibnamefont {Monga}}, \bibinfo {author} {\bibfnamefont
  {S.}~\bibnamefont {Moore}}, \bibinfo {author} {\bibfnamefont
  {D.}~\bibnamefont {Murray}}, \bibinfo {author} {\bibfnamefont
  {C.}~\bibnamefont {Olah}}, \bibinfo {author} {\bibfnamefont {M.}~\bibnamefont
  {Schuster}}, \bibinfo {author} {\bibfnamefont {J.}~\bibnamefont {Shlens}},
  \bibinfo {author} {\bibfnamefont {B.}~\bibnamefont {Steiner}}, \bibinfo
  {author} {\bibfnamefont {I.}~\bibnamefont {Sutskever}}, \bibinfo {author}
  {\bibfnamefont {K.}~\bibnamefont {Talwar}}, \bibinfo {author} {\bibfnamefont
  {P.}~\bibnamefont {Tucker}}, \bibinfo {author} {\bibfnamefont
  {V.}~\bibnamefont {Vanhoucke}}, \bibinfo {author} {\bibfnamefont
  {V.}~\bibnamefont {Vasudevan}}, \bibinfo {author} {\bibfnamefont
  {F.}~\bibnamefont {Vi\'{e}gas}}, \bibinfo {author} {\bibfnamefont
  {O.}~\bibnamefont {Vinyals}}, \bibinfo {author} {\bibfnamefont
  {P.}~\bibnamefont {Warden}}, \bibinfo {author} {\bibfnamefont
  {M.}~\bibnamefont {Wattenberg}}, \bibinfo {author} {\bibfnamefont
  {M.}~\bibnamefont {Wicke}}, \bibinfo {author} {\bibfnamefont
  {Y.}~\bibnamefont {Yu}}, \ and\ \bibinfo {author} {\bibfnamefont
  {X.}~\bibnamefont {Zheng}},\ }\href {http://tensorflow.org/} {\enquote
  {\bibinfo {title} {{TensorFlow}: Large-scale machine learning on
  heterogeneous systems},}\ } (\bibinfo {year} {2015}),\ \bibinfo {note}
  {software available from tensorflow.org}\BibitemShut {NoStop}%
\bibitem [{\citenamefont {{Buchner}}(2016)}]{2016ascl.soft06005B}%
  \BibitemOpen
  \bibfield  {author} {\bibinfo {author} {\bibfnamefont {J.}~\bibnamefont
  {{Buchner}}},\ }\href@noop {} {\enquote {\bibinfo {title} {{PyMultiNest:
  Python interface for MultiNest}},}\ }\bibinfo {howpublished} {Astrophysics
  Source Code Library, record ascl:1606.005} (\bibinfo {year} {2016}),\ \Eprint
  {http://arxiv.org/abs/1606.005} {ascl:1606.005} \BibitemShut {NoStop}%
\bibitem [{\citenamefont {Dembinski}\ and\ \citenamefont
  {et~al.}(2020)}]{iminuit}%
  \BibitemOpen
  \bibfield  {author} {\bibinfo {author} {\bibfnamefont {H.}~\bibnamefont
  {Dembinski}}\ and\ \bibinfo {author} {\bibfnamefont {P.~O.}\ \bibnamefont
  {et~al.}},\ }\href {\doibase 10.5281/zenodo.3949207} {\  (\bibinfo {year}
  {2020}),\ 10.5281/zenodo.3949207}\BibitemShut {NoStop}%
\bibitem [{\citenamefont {Dalcín}\ \emph {et~al.}(2005)\citenamefont
  {Dalcín}, \citenamefont {Paz},\ and\ \citenamefont
  {Storti}}]{DALCIN20051108}%
  \BibitemOpen
  \bibfield  {author} {\bibinfo {author} {\bibfnamefont {L.}~\bibnamefont
  {Dalcín}}, \bibinfo {author} {\bibfnamefont {R.}~\bibnamefont {Paz}}, \ and\
  \bibinfo {author} {\bibfnamefont {M.}~\bibnamefont {Storti}},\ }\href
  {\doibase https://doi.org/10.1016/j.jpdc.2005.03.010} {\bibfield  {journal}
  {\bibinfo  {journal} {Journal of Parallel and Distributed Computing}\
  }\textbf {\bibinfo {volume} {65}},\ \bibinfo {pages} {1108} (\bibinfo {year}
  {2005})}\BibitemShut {NoStop}%
\bibitem [{\citenamefont {{Schlegel}}\ \emph {et~al.}(1998)\citenamefont
  {{Schlegel}}, \citenamefont {{Finkbeiner}},\ and\ \citenamefont
  {{Davis}}}]{1998ApJ...500..525S}%
  \BibitemOpen
  \bibfield  {author} {\bibinfo {author} {\bibfnamefont {D.~J.}\ \bibnamefont
  {{Schlegel}}}, \bibinfo {author} {\bibfnamefont {D.~P.}\ \bibnamefont
  {{Finkbeiner}}}, \ and\ \bibinfo {author} {\bibfnamefont {M.}~\bibnamefont
  {{Davis}}},\ }\href {\doibase 10.1086/305772} {\bibfield  {journal} {\bibinfo
   {journal} {\apj}\ }\textbf {\bibinfo {volume} {500}},\ \bibinfo {pages}
  {525} (\bibinfo {year} {1998})},\ \Eprint
  {http://arxiv.org/abs/astro-ph/9710327} {arXiv:astro-ph/9710327 [astro-ph]}
  \BibitemShut {NoStop}%
\bibitem [{\citenamefont {{Neugebauer}}\ \emph {et~al.}(1984)\citenamefont
  {{Neugebauer}}, \citenamefont {{Habing}}, \citenamefont {{van Duinen}},
  \citenamefont {{Aumann}}, \citenamefont {{Baud}}, \citenamefont {{Beichman}},
  \citenamefont {{Beintema}}, \citenamefont {{Boggess}},\ and\ \citenamefont
  {{et al.}}}]{1984ApJ...278L...1N}%
  \BibitemOpen
  \bibfield  {author} {\bibinfo {author} {\bibfnamefont {G.}~\bibnamefont
  {{Neugebauer}}}, \bibinfo {author} {\bibfnamefont {H.}~\bibnamefont
  {{Habing}}}, \bibinfo {author} {\bibfnamefont {R.}~\bibnamefont {{van
  Duinen}}}, \bibinfo {author} {\bibfnamefont {H.}~\bibnamefont {{Aumann}}},
  \bibinfo {author} {\bibfnamefont {B.}~\bibnamefont {{Baud}}}, \bibinfo
  {author} {\bibfnamefont {C.}~\bibnamefont {{Beichman}}}, \bibinfo {author}
  {\bibfnamefont {D.}~\bibnamefont {{Beintema}}}, \bibinfo {author}
  {\bibfnamefont {N.}~\bibnamefont {{Boggess}}}, \ and\ \bibinfo {author}
  {\bibnamefont {{et al.}}},\ }\href {\doibase 10.1086/184209} {\bibfield
  {journal} {\bibinfo  {journal} {\apjl}\ }\textbf {\bibinfo {volume} {278}},\
  \bibinfo {pages} {L1} (\bibinfo {year} {1984})}\BibitemShut {NoStop}%
\bibitem [{\citenamefont {{Silverberg}}\ \emph {et~al.}(1993)\citenamefont
  {{Silverberg}}, \citenamefont {{Hauser}}, \citenamefont {{Boggess}},
  \citenamefont {{Kelsall}}, \citenamefont {{Moseley}},\ and\ \citenamefont
  {{Murdock}}}]{1993SPIE.2019..180S}%
  \BibitemOpen
  \bibfield  {author} {\bibinfo {author} {\bibfnamefont {R.~F.}\ \bibnamefont
  {{Silverberg}}}, \bibinfo {author} {\bibfnamefont {M.~G.}\ \bibnamefont
  {{Hauser}}}, \bibinfo {author} {\bibfnamefont {N.~W.}\ \bibnamefont
  {{Boggess}}}, \bibinfo {author} {\bibfnamefont {T.~J.}\ \bibnamefont
  {{Kelsall}}}, \bibinfo {author} {\bibfnamefont {S.~H.}\ \bibnamefont
  {{Moseley}}}, \ and\ \bibinfo {author} {\bibfnamefont {T.~L.}\ \bibnamefont
  {{Murdock}}},\ }in\ \href {\doibase 10.1117/12.157825} {\emph {\bibinfo
  {booktitle} {Infrared Spaceborne Remote Sensing}}},\ \bibinfo {series}
  {Society of Photo-Optical Instrumentation Engineers (SPIE) Conference
  Series}, Vol.\ \bibinfo {volume} {2019},\ \bibinfo {editor} {edited by\
  \bibinfo {editor} {\bibfnamefont {M.~S.}\ \bibnamefont {{Scholl}}}}\
  (\bibinfo {year} {1993})\ pp.\ \bibinfo {pages} {180--189}\BibitemShut
  {NoStop}%
\bibitem [{\citenamefont {{Yahata}}\ \emph {et~al.}(2007)\citenamefont
  {{Yahata}}, \citenamefont {{Yonehara}}, \citenamefont {{Suto}}, \citenamefont
  {{Turner}}, \citenamefont {{Broadhurst}},\ and\ \citenamefont
  {{Finkbeiner}}}]{2007PASJ...59..205Y}%
  \BibitemOpen
  \bibfield  {author} {\bibinfo {author} {\bibfnamefont {K.}~\bibnamefont
  {{Yahata}}}, \bibinfo {author} {\bibfnamefont {A.}~\bibnamefont
  {{Yonehara}}}, \bibinfo {author} {\bibfnamefont {Y.}~\bibnamefont {{Suto}}},
  \bibinfo {author} {\bibfnamefont {E.}~\bibnamefont {{Turner}}}, \bibinfo
  {author} {\bibfnamefont {T.}~\bibnamefont {{Broadhurst}}}, \ and\ \bibinfo
  {author} {\bibfnamefont {D.}~\bibnamefont {{Finkbeiner}}},\ }\href {\doibase
  10.1093/pasj/59.1.205} {\bibfield  {journal} {\bibinfo  {journal} {\pasj}\
  }\textbf {\bibinfo {volume} {59}},\ \bibinfo {pages} {205} (\bibinfo {year}
  {2007})},\ \Eprint {http://arxiv.org/abs/astro-ph/0607098}
  {arXiv:astro-ph/0607098 [astro-ph]} \BibitemShut {NoStop}%
\bibitem [{\citenamefont {{Schlafly}}\ \emph {et~al.}(2014)\citenamefont
  {{Schlafly}}, \citenamefont {{Green}}, \citenamefont {{Finkbeiner}},
  \citenamefont {{Rix}}, \citenamefont {{Bell}}, \citenamefont {{Burgett}},
  \citenamefont {{Chambers}}, \citenamefont {{Draper}},\ and\ \citenamefont
  {{et al.}}}]{2014ApJ...786...29S}%
  \BibitemOpen
  \bibfield  {author} {\bibinfo {author} {\bibfnamefont {E.~F.}\ \bibnamefont
  {{Schlafly}}}, \bibinfo {author} {\bibfnamefont {G.}~\bibnamefont {{Green}}},
  \bibinfo {author} {\bibfnamefont {D.~P.}\ \bibnamefont {{Finkbeiner}}},
  \bibinfo {author} {\bibfnamefont {H.~W.}\ \bibnamefont {{Rix}}}, \bibinfo
  {author} {\bibfnamefont {E.~F.}\ \bibnamefont {{Bell}}}, \bibinfo {author}
  {\bibfnamefont {W.~S.}\ \bibnamefont {{Burgett}}}, \bibinfo {author}
  {\bibfnamefont {K.~C.}\ \bibnamefont {{Chambers}}}, \bibinfo {author}
  {\bibfnamefont {P.~W.}\ \bibnamefont {{Draper}}}, \ and\ \bibinfo {author}
  {\bibnamefont {{et al.}}},\ }\href {\doibase 10.1088/0004-637X/786/1/29}
  {\bibfield  {journal} {\bibinfo  {journal} {\apj}\ }\textbf {\bibinfo
  {volume} {786}},\ \bibinfo {eid} {29} (\bibinfo {year} {2014})},\ \Eprint
  {http://arxiv.org/abs/1403.3393} {arXiv:1403.3393 [astro-ph.GA]} \BibitemShut
  {NoStop}%
\bibitem [{\citenamefont {Korsmeier}\ and\ \citenamefont
  {Cuoco}(2022)}]{Korsmeier:2021bkw}%
  \BibitemOpen
  \bibfield  {author} {\bibinfo {author} {\bibfnamefont {M.}~\bibnamefont
  {Korsmeier}}\ and\ \bibinfo {author} {\bibfnamefont {A.}~\bibnamefont
  {Cuoco}},\ }\href {\doibase 10.1103/PhysRevD.105.103033} {\bibfield
  {journal} {\bibinfo  {journal} {Phys. Rev. D}\ }\textbf {\bibinfo {volume}
  {105}},\ \bibinfo {pages} {103033} (\bibinfo {year} {2022})},\ \Eprint
  {http://arxiv.org/abs/2112.08381} {arXiv:2112.08381 [astro-ph.HE]}
  \BibitemShut {NoStop}%
\bibitem [{\citenamefont {Weinrich}\ \emph {et~al.}(2020)\citenamefont
  {Weinrich}, \citenamefont {Boudaud}, \citenamefont {Derome}, \citenamefont
  {Genolini}, \citenamefont {Lavalle}, \citenamefont {Maurin}, \citenamefont
  {Salati}, \citenamefont {Serpico},\ and\ \citenamefont
  {Weymann-Despres}}]{Weinrich:2020ftb}%
  \BibitemOpen
  \bibfield  {author} {\bibinfo {author} {\bibfnamefont {N.}~\bibnamefont
  {Weinrich}}, \bibinfo {author} {\bibfnamefont {M.}~\bibnamefont {Boudaud}},
  \bibinfo {author} {\bibfnamefont {L.}~\bibnamefont {Derome}}, \bibinfo
  {author} {\bibfnamefont {Y.}~\bibnamefont {Genolini}}, \bibinfo {author}
  {\bibfnamefont {J.}~\bibnamefont {Lavalle}}, \bibinfo {author} {\bibfnamefont
  {D.}~\bibnamefont {Maurin}}, \bibinfo {author} {\bibfnamefont
  {P.}~\bibnamefont {Salati}}, \bibinfo {author} {\bibfnamefont
  {P.}~\bibnamefont {Serpico}}, \ and\ \bibinfo {author} {\bibfnamefont
  {G.}~\bibnamefont {Weymann-Despres}},\ }\href {\doibase
  10.1051/0004-6361/202038064} {\bibfield  {journal} {\bibinfo  {journal}
  {Astron. Astrophys.}\ }\textbf {\bibinfo {volume} {639}},\ \bibinfo {pages}
  {A74} (\bibinfo {year} {2020})},\ \Eprint {http://arxiv.org/abs/2004.00441}
  {arXiv:2004.00441 [astro-ph.HE]} \BibitemShut {NoStop}%
\bibitem [{\citenamefont {Drury}\ and\ \citenamefont
  {Strong}(2017)}]{Drury:2016ubm}%
  \BibitemOpen
  \bibfield  {author} {\bibinfo {author} {\bibfnamefont {L.~O.}\ \bibnamefont
  {Drury}}\ and\ \bibinfo {author} {\bibfnamefont {A.~W.}\ \bibnamefont
  {Strong}},\ }\href {\doibase 10.1051/0004-6361/201629526} {\bibfield
  {journal} {\bibinfo  {journal} {Astron. Astrophys.}\ }\textbf {\bibinfo
  {volume} {597}},\ \bibinfo {pages} {A117} (\bibinfo {year} {2017})},\ \Eprint
  {http://arxiv.org/abs/1608.04227} {arXiv:1608.04227 [astro-ph.HE]}
  \BibitemShut {NoStop}%
\bibitem [{\citenamefont {Derome}\ \emph {et~al.}(2019)\citenamefont {Derome},
  \citenamefont {Maurin}, \citenamefont {Salati}, \citenamefont {Boudaud},
  \citenamefont {G\'enolini},\ and\ \citenamefont {Kunz\'e}}]{Derome:2019jfs}%
  \BibitemOpen
  \bibfield  {author} {\bibinfo {author} {\bibfnamefont {L.}~\bibnamefont
  {Derome}}, \bibinfo {author} {\bibfnamefont {D.}~\bibnamefont {Maurin}},
  \bibinfo {author} {\bibfnamefont {P.}~\bibnamefont {Salati}}, \bibinfo
  {author} {\bibfnamefont {M.}~\bibnamefont {Boudaud}}, \bibinfo {author}
  {\bibfnamefont {Y.}~\bibnamefont {G\'enolini}}, \ and\ \bibinfo {author}
  {\bibfnamefont {P.}~\bibnamefont {Kunz\'e}},\ }\href {\doibase
  10.1051/0004-6361/201935717} {\bibfield  {journal} {\bibinfo  {journal}
  {Astron. Astrophys.}\ }\textbf {\bibinfo {volume} {627}},\ \bibinfo {pages}
  {A158} (\bibinfo {year} {2019})},\ \Eprint {http://arxiv.org/abs/1904.08210}
  {arXiv:1904.08210 [astro-ph.HE]} \BibitemShut {NoStop}%
\bibitem [{\citenamefont {Aguilar}\ \emph {et~al.}(2021)\citenamefont {Aguilar}
  \emph {et~al.}}]{Aguilar:2021tos}%
  \BibitemOpen
  \bibfield  {author} {\bibinfo {author} {\bibfnamefont {M.}~\bibnamefont
  {Aguilar}} \emph {et~al.} (\bibinfo {collaboration} {AMS}),\ }\href {\doibase
  10.1016/j.physrep.2020.09.003} {\bibfield  {journal} {\bibinfo  {journal}
  {Phys. Rept.}\ }\textbf {\bibinfo {volume} {894}},\ \bibinfo {pages} {1}
  (\bibinfo {year} {2021})}\BibitemShut {NoStop}%
\bibitem [{\citenamefont {Korsmeier}\ and\ \citenamefont
  {Cuoco}(2016)}]{Korsmeier:2016kha}%
  \BibitemOpen
  \bibfield  {author} {\bibinfo {author} {\bibfnamefont {M.}~\bibnamefont
  {Korsmeier}}\ and\ \bibinfo {author} {\bibfnamefont {A.}~\bibnamefont
  {Cuoco}},\ }\href {\doibase 10.1103/PhysRevD.94.123019} {\bibfield  {journal}
  {\bibinfo  {journal} {Phys. Rev. D}\ }\textbf {\bibinfo {volume} {94}},\
  \bibinfo {pages} {123019} (\bibinfo {year} {2016})},\ \Eprint
  {http://arxiv.org/abs/1607.06093} {arXiv:1607.06093 [astro-ph.HE]}
  \BibitemShut {NoStop}%
\bibitem [{\citenamefont {Tomassetti}(2017)}]{Tomassetti:2017hbe}%
  \BibitemOpen
  \bibfield  {author} {\bibinfo {author} {\bibfnamefont {N.}~\bibnamefont
  {Tomassetti}},\ }\href {\doibase 10.1103/PhysRevD.96.103005} {\bibfield
  {journal} {\bibinfo  {journal} {Phys. Rev. D}\ }\textbf {\bibinfo {volume}
  {96}},\ \bibinfo {pages} {103005} (\bibinfo {year} {2017})},\ \Eprint
  {http://arxiv.org/abs/1707.06917} {arXiv:1707.06917 [astro-ph.HE]}
  \BibitemShut {NoStop}%
\bibitem [{\citenamefont {Liu}\ \emph {et~al.}(2018)\citenamefont {Liu},
  \citenamefont {Yao},\ and\ \citenamefont {Guo}}]{Liu:2018ujp}%
  \BibitemOpen
  \bibfield  {author} {\bibinfo {author} {\bibfnamefont {W.}~\bibnamefont
  {Liu}}, \bibinfo {author} {\bibfnamefont {Y.-h.}\ \bibnamefont {Yao}}, \ and\
  \bibinfo {author} {\bibfnamefont {Y.-Q.}\ \bibnamefont {Guo}},\ }\href
  {\doibase 10.3847/1538-4357/aaef39} {\bibfield  {journal} {\bibinfo
  {journal} {Astrophys. J.}\ }\textbf {\bibinfo {volume} {869}},\ \bibinfo
  {pages} {176} (\bibinfo {year} {2018})},\ \Eprint
  {http://arxiv.org/abs/1802.03602} {arXiv:1802.03602 [astro-ph.HE]}
  \BibitemShut {NoStop}%
\bibitem [{\citenamefont {G\'enolini}\ \emph {et~al.}(2019)\citenamefont
  {G\'enolini} \emph {et~al.}}]{Genolini:2019ewc}%
  \BibitemOpen
  \bibfield  {author} {\bibinfo {author} {\bibfnamefont {Y.}~\bibnamefont
  {G\'enolini}} \emph {et~al.},\ }\href {\doibase 10.1103/PhysRevD.99.123028}
  {\bibfield  {journal} {\bibinfo  {journal} {Phys. Rev. D}\ }\textbf {\bibinfo
  {volume} {99}},\ \bibinfo {pages} {123028} (\bibinfo {year} {2019})},\
  \Eprint {http://arxiv.org/abs/1904.08917} {arXiv:1904.08917 [astro-ph.HE]}
  \BibitemShut {NoStop}%
\bibitem [{\citenamefont {Evoli}\ \emph {et~al.}(2019)\citenamefont {Evoli},
  \citenamefont {Aloisio},\ and\ \citenamefont {Blasi}}]{Evoli:2019wwu}%
  \BibitemOpen
  \bibfield  {author} {\bibinfo {author} {\bibfnamefont {C.}~\bibnamefont
  {Evoli}}, \bibinfo {author} {\bibfnamefont {R.}~\bibnamefont {Aloisio}}, \
  and\ \bibinfo {author} {\bibfnamefont {P.}~\bibnamefont {Blasi}},\ }\href
  {\doibase 10.1103/PhysRevD.99.103023} {\bibfield  {journal} {\bibinfo
  {journal} {Phys. Rev. D}\ }\textbf {\bibinfo {volume} {99}},\ \bibinfo
  {pages} {103023} (\bibinfo {year} {2019})},\ \Eprint
  {http://arxiv.org/abs/1904.10220} {arXiv:1904.10220 [astro-ph.HE]}
  \BibitemShut {NoStop}%
\bibitem [{\citenamefont {Evoli}\ \emph {et~al.}(2020)\citenamefont {Evoli},
  \citenamefont {Morlino}, \citenamefont {Blasi},\ and\ \citenamefont
  {Aloisio}}]{Evoli:2019iih}%
  \BibitemOpen
  \bibfield  {author} {\bibinfo {author} {\bibfnamefont {C.}~\bibnamefont
  {Evoli}}, \bibinfo {author} {\bibfnamefont {G.}~\bibnamefont {Morlino}},
  \bibinfo {author} {\bibfnamefont {P.}~\bibnamefont {Blasi}}, \ and\ \bibinfo
  {author} {\bibfnamefont {R.}~\bibnamefont {Aloisio}},\ }\href {\doibase
  10.1103/PhysRevD.101.023013} {\bibfield  {journal} {\bibinfo  {journal}
  {Phys. Rev. D}\ }\textbf {\bibinfo {volume} {101}},\ \bibinfo {pages}
  {023013} (\bibinfo {year} {2020})},\ \Eprint
  {http://arxiv.org/abs/1910.04113} {arXiv:1910.04113 [astro-ph.HE]}
  \BibitemShut {NoStop}%
\bibitem [{\citenamefont {Boschini}\ \emph {et~al.}(2020)\citenamefont
  {Boschini} \emph {et~al.}}]{Boschini:2020jty}%
  \BibitemOpen
  \bibfield  {author} {\bibinfo {author} {\bibfnamefont {M.~J.}\ \bibnamefont
  {Boschini}} \emph {et~al.},\ }\href {\doibase 10.3847/1538-4365/aba901}
  {\bibfield  {journal} {\bibinfo  {journal} {Astrophys. J. Suppl.}\ }\textbf
  {\bibinfo {volume} {250}},\ \bibinfo {pages} {27} (\bibinfo {year} {2020})},\
  \Eprint {http://arxiv.org/abs/2006.01337} {arXiv:2006.01337 [astro-ph.HE]}
  \BibitemShut {NoStop}%
\bibitem [{\citenamefont {Luque}\ \emph {et~al.}(2021)\citenamefont {Luque},
  \citenamefont {Mazziotta}, \citenamefont {Loparco}, \citenamefont {Gargano},\
  and\ \citenamefont {Serini}}]{Luque:2021nxb}%
  \BibitemOpen
  \bibfield  {author} {\bibinfo {author} {\bibfnamefont {P.~D. L.~T.}\
  \bibnamefont {Luque}}, \bibinfo {author} {\bibfnamefont {M.~N.}\ \bibnamefont
  {Mazziotta}}, \bibinfo {author} {\bibfnamefont {F.}~\bibnamefont {Loparco}},
  \bibinfo {author} {\bibfnamefont {F.}~\bibnamefont {Gargano}}, \ and\
  \bibinfo {author} {\bibfnamefont {D.}~\bibnamefont {Serini}},\ }\href
  {\doibase 10.1088/1475-7516/2021/07/010} {\bibfield  {journal} {\bibinfo
  {journal} {JCAP}\ }\textbf {\bibinfo {volume} {07}},\ \bibinfo {pages} {010}
  (\bibinfo {year} {2021})},\ \Eprint {http://arxiv.org/abs/2102.13238}
  {arXiv:2102.13238 [astro-ph.HE]} \BibitemShut {NoStop}%
\bibitem [{\citenamefont {De~La Torre~Luque}\ \emph {et~al.}(2021)\citenamefont
  {De~La Torre~Luque}, \citenamefont {Mazziotta}, \citenamefont {Loparco},
  \citenamefont {Gargano},\ and\ \citenamefont
  {Serini}}]{DeLaTorreLuque:2021yfq}%
  \BibitemOpen
  \bibfield  {author} {\bibinfo {author} {\bibfnamefont {P.}~\bibnamefont
  {De~La Torre~Luque}}, \bibinfo {author} {\bibfnamefont {M.~N.}\ \bibnamefont
  {Mazziotta}}, \bibinfo {author} {\bibfnamefont {F.}~\bibnamefont {Loparco}},
  \bibinfo {author} {\bibfnamefont {F.}~\bibnamefont {Gargano}}, \ and\
  \bibinfo {author} {\bibfnamefont {D.}~\bibnamefont {Serini}},\ }\href
  {\doibase 10.1088/1475-7516/2021/03/099} {\bibfield  {journal} {\bibinfo
  {journal} {JCAP}\ }\textbf {\bibinfo {volume} {03}},\ \bibinfo {pages} {099}
  (\bibinfo {year} {2021})},\ \Eprint {http://arxiv.org/abs/2101.01547}
  {arXiv:2101.01547 [astro-ph.HE]} \BibitemShut {NoStop}%
\bibitem [{\citenamefont {Schroer}\ \emph {et~al.}(2021)\citenamefont
  {Schroer}, \citenamefont {Evoli},\ and\ \citenamefont
  {Blasi}}]{Schroer:2021ojh}%
  \BibitemOpen
  \bibfield  {author} {\bibinfo {author} {\bibfnamefont {B.}~\bibnamefont
  {Schroer}}, \bibinfo {author} {\bibfnamefont {C.}~\bibnamefont {Evoli}}, \
  and\ \bibinfo {author} {\bibfnamefont {P.}~\bibnamefont {Blasi}},\ }\href
  {\doibase 10.1103/PhysRevD.103.123010} {\bibfield  {journal} {\bibinfo
  {journal} {Phys. Rev. D}\ }\textbf {\bibinfo {volume} {103}},\ \bibinfo
  {pages} {123010} (\bibinfo {year} {2021})},\ \Eprint
  {http://arxiv.org/abs/2102.12576} {arXiv:2102.12576 [astro-ph.HE]}
  \BibitemShut {NoStop}%
\bibitem [{\citenamefont {Zhao}\ \emph {et~al.}(2021)\citenamefont {Zhao},
  \citenamefont {Fang},\ and\ \citenamefont {Bi}}]{Zhao:2021yzf}%
  \BibitemOpen
  \bibfield  {author} {\bibinfo {author} {\bibfnamefont {M.-J.}\ \bibnamefont
  {Zhao}}, \bibinfo {author} {\bibfnamefont {K.}~\bibnamefont {Fang}}, \ and\
  \bibinfo {author} {\bibfnamefont {X.-J.}\ \bibnamefont {Bi}},\ }\href
  {\doibase 10.1103/PhysRevD.104.123001} {\bibfield  {journal} {\bibinfo
  {journal} {Phys. Rev. D}\ }\textbf {\bibinfo {volume} {104}},\ \bibinfo
  {pages} {123001} (\bibinfo {year} {2021})},\ \Eprint
  {http://arxiv.org/abs/2109.04112} {arXiv:2109.04112 [astro-ph.HE]}
  \BibitemShut {NoStop}%
\bibitem [{\citenamefont {Coste}\ \emph {et~al.}(2012)\citenamefont {Coste},
  \citenamefont {Derome}, \citenamefont {Maurin},\ and\ \citenamefont
  {Putze}}]{Coste:2011jc}%
  \BibitemOpen
  \bibfield  {author} {\bibinfo {author} {\bibfnamefont {B.}~\bibnamefont
  {Coste}}, \bibinfo {author} {\bibfnamefont {L.}~\bibnamefont {Derome}},
  \bibinfo {author} {\bibfnamefont {D.}~\bibnamefont {Maurin}}, \ and\ \bibinfo
  {author} {\bibfnamefont {A.}~\bibnamefont {Putze}},\ }\href {\doibase
  10.1051/0004-6361/201117927} {\bibfield  {journal} {\bibinfo  {journal}
  {Astron. Astrophys.}\ }\textbf {\bibinfo {volume} {539}},\ \bibinfo {pages}
  {A88} (\bibinfo {year} {2012})},\ \Eprint {http://arxiv.org/abs/1108.4349}
  {arXiv:1108.4349 [astro-ph.GA]} \BibitemShut {NoStop}%
\bibitem [{\citenamefont {Wu}\ and\ \citenamefont {Chen}(2019)}]{Wu:2018lqu}%
  \BibitemOpen
  \bibfield  {author} {\bibinfo {author} {\bibfnamefont {J.}~\bibnamefont
  {Wu}}\ and\ \bibinfo {author} {\bibfnamefont {H.}~\bibnamefont {Chen}},\
  }\href {\doibase 10.1016/j.physletb.2018.11.052} {\bibfield  {journal}
  {\bibinfo  {journal} {Phys. Lett. B}\ }\textbf {\bibinfo {volume} {789}},\
  \bibinfo {pages} {292} (\bibinfo {year} {2019})},\ \Eprint
  {http://arxiv.org/abs/1809.04905} {arXiv:1809.04905 [astro-ph.HE]}
  \BibitemShut {NoStop}%
\bibitem [{\citenamefont {Aguilar}\ \emph {et~al.}(2018)\citenamefont {Aguilar}
  \emph {et~al.}}]{Aguilar:2018njt}%
  \BibitemOpen
  \bibfield  {author} {\bibinfo {author} {\bibfnamefont {M.}~\bibnamefont
  {Aguilar}} \emph {et~al.} (\bibinfo {collaboration} {AMS}),\ }\href {\doibase
  10.1103/PhysRevLett.120.021101} {\bibfield  {journal} {\bibinfo  {journal}
  {Phys. Rev. Lett.}\ }\textbf {\bibinfo {volume} {120}},\ \bibinfo {pages}
  {021101} (\bibinfo {year} {2018})}\BibitemShut {NoStop}%
\bibitem [{\citenamefont {Derome}(2021)}]{DeromeICRC2021}%
  \BibitemOpen
  \bibfield  {author} {\bibinfo {author} {\bibfnamefont {L.}~\bibnamefont
  {Derome}},\ }\href@noop {} {\ \textbf {\bibinfo {volume} {ICRC2021}}
  (\bibinfo {year} {2021})},\ \bibinfo {note}
  {{\url{https://video.desy.de/video/Cosmic-Ray-Isotopes-with-the-Alpha-Magnetic-Spectrometer/c1e42f1de4cdb29069278dffd612add1}}}\BibitemShut
  {NoStop}%
\bibitem [{\citenamefont {Genolini}\ \emph {et~al.}(2018)\citenamefont
  {Genolini}, \citenamefont {Maurin}, \citenamefont {Moskalenko},\ and\
  \citenamefont {Unger}}]{Genolini:2018ekk}%
  \BibitemOpen
  \bibfield  {author} {\bibinfo {author} {\bibfnamefont {Y.}~\bibnamefont
  {Genolini}}, \bibinfo {author} {\bibfnamefont {D.}~\bibnamefont {Maurin}},
  \bibinfo {author} {\bibfnamefont {I.~V.}\ \bibnamefont {Moskalenko}}, \ and\
  \bibinfo {author} {\bibfnamefont {M.}~\bibnamefont {Unger}},\ }\href
  {\doibase 10.1103/PhysRevC.98.034611} {\bibfield  {journal} {\bibinfo
  {journal} {Phys. Rev. C}\ }\textbf {\bibinfo {volume} {98}},\ \bibinfo
  {pages} {034611} (\bibinfo {year} {2018})},\ \Eprint
  {http://arxiv.org/abs/1803.04686} {arXiv:1803.04686 [astro-ph.HE]}
  \BibitemShut {NoStop}%
\bibitem [{\citenamefont {Fisk}(1976)}]{Fisk:1976aw}%
  \BibitemOpen
  \bibfield  {author} {\bibinfo {author} {\bibfnamefont {L.~A.}\ \bibnamefont
  {Fisk}},\ }\href {\doibase 10.1086/154387} {\bibfield  {journal} {\bibinfo
  {journal} {Astrophys. J.}\ }\textbf {\bibinfo {volume} {206}},\ \bibinfo
  {pages} {333} (\bibinfo {year} {1976})}\BibitemShut {NoStop}%
\bibitem [{\citenamefont {Green}(2015)}]{Green:2015isa}%
  \BibitemOpen
  \bibfield  {author} {\bibinfo {author} {\bibfnamefont {D.~A.}\ \bibnamefont
  {Green}},\ }\href {\doibase 10.1093/mnras/stv1885} {\bibfield  {journal}
  {\bibinfo  {journal} {Mon. Not. Roy. Astron. Soc.}\ }\textbf {\bibinfo
  {volume} {454}},\ \bibinfo {pages} {1517} (\bibinfo {year} {2015})},\ \Eprint
  {http://arxiv.org/abs/1508.02931} {arXiv:1508.02931 [astro-ph.HE]}
  \BibitemShut {NoStop}%
\bibitem [{\citenamefont {Lorimer}(2004)}]{astro-ph/0308501}%
  \BibitemOpen
  \bibfield  {author} {\bibinfo {author} {\bibfnamefont {D.~R.}\ \bibnamefont
  {Lorimer}},\ }\href@noop {} {\bibfield  {journal} {\bibinfo  {journal} {IAU
  Symp.}\ }\textbf {\bibinfo {volume} {218}},\ \bibinfo {pages} {105} (\bibinfo
  {year} {2004})},\ \Eprint {http://arxiv.org/abs/astro-ph/0308501}
  {arXiv:astro-ph/0308501} \BibitemShut {NoStop}%
\bibitem [{\citenamefont {Hooper}\ \emph {et~al.}(2009)\citenamefont {Hooper},
  \citenamefont {Blasi},\ and\ \citenamefont {Serpico}}]{Hooper:2008kg}%
  \BibitemOpen
  \bibfield  {author} {\bibinfo {author} {\bibfnamefont {D.}~\bibnamefont
  {Hooper}}, \bibinfo {author} {\bibfnamefont {P.}~\bibnamefont {Blasi}}, \
  and\ \bibinfo {author} {\bibfnamefont {P.~D.}\ \bibnamefont {Serpico}},\
  }\href {\doibase 10.1088/1475-7516/2009/01/025} {\bibfield  {journal}
  {\bibinfo  {journal} {JCAP}\ }\textbf {\bibinfo {volume} {01}},\ \bibinfo
  {pages} {025} (\bibinfo {year} {2009})},\ \Eprint
  {http://arxiv.org/abs/0810.1527} {arXiv:0810.1527 [astro-ph]} \BibitemShut
  {NoStop}%
\bibitem [{\citenamefont {Aguilar}\ \emph {et~al.}(2015)\citenamefont {Aguilar}
  \emph {et~al.}}]{Aguilar:2015ooa}%
  \BibitemOpen
  \bibfield  {author} {\bibinfo {author} {\bibfnamefont {M.}~\bibnamefont
  {Aguilar}} \emph {et~al.} (\bibinfo {collaboration} {AMS}),\ }\href {\doibase
  10.1103/PhysRevLett.114.171103} {\bibfield  {journal} {\bibinfo  {journal}
  {Phys. Rev. Lett.}\ }\textbf {\bibinfo {volume} {114}},\ \bibinfo {pages}
  {171103} (\bibinfo {year} {2015})}\BibitemShut {NoStop}%
\bibitem [{\citenamefont {{Stone}}\ \emph {et~al.}(2013)\citenamefont
  {{Stone}}, \citenamefont {{Cummings}}, \citenamefont {{McDonald}},
  \citenamefont {{Heikkila}}, \citenamefont {{Lal}},\ and\ \citenamefont
  {{Webber}}}]{2013Sci...341..150S}%
  \BibitemOpen
  \bibfield  {author} {\bibinfo {author} {\bibfnamefont {E.~C.}\ \bibnamefont
  {{Stone}}}, \bibinfo {author} {\bibfnamefont {A.~C.}\ \bibnamefont
  {{Cummings}}}, \bibinfo {author} {\bibfnamefont {F.~B.}\ \bibnamefont
  {{McDonald}}}, \bibinfo {author} {\bibfnamefont {B.~C.}\ \bibnamefont
  {{Heikkila}}}, \bibinfo {author} {\bibfnamefont {N.}~\bibnamefont {{Lal}}}, \
  and\ \bibinfo {author} {\bibfnamefont {W.~R.}\ \bibnamefont {{Webber}}},\
  }\href {\doibase 10.1126/science.1236408} {\bibfield  {journal} {\bibinfo
  {journal} {Science}\ }\textbf {\bibinfo {volume} {341}},\ \bibinfo {pages}
  {150} (\bibinfo {year} {2013})}\BibitemShut {NoStop}%
\bibitem [{\citenamefont {Aguilar}\ \emph {et~al.}(2017)\citenamefont {Aguilar}
  \emph {et~al.}}]{Aguilar:2017hno}%
  \BibitemOpen
  \bibfield  {author} {\bibinfo {author} {\bibfnamefont {M.}~\bibnamefont
  {Aguilar}} \emph {et~al.} (\bibinfo {collaboration} {AMS}),\ }\href {\doibase
  10.1103/PhysRevLett.119.251101} {\bibfield  {journal} {\bibinfo  {journal}
  {Phys. Rev. Lett.}\ }\textbf {\bibinfo {volume} {119}},\ \bibinfo {pages}
  {251101} (\bibinfo {year} {2017})}\BibitemShut {NoStop}%
\bibitem [{\citenamefont {Aguilar}\ \emph
  {et~al.}(2019{\natexlab{a}})\citenamefont {Aguilar} \emph
  {et~al.}}]{Aguilar:2019eiz}%
  \BibitemOpen
  \bibfield  {author} {\bibinfo {author} {\bibfnamefont {M.}~\bibnamefont
  {Aguilar}} \emph {et~al.} (\bibinfo {collaboration} {AMS}),\ }\href {\doibase
  10.1103/PhysRevLett.123.181102} {\bibfield  {journal} {\bibinfo  {journal}
  {Phys. Rev. Lett.}\ }\textbf {\bibinfo {volume} {123}},\ \bibinfo {pages}
  {181102} (\bibinfo {year} {2019}{\natexlab{a}})}\BibitemShut {NoStop}%
\bibitem [{\citenamefont {Aguilar}\ \emph {et~al.}(2016)\citenamefont {Aguilar}
  \emph {et~al.}}]{Aguilar:2016kjl}%
  \BibitemOpen
  \bibfield  {author} {\bibinfo {author} {\bibfnamefont {M.}~\bibnamefont
  {Aguilar}} \emph {et~al.} (\bibinfo {collaboration} {AMS}),\ }\href {\doibase
  10.1103/PhysRevLett.117.091103} {\bibfield  {journal} {\bibinfo  {journal}
  {Phys. Rev. Lett.}\ }\textbf {\bibinfo {volume} {117}},\ \bibinfo {pages}
  {091103} (\bibinfo {year} {2016})}\BibitemShut {NoStop}%
\bibitem [{\citenamefont {Aguilar}\ \emph {et~al.}(2014)\citenamefont {Aguilar}
  \emph {et~al.}}]{AMS:2014xys}%
  \BibitemOpen
  \bibfield  {author} {\bibinfo {author} {\bibfnamefont {M.}~\bibnamefont
  {Aguilar}} \emph {et~al.} (\bibinfo {collaboration} {AMS}),\ }\href {\doibase
  10.1103/PhysRevLett.113.121102} {\bibfield  {journal} {\bibinfo  {journal}
  {Phys. Rev. Lett.}\ }\textbf {\bibinfo {volume} {113}},\ \bibinfo {pages}
  {121102} (\bibinfo {year} {2014})}\BibitemShut {NoStop}%
\bibitem [{\citenamefont {Aguilar}\ \emph
  {et~al.}(2019{\natexlab{b}})\citenamefont {Aguilar} \emph
  {et~al.}}]{AMS:2019rhg}%
  \BibitemOpen
  \bibfield  {author} {\bibinfo {author} {\bibfnamefont {M.}~\bibnamefont
  {Aguilar}} \emph {et~al.} (\bibinfo {collaboration} {AMS}),\ }\href {\doibase
  10.1103/PhysRevLett.122.041102} {\bibfield  {journal} {\bibinfo  {journal}
  {Phys. Rev. Lett.}\ }\textbf {\bibinfo {volume} {122}},\ \bibinfo {pages}
  {041102} (\bibinfo {year} {2019}{\natexlab{b}})}\BibitemShut {NoStop}%
\bibitem [{\citenamefont {Feroz}\ \emph {et~al.}(2009)\citenamefont {Feroz},
  \citenamefont {Hobson},\ and\ \citenamefont {Bridges}}]{Feroz:2008xx}%
  \BibitemOpen
  \bibfield  {author} {\bibinfo {author} {\bibfnamefont {F.}~\bibnamefont
  {Feroz}}, \bibinfo {author} {\bibfnamefont {M.~P.}\ \bibnamefont {Hobson}}, \
  and\ \bibinfo {author} {\bibfnamefont {M.}~\bibnamefont {Bridges}},\ }\href
  {\doibase 10.1111/j.1365-2966.2009.14548.x} {\bibfield  {journal} {\bibinfo
  {journal} {Mon. Not. Roy. Astron. Soc.}\ }\textbf {\bibinfo {volume} {398}},\
  \bibinfo {pages} {1601} (\bibinfo {year} {2009})},\ \Eprint
  {http://arxiv.org/abs/0809.3437} {arXiv:0809.3437 [astro-ph]} \BibitemShut
  {NoStop}%
\bibitem [{\citenamefont {Acero}\ \emph {et~al.}(2016)\citenamefont {Acero}
  \emph {et~al.}}]{Fermi-LAT:2016zaq}%
  \BibitemOpen
  \bibfield  {author} {\bibinfo {author} {\bibfnamefont {F.}~\bibnamefont
  {Acero}} \emph {et~al.} (\bibinfo {collaboration} {Fermi-LAT}),\ }\href
  {\doibase 10.3847/0067-0049/223/2/26} {\bibfield  {journal} {\bibinfo
  {journal} {Astrophys. J. Suppl.}\ }\textbf {\bibinfo {volume} {223}},\
  \bibinfo {pages} {26} (\bibinfo {year} {2016})},\ \Eprint
  {http://arxiv.org/abs/1602.07246} {arXiv:1602.07246 [astro-ph.HE]}
  \BibitemShut {NoStop}%
\bibitem [{\citenamefont {Yang}\ \emph {et~al.}(2016)\citenamefont {Yang},
  \citenamefont {Aharonian},\ and\ \citenamefont {Evoli}}]{Yang:2016jda}%
  \BibitemOpen
  \bibfield  {author} {\bibinfo {author} {\bibfnamefont {R.}~\bibnamefont
  {Yang}}, \bibinfo {author} {\bibfnamefont {F.}~\bibnamefont {Aharonian}}, \
  and\ \bibinfo {author} {\bibfnamefont {C.}~\bibnamefont {Evoli}},\ }\href
  {\doibase 10.1103/PhysRevD.93.123007} {\bibfield  {journal} {\bibinfo
  {journal} {Phys. Rev. D}\ }\textbf {\bibinfo {volume} {93}},\ \bibinfo
  {pages} {123007} (\bibinfo {year} {2016})},\ \Eprint
  {http://arxiv.org/abs/1602.04710} {arXiv:1602.04710 [astro-ph.HE]}
  \BibitemShut {NoStop}%
\bibitem [{\citenamefont {Pothast}\ \emph {et~al.}(2018)\citenamefont
  {Pothast}, \citenamefont {Gaggero}, \citenamefont {Storm},\ and\
  \citenamefont {Weniger}}]{Pothast:2018bvh}%
  \BibitemOpen
  \bibfield  {author} {\bibinfo {author} {\bibfnamefont {M.}~\bibnamefont
  {Pothast}}, \bibinfo {author} {\bibfnamefont {D.}~\bibnamefont {Gaggero}},
  \bibinfo {author} {\bibfnamefont {E.}~\bibnamefont {Storm}}, \ and\ \bibinfo
  {author} {\bibfnamefont {C.}~\bibnamefont {Weniger}},\ }\href {\doibase
  10.1088/1475-7516/2018/10/045} {\bibfield  {journal} {\bibinfo  {journal}
  {JCAP}\ }\textbf {\bibinfo {volume} {10}},\ \bibinfo {pages} {045} (\bibinfo
  {year} {2018})},\ \Eprint {http://arxiv.org/abs/1807.04554} {arXiv:1807.04554
  [astro-ph.HE]} \BibitemShut {NoStop}%
\bibitem [{\citenamefont {Ajello}\ \emph {et~al.}(2016)\citenamefont {Ajello}
  \emph {et~al.}}]{Fermi-LAT:2015sau}%
  \BibitemOpen
  \bibfield  {author} {\bibinfo {author} {\bibfnamefont {M.}~\bibnamefont
  {Ajello}} \emph {et~al.} (\bibinfo {collaboration} {Fermi-LAT}),\ }\href
  {\doibase 10.3847/0004-637X/819/1/44} {\bibfield  {journal} {\bibinfo
  {journal} {Astrophys. J.}\ }\textbf {\bibinfo {volume} {819}},\ \bibinfo
  {pages} {44} (\bibinfo {year} {2016})},\ \Eprint
  {http://arxiv.org/abs/1511.02938} {arXiv:1511.02938 [astro-ph.HE]}
  \BibitemShut {NoStop}%
\bibitem [{\citenamefont {Carlson}\ \emph {et~al.}(2016)\citenamefont
  {Carlson}, \citenamefont {Linden},\ and\ \citenamefont
  {Profumo}}]{Carlson:2016iis}%
  \BibitemOpen
  \bibfield  {author} {\bibinfo {author} {\bibfnamefont {E.}~\bibnamefont
  {Carlson}}, \bibinfo {author} {\bibfnamefont {T.}~\bibnamefont {Linden}}, \
  and\ \bibinfo {author} {\bibfnamefont {S.}~\bibnamefont {Profumo}},\ }\href
  {\doibase 10.1103/PhysRevD.94.063504} {\bibfield  {journal} {\bibinfo
  {journal} {Phys. Rev. D}\ }\textbf {\bibinfo {volume} {94}},\ \bibinfo
  {pages} {063504} (\bibinfo {year} {2016})},\ \Eprint
  {http://arxiv.org/abs/1603.06584} {arXiv:1603.06584 [astro-ph.HE]}
  \BibitemShut {NoStop}%
\bibitem [{\citenamefont {Bissantz}\ \emph {et~al.}(2004)\citenamefont
  {Bissantz}, \citenamefont {Debattista},\ and\ \citenamefont
  {Gerhard}}]{Bissantz:2003ak}%
  \BibitemOpen
  \bibfield  {author} {\bibinfo {author} {\bibfnamefont {N.}~\bibnamefont
  {Bissantz}}, \bibinfo {author} {\bibfnamefont {V.~P.}\ \bibnamefont
  {Debattista}}, \ and\ \bibinfo {author} {\bibfnamefont {O.}~\bibnamefont
  {Gerhard}},\ }\href {\doibase 10.1086/382043} {\bibfield  {journal} {\bibinfo
   {journal} {Astrophys. J. Lett.}\ }\textbf {\bibinfo {volume} {601}},\
  \bibinfo {pages} {L155} (\bibinfo {year} {2004})},\ \Eprint
  {http://arxiv.org/abs/astro-ph/0312461} {arXiv:astro-ph/0312461} \BibitemShut
  {NoStop}%
\bibitem [{\citenamefont {Pohl}\ \emph {et~al.}(2008)\citenamefont {Pohl},
  \citenamefont {Englmaier},\ and\ \citenamefont {Bissantz}}]{Pohl:2007dz}%
  \BibitemOpen
  \bibfield  {author} {\bibinfo {author} {\bibfnamefont {M.}~\bibnamefont
  {Pohl}}, \bibinfo {author} {\bibfnamefont {P.}~\bibnamefont {Englmaier}}, \
  and\ \bibinfo {author} {\bibfnamefont {N.}~\bibnamefont {Bissantz}},\ }\href
  {\doibase 10.1086/529004} {\bibfield  {journal} {\bibinfo  {journal}
  {Astrophys. J.}\ }\textbf {\bibinfo {volume} {677}},\ \bibinfo {pages} {283}
  (\bibinfo {year} {2008})},\ \Eprint {http://arxiv.org/abs/0712.4264}
  {arXiv:0712.4264 [astro-ph]} \BibitemShut {NoStop}%
\bibitem [{\citenamefont {Mertsch}\ and\ \citenamefont
  {Phan}(2022)}]{Mertsch:2022oee}%
  \BibitemOpen
  \bibfield  {author} {\bibinfo {author} {\bibfnamefont {P.}~\bibnamefont
  {Mertsch}}\ and\ \bibinfo {author} {\bibfnamefont {V.~H.~M.}\ \bibnamefont
  {Phan}},\ }\href@noop {} {\  (\bibinfo {year} {2022})},\ \Eprint
  {http://arxiv.org/abs/2202.02341} {arXiv:2202.02341 [astro-ph.GA]}
  \BibitemShut {NoStop}%
\bibitem [{\citenamefont {Abdollahi}\ \emph {et~al.}(2022)\citenamefont
  {Abdollahi} \emph {et~al.}}]{Fermi-LAT:2022byn}%
  \BibitemOpen
  \bibfield  {author} {\bibinfo {author} {\bibfnamefont {S.}~\bibnamefont
  {Abdollahi}} \emph {et~al.} (\bibinfo {collaboration} {Fermi-LAT}),\ }\href
  {\doibase 10.3847/1538-4365/ac6751} {\bibfield  {journal} {\bibinfo
  {journal} {Astrophys. J. Supp.}\ }\textbf {\bibinfo {volume} {260}},\
  \bibinfo {pages} {53} (\bibinfo {year} {2022})},\ \Eprint
  {http://arxiv.org/abs/2201.11184} {arXiv:2201.11184 [astro-ph.HE]}
  \BibitemShut {NoStop}%
\bibitem [{\citenamefont {Kamae}\ \emph {et~al.}(2005)\citenamefont {Kamae},
  \citenamefont {Abe},\ and\ \citenamefont {Koi}}]{Kamae:2004xx}%
  \BibitemOpen
  \bibfield  {author} {\bibinfo {author} {\bibfnamefont {T.}~\bibnamefont
  {Kamae}}, \bibinfo {author} {\bibfnamefont {T.}~\bibnamefont {Abe}}, \ and\
  \bibinfo {author} {\bibfnamefont {T.}~\bibnamefont {Koi}},\ }\href {\doibase
  10.1086/426935} {\bibfield  {journal} {\bibinfo  {journal} {Astrophys. J.}\
  }\textbf {\bibinfo {volume} {620}},\ \bibinfo {pages} {244} (\bibinfo {year}
  {2005})},\ \Eprint {http://arxiv.org/abs/astro-ph/0410617}
  {arXiv:astro-ph/0410617} \BibitemShut {NoStop}%
\bibitem [{\citenamefont {Kamae}\ \emph {et~al.}(2006)\citenamefont {Kamae},
  \citenamefont {Karlsson}, \citenamefont {Mizuno}, \citenamefont {Abe},\ and\
  \citenamefont {Koi}}]{Kamae:2006bf}%
  \BibitemOpen
  \bibfield  {author} {\bibinfo {author} {\bibfnamefont {T.}~\bibnamefont
  {Kamae}}, \bibinfo {author} {\bibfnamefont {N.}~\bibnamefont {Karlsson}},
  \bibinfo {author} {\bibfnamefont {T.}~\bibnamefont {Mizuno}}, \bibinfo
  {author} {\bibfnamefont {T.}~\bibnamefont {Abe}}, \ and\ \bibinfo {author}
  {\bibfnamefont {T.}~\bibnamefont {Koi}},\ }\href {\doibase 10.1086/513602}
  {\bibfield  {journal} {\bibinfo  {journal} {Astrophys. J.}\ }\textbf
  {\bibinfo {volume} {647}},\ \bibinfo {pages} {692} (\bibinfo {year}
  {2006})},\ \bibinfo {note} {[Erratum: Astrophys.J. 662, 779 (2007)]},\
  \Eprint {http://arxiv.org/abs/astro-ph/0605581} {arXiv:astro-ph/0605581}
  \BibitemShut {NoStop}%
\bibitem [{\citenamefont {Sjostrand}\ \emph {et~al.}(2001)\citenamefont
  {Sjostrand}, \citenamefont {Lonnblad},\ and\ \citenamefont
  {Mrenna}}]{Sjostrand:2001yu}%
  \BibitemOpen
  \bibfield  {author} {\bibinfo {author} {\bibfnamefont {T.}~\bibnamefont
  {Sjostrand}}, \bibinfo {author} {\bibfnamefont {L.}~\bibnamefont {Lonnblad}},
  \ and\ \bibinfo {author} {\bibfnamefont {S.}~\bibnamefont {Mrenna}},\
  }\href@noop {} {\  (\bibinfo {year} {2001})},\ \Eprint
  {http://arxiv.org/abs/hep-ph/0108264} {arXiv:hep-ph/0108264} \BibitemShut
  {NoStop}%
\bibitem [{\citenamefont {Kachelrie\ss{}}\ \emph {et~al.}(2019)\citenamefont
  {Kachelrie\ss{}}, \citenamefont {Moskalenko},\ and\ \citenamefont
  {Ostapchenko}}]{Kachelriess:2019ifk}%
  \BibitemOpen
  \bibfield  {author} {\bibinfo {author} {\bibfnamefont {M.}~\bibnamefont
  {Kachelrie\ss{}}}, \bibinfo {author} {\bibfnamefont {I.~V.}\ \bibnamefont
  {Moskalenko}}, \ and\ \bibinfo {author} {\bibfnamefont {S.}~\bibnamefont
  {Ostapchenko}},\ }\href {\doibase 10.1016/j.cpc.2019.08.001} {\bibfield
  {journal} {\bibinfo  {journal} {Comput. Phys. Commun.}\ }\textbf {\bibinfo
  {volume} {245}},\ \bibinfo {pages} {106846} (\bibinfo {year} {2019})},\
  \Eprint {http://arxiv.org/abs/1904.05129} {arXiv:1904.05129 [hep-ph]}
  \BibitemShut {NoStop}%
\bibitem [{\citenamefont {Ostapchenko}(2011)}]{Ostapchenko:2010vb}%
  \BibitemOpen
  \bibfield  {author} {\bibinfo {author} {\bibfnamefont {S.}~\bibnamefont
  {Ostapchenko}},\ }\href {\doibase 10.1103/PhysRevD.83.014018} {\bibfield
  {journal} {\bibinfo  {journal} {Phys. Rev. D}\ }\textbf {\bibinfo {volume}
  {83}},\ \bibinfo {pages} {014018} (\bibinfo {year} {2011})},\ \Eprint
  {http://arxiv.org/abs/1010.1869} {arXiv:1010.1869 [hep-ph]} \BibitemShut
  {NoStop}%
\bibitem [{\citenamefont {Moskalenko}\ and\ \citenamefont
  {Strong}(1998)}]{Moskalenko:1997gh}%
  \BibitemOpen
  \bibfield  {author} {\bibinfo {author} {\bibfnamefont {I.~V.}\ \bibnamefont
  {Moskalenko}}\ and\ \bibinfo {author} {\bibfnamefont {A.~W.}\ \bibnamefont
  {Strong}},\ }\href {\doibase 10.1086/305152} {\bibfield  {journal} {\bibinfo
  {journal} {Astrophys. J.}\ }\textbf {\bibinfo {volume} {493}},\ \bibinfo
  {pages} {694} (\bibinfo {year} {1998})},\ \Eprint
  {http://arxiv.org/abs/astro-ph/9710124} {arXiv:astro-ph/9710124} \BibitemShut
  {NoStop}%
\bibitem [{\citenamefont {{Stephens}}\ and\ \citenamefont
  {{Badhwar}}(1981)}]{1981Ap&SS..76..213S}%
  \BibitemOpen
  \bibfield  {author} {\bibinfo {author} {\bibfnamefont {S.~A.}\ \bibnamefont
  {{Stephens}}}\ and\ \bibinfo {author} {\bibfnamefont {G.~D.}\ \bibnamefont
  {{Badhwar}}},\ }\href {\doibase 10.1007/BF00651256} {\bibfield  {journal}
  {\bibinfo  {journal} {\apss}\ }\textbf {\bibinfo {volume} {76}},\ \bibinfo
  {pages} {213} (\bibinfo {year} {1981})}\BibitemShut {NoStop}%
\bibitem [{\citenamefont {Efron}(1982)}]{efron1982jackknife}%
  \BibitemOpen
  \bibfield  {author} {\bibinfo {author} {\bibfnamefont {B.}~\bibnamefont
  {Efron}},\ }\href@noop {} {\emph {\bibinfo {title} {The jackknife, the
  bootstrap and other resampling plans}}}\ (\bibinfo  {publisher} {Society for
  industrial and applied mathematics},\ \bibinfo {year} {1982})\BibitemShut
  {NoStop}%
\bibitem [{\citenamefont {Schutz}\ \emph {et~al.}(2018)\citenamefont {Schutz},
  \citenamefont {Lin}, \citenamefont {Safdi},\ and\ \citenamefont
  {Wu}}]{Schutz:2017tfp}%
  \BibitemOpen
  \bibfield  {author} {\bibinfo {author} {\bibfnamefont {K.}~\bibnamefont
  {Schutz}}, \bibinfo {author} {\bibfnamefont {T.}~\bibnamefont {Lin}},
  \bibinfo {author} {\bibfnamefont {B.~R.}\ \bibnamefont {Safdi}}, \ and\
  \bibinfo {author} {\bibfnamefont {C.-L.}\ \bibnamefont {Wu}},\ }\href
  {\doibase 10.1103/PhysRevLett.121.081101} {\bibfield  {journal} {\bibinfo
  {journal} {Phys. Rev. Lett.}\ }\textbf {\bibinfo {volume} {121}},\ \bibinfo
  {pages} {081101} (\bibinfo {year} {2018})},\ \Eprint
  {http://arxiv.org/abs/1711.03103} {arXiv:1711.03103 [astro-ph.GA]}
  \BibitemShut {NoStop}%
\end{thebibliography}%

\end{document}